\documentclass[useAMS,usenatbib]{mn2e}

\usepackage{epsfig,graphicx}
\usepackage{amssymb,subeqnarray}
\usepackage[letterpaper,margin=1in]{geometry}
\usepackage{array}
\usepackage{rotating}

\begin{document}

\newcommand{\ns}{n_{\rm s}}
\newcommand{\ha}{{\rm H}\alpha}
\newcommand{\hb}{{\rm H}\beta}
\newcommand{\rphys}{R_{50,{\rm phys}}}
\newcommand{\kms}{{\rm km\, s}^{-1}}
\newcommand{\fobs}{f_{\rm obs}(\ha)}
\newcommand{\vc}{V_{\rm c}}
\newcommand{\vcobs}{V_{\rm c,obs}}
\newcommand{\vrot}{V_{\rm rot}}
\newcommand{\rto}{R_{\rm TO}}
\newcommand{\rd}{R_{\rm d}}
\newcommand{\qd}{q_{\rm d}}
\newcommand{\msyn}{M_{\rm syn}}
\newcommand{\qiso}{q_{\rm iso}}
\newcommand{\rb}{R_{\rm b}}
\newcommand{\qb}{q_{\rm b}}
\newcommand{\mgas}{M_{\rm gas}}
\newcommand{\mbar}{M_{\rm bar}}
\newcommand{\mmpa}{M_{\rm *,MPA}}
\newcommand{\mbell}{M_{\rm *,Bell}}
\newcommand{\tildesig}{\tilde{\sigma}}
\newcommand{\sigmeas}{\sigma_{\rm meas}}
\newcommand{\minc}{M_i^{\rm NC}}
\newcommand{\mdyn}{M_{\rm dyn}}
\newcommand{\mdynopt}{M_{\rm dyn,opt}}
\newcommand{\mstropt}{M_{\rm *,opt}}
\newcommand{\mdynratio}{M_{\rm dyn}/M_*}
\newcommand{\ropt}{R_{\rm opt}}
\newcommand{\sigmaopt}{\Sigma_{\rm *,opt}}
\newcommand{\fgeom}{f_{\rm geom}}
\newcommand{\vdiskratio}{V_{\rm d}/\vrot}
\newcommand{\vvir}{V_{\rm vir}}
\newcommand{\rvir}{R_{\rm vir}}
\newcommand{\sigmabar}{\Sigma_{\rm bar,opt}}
\newcommand{\mdynratiobar}{M_{\rm dyn}/M_{\rm bar}}
\newcommand{\mdynratioopt}{(M_{\rm dyn}/M_*)_{\rm opt}}
\newcommand{\mdynratiobaropt}{(M_{\rm dyn}/M_{\rm bar})_{\rm opt}}

\newcommand{\beq}{\begin{equation}}
\newcommand{\eeq}{\end{equation}}
\newcommand{\bef}{\begin{figure}}
\newcommand{\eef}{\end{figure}}
\newcommand{\bec}{\begin{center}}
\newcommand{\eec}{\end{center}}
\newcommand{\beqa}{\begin{eqnarray}}
\newcommand{\eeqa}{\end{eqnarray}}

\newcommand{\aaps}{{A\&AS}}
\newcommand{\araa}{{ARA\&A}}
\newcommand{\aap}{{A\&A}}
\newcommand{\apj}{{ApJ}}
\newcommand{\apjl}{{ApJL}}
\newcommand{\apjs}{{ApJS}}
\newcommand{\aj}{{AJ}}
\newcommand{\prd}{{PRD}}
\newcommand{\pasp}{{PASP}}
\newcommand{\mnras}{{MNRAS}}
\newcommand{\nat}{{Nature}}
\newcommand{\physrep}{Phys. Rep.}
\newcommand{\hMsun}{h^{-1}M_{\odot}}

\def\imagetop#1{\vtop{\null\hbox{#1}}}

\title{Calibrated Tully-Fisher relations for improved estimates of disk rotation velocities}
\author[Reyes et al.]{
  R. Reyes$^1$\thanks{\tt rreyes@astro.princeton.edu}, 
  R. Mandelbaum$^1$,
  J. E. Gunn$^1$,
  J. Pizagno$^2$,
  C.~N. Lackner$^1$
\\$^1$Peyton Hall Observatory, Princeton University, 
Peyton Hall, Princeton, NJ 08544, USA 
\\$^2$Kapteyn Astronomical Institute, University of Groningen, PO Box 800, 9700 AV Groningen, The Netherlands
}


\bibliographystyle{mn2e}
\maketitle

\label{firstpage}
\begin{abstract}

In this paper, we derive scaling relations between photometric observable quantities and disk galaxy rotation velocity $\vrot$, or Tully-Fisher relations (TFRs). Our methodology is dictated by our purpose of obtaining purely photometric, minimal-scatter estimators of $\vrot$ applicable to large galaxy samples from imaging surveys. To achieve this goal, we have constructed a sample of 189 disk galaxies at redshifts $z<0.1$ with long-slit H$\alpha$ spectroscopy from Pizagno et~al. (2007) and new observations. By construction, this sample is a fair subsample of a large, well-defined parent disk sample of $\sim 170~000$ galaxies selected from the Sloan Digital Sky Survey Data Release 7 (SDSS DR7). The optimal photometric estimator of $\vrot$ we find is stellar mass $M_*$ from Bell et~al. (2003), based on the linear combination of a luminosity and a colour. Assuming a Kroupa IMF, we find: $\log [V_{80}/(\kms)] = (2.142\pm 0.004)+(0.278\pm 0.010)[\log (M_*/M_\odot)-10.10]$, where $V_{80}$ is the rotation velocity measured at the radius $R_{80}$ containing 80 per cent of the $i$-band galaxy light.  This relation has an intrinsic Gaussian scatter $\tildesig=0.036\pm 0.005$ dex and a measured scatter $\sigmeas=0.056$ dex in $\log V_{80}$. For a fixed IMF, we find that the dynamical-to-stellar mass ratios within $R_{80}$, $(\mdynratio)(R_{80})$, decrease from approximately 10 to 3, as stellar mass increases from $M_* \approx 10^{9}$ to $10^{11} M_\odot$. At a fixed stellar mass, $(\mdynratio)(R_{80})$ increases with disk size, so that it correlates more tightly with stellar surface density than with stellar mass or disk size alone. We interpret the observed variation in $(\mdynratio)(R_{80})$ with disk size as a reflection of the fact that disk size dictates the radius at which $\mdynratio$ is measured, and consequently, the fraction of the dark matter ``seen'' by the gas at that radius. For the lowest $M_*$ galaxies, we find a positive correlation between TFR residuals and disk sizes, indicating that the total density profile is dominated by dark matter on these scales.  For the highest $M_*$ galaxies, we find instead a weak negative correlation, indicating a larger contribution of stars to the total density profile. This change in the sense of the correlation (from positive to negative) is consistent with the decreasing trend in $(\mdynratio)(R_{80})$ with stellar mass. In future work, we will use these results to study disk galaxy formation and evolution, and perform a fair, statistical analysis of the dynamics and masses of a photometrically-selected sample of disk galaxies.
\end{abstract}


\section{Introduction}
\label{sec:intr}

In the standard cold dark matter (CDM) based model for disk galaxy formation, gas cools out of a hot gaseous halo, maintaining its specific angular momentum, and forms a disk at the center of the potential well of a dark matter (DM) halo \citep{1980MNRAS.193..189F}. Although the basic picture has long been in place, a variety of physical processes underlying galaxy formation (such as star formation, feedback, angular momentum transfer, and mergers) is still poorly understood. 

It is thought that scaling relations between halo properties (e.g., halo mass and circular velocity) translate into scaling relations between observed galaxy properties (e.g., stellar mass and disk rotation velocity). In detail, however, what one expects for the final relations (including the slope, zero-point, and scatter) depends on still-uncertain aspects of galaxy formation.

Therefore, observed scaling relations for disk galaxies are expected to provide strong constraints on galaxy formation scenarios. In particular, the well-established relation between disk rotation velocity and galaxy luminosity \citep{1977A&A....54..661T} is arguably surprisingly tight, and hence, deserves special attention. In this work, we refer to scaling relations between disk rotation velocity and various galaxy properties (such as luminosity and stellar mass) generally as Tully-Fisher relations (TFRs). The purpose of studying these scaling relations has changed over the years, and the philosophy of sample selection and calibration has varied accordingly. 

Originally, TFRs were used to determine galaxy distances and measure deviations from the cosmic flow \citep[e.g.,][]{1977A&A....54..661T,1995PhR...261..271S,1997AJ....113...53G,2000ApJ...544..636C,2009yCat..21720599S}. Accordingly, galaxy samples were pruned on the basis of morphological type to minimize the scatter in the relation, and ensure the smallest possible error in magnitude and distance.

More recently, TFRs have been derived from galaxy samples including a broad range of morphological types \citep[e.g.,][hereafter, P07]{2007ApJ...671..203C,2007AJ....134..945P}, for the purpose of testing galaxy formation and evolution models \citep[e.g.,][]{1997ApJ...482..659D,1998MNRAS.295..319M,2007ApJ...654...27D,2007ApJ...671.1115G}. These TFR studies attempt to constrain the theoretical models, e.g., by quantifying, rather than minimizing, the scatter in the TFR. 

A third application of TFRs is connecting disk galaxies with their dark matter haloes through dynamics \citep[e.g.,][]{2010MNRAS.407....2D,2010arXiv1005.1289T}. Because long-slit spectroscopy is more expensive than imaging, one can exploit existing large imaging datasets by using TFRs to provide photometric estimates of disk rotation velocities $\vrot$. Unlike previous TFR studies, we tailor our sample selection and calibration explicitly for this purpose. We consider several different photometric quantities ${\cal O}$--- including absolute magnitudes, synthetic magnitudes, stellar mass estimates, and baryonic mass estimates--- to determine the one that yields the tightest relation with $\vrot$, and hence, is the optimal estimator of $\vrot$.

In a future paper, we will combine results from this optimal TFR with weak lensing measurements to constrain the ratio of rotation velocities at the optical and virial radii, $\vrot/\vvir$. To avoid sample selection issues, we want to use the same sample, or--- in practice--- similarly-defined samples, for the two analyses. The ratio $\vrot/\vvir$ provides a direct measurement of the slope of the total (dark + baryonic) mass profile within the virial radius. Measurement of $\vrot/\vvir$ can thus  constrain halo structure and indicate whether baryons have significantly modified the halo potential well \citep{2002MNRAS.334..797S,2007ApJ...654...27D}. 

\citet{2002MNRAS.334..797S} combined early TFR and weak lensing measurements and inferred $\vrot/\vvir =1.8$ with a 2$\sigma$ lower limit of 1.4, for $L^*$ late-type galaxies. They found this result to be consistent with the prediction of a model with adiabatic contraction of the dark matter halo due to baryonic infall as described in \citet{1986ApJ...301...27B}. More recently, \citet{2010MNRAS.407....2D} combined the TFR from P07 with halo mass measurements from different techniques (including weak lensing and satellite kinematics) and found $\vrot/\vvir \simeq 1$, for disk galaxies with stellar masses $10^{9.4}$--$10^{11.5} h^{-2} M_\odot$.
We aim to improve upon these measurements by carefully combining results from our TFR sample with weak lensing measurements on a similarly-selected lens galaxy sample, leading to a result that is not as severely affected by systematic effects due to differences in sample selection, among others.

We take advantage of the large, homogeneous dataset from the Sloan Digital Sky Survey \citep[SDSS;][]{2000AJ....120.1579Y} with well-defined photometry and available fibre spectroscopy. For the stacked weak lensing analysis, we will use a lens sample of $\sim 10^5$ disk galaxies at redshifts $z<0.1$; this is large enough to obtain decent $S/N$ measurements. For the TFR analysis, we use a {fair} subsample of the lens sample, spanning the region in the parameter space of luminosity, galaxy colour, and size that the lens sample occupies. Our TFR sample consists of a set of galaxies with existing rotation curve measurements from P07, augmented by a comparable number of galaxies for which we have obtained new rotation curve measurements. By construction, our TFR sample matches the stellar mass function of the parent disk sample, and extends out to the same maximum redshift of 0.1 (unlike that used in P07).

In addition to determining the slope, zero-point, and scatter in the TFR, we also study the residuals from the TFR, and their correlation with other galaxy properties, such as disk size. \citet{1999ApJ...513..561C} argued that the lack of correlation between TFR residuals and disk size indicates that stars do not dominate the potential well in the optical regions of disk galaxies \citep[but also see][]{2007ApJ...654...27D}. Both \citet{2007ApJ...671..203C} and P07 confirmed the lack of residual correlations in their galaxy samples. In this work, we go one step further and investigate residual correlations within different bins in stellar mass.

We also constrain the ratio of dynamical to stellar masses within the optical radius of disk galaxies, $(\mdynratio)(\ropt)$. We measure $\mdyn(\ropt)$ directly from the rotation velocity measurements, and infer $M_*$ from photometric estimates based on galaxy luminosity and colour \citep{2003ApJS..149..289B}.

This paper is organized as follows: in \S\ref{sec:data}, we describe previously existing data used in this work from SDSS (\S\ref{subsec:data_sdss}) and from P07 measurements (\S\ref{subsec:data_piz}). In \S\ref{sec:samp_sele}, we describe our sample selection. In the next three sections, we describe our derivation of the photometric and kinematic quantities used in the TFRs. In \S\ref{sec:bdfit}, we describe the bulge-disk decomposition fits from which we obtain disk parameters, such as disk size and axis ratio. In \S\ref{sec:phot}, we define the photometric quantities ${\cal O}$ that we consider as estimators of disk rotation velocity. In \S\ref{sec:long}, we describe the steps in the derivation of the kinematic quantities $\vrot$ from the long-slit spectroscopy observations. In \S\ref{subsec:long_syst}, we check for various systematic effects that may be affecting these measurements. 

In the next three sections, we analyse trends in these data. In \S\ref{sec:tfr_derive}, we describe the modelling, fitting, and interpreting the TFRs. In \S\ref{sec:itfr_calib}, we present calibrated TFRs for the different photometric quantities considered. In \S\ref{sec:alt_fits}, we present alternative fits to the TFRs. In \S\ref{sec:tfr_correl}, we study correlations between residuals from the TFR and galaxy properties, such as disk axis ratio, galaxy colour, and disk size. In \S\ref{sec:mass_ratios}, we calculate the dynamical-to-stellar mass ratios within the optical radius of the galaxies, and study its correlations with stellar mass, disk size, and stellar surface density. Finally, we summarize and discuss our main results in \S\ref{sec:summ}.

Throughout, we adopt a cosmology with $\Omega_{\rm m}=0.3$, $\Omega_\Lambda=0.7$, and $h\equiv H/(100\ \kms {\rm Mpc}^{-1})=0.7$, and express all lengths in physical (not comoving) coordinates. 

\section{Previously-existing Data}
\label{sec:data}

We use publicly-available data from the SDSS, and augment long-slit spectroscopy data obtained by P07 with a set of new observations (described in \S\ref{subsec:long_obs}). In \S\ref{subsec:data_sdss}, we describe SDSS imaging and spectroscopy, and define the SDSS photometric quantities used throughout this work. Then, in \S\ref{subsec:data_piz}, we describe the sample and long-slit spectroscopy data of P07.

\subsection{SDSS data}
\label{subsec:data_sdss}
The seventh SDSS data release \citep[DR7;][]{2009ApJS..182..543A} marks the completion of the SDSS-II phase of the survey and includes 11,663 deg$^2$ of imaging data, and over 9380 deg$^2$ of spectroscopy. The imaging was carried out by drift-scanning the sky in photometric conditions \citep{2001AJ....122.2129H, 2004AN....325..583I}, in five bands ($ugriz$) \citep{1996AJ....111.1748F,2002AJ....123.2121S} using a specially-designed wide-field camera \citep{1998AJ....116.3040G}. All of these data were processed by automated pipelines that detect and measure photometric properties of sources, and astrometrically calibrate the data \citep{2001adass..10..269L,2003AJ....125.1559P,2006AN....327..821T}. 

Objects are targeted for spectroscopy using the imaging data \citep{2003AJ....125.2276B}. Main galaxy sample targets are selected as described by \citet{2002AJ....124.1810S}. The Main galaxy sample target selection includes a \citet{1976ApJ...209L...1P} apparent magnitude cut of $r_{\rm P}=17.77$ mag, with slight variation in this cut across the survey area. Targets are observed with a 640-fibre spectrograph on the same telescope \citep{2006AJ....131.2332G}. These spectra are obtained through 3\arcsec-diameter fibres. In this work, SDSS fibre spectra are used only for sample selection (\S\ref{subsec:samp_parent}); the extraction of kinematic information from these spectra will be explored in future work.

In this work, we selected galaxy samples from the SDSS DR7 NYU-Value Added Galaxy Catalog \citep[NYU-VAGC;][]{2005AJ....129.2562B}. The selection criteria for inclusion in the VAGC is very similar to that for the Main galaxy sample. The imaging reductions used is a recent version, v5\_4 \citep[described in][]{2004AJ....128..502A}. The photometric calibration used is ``uber-calibration'', which utilizes overlaps of SDSS runs to obtain a more consistent large-scale photometric calibration of the survey \citep{2008ApJ...674.1217P}. 

We derive photometric quantities based on measurements from the SDSS imaging pipeline. These measurements include Petrosian and model apparent magnitudes, Petrosian half-light radii $R_{50}$, and isophotal axis ratios $\qiso$. We note that both $R_{50}$ and $\qiso$ are not corrected for seeing. $\qiso\equiv b_{25}/a_{25}$ or {\verb ISO_B/ISO_A } are measured at an isophote of 25 mag arcsec$^{-2}$. They provide a measure of the galaxy shape in its outer regions\footnote{The semimajor axis of the 25 mag arcsec$^{-2}$ isophote $a_{25}$ is around 2 to 3 times the effective radius of the galaxy \citep{2004ApJ...601..214R}.} and therefore a good initial estimate of disk axis ratios.
\footnote{In contrast, axis ratios from SDSS model fits are corrected for seeing, but are more severely affected by the light from the bulge than isophotal axis ratios.}
We will only use isophotal axis ratios for sample selection (c.f. \S\ref{subsec:samp_child}) and use seeing-corrected estimates of disk axis ratios and sizes from two-dimensional bulge-disk decompositions (described in \S\ref{sec:bdfit}) for the TFR analysis, in particular, for deriving inclination corrections to the rotation velocities. 

Absolute magnitudes are based on Petrosian apparent magnitudes, defined in \citet{2002AJ....124.1810S}, following the original proposal by \citet{1976ApJ...209L...1P}. The essential feature of Petrosian magnitudes is that in the absence of seeing, they measure a constant fraction of a galaxy's light regardless of distance (or size).
 
Galaxy colours are based on model apparent magnitudes, described in \citet{2002AJ....123..485S} and \citet{2004AJ....128..502A}. These magnitudes are based on fitting the two-dimensional PSF-convolved galaxy image with either a pure deVaucouleurs or a pure exponential profile, depending on which model has a higher likelihood based on the $r$-band galaxy image. The fits to the other bands use the same model parameters--- size, axis ratio, and position angle--- as in the $r$-band galaxy image, to get stable colours. 

We correct both absolute magnitudes and colours for Galactic extinction using the dust maps of \citet{1998ApJ...500..525S}, internal extinction (as described in \S\ref{subsec:int_ext_corr}), and $k$-corrections to $z=0$ using the {\verb kcorrect } product version {\verb v4_1_4 } of \citet{2007AJ....133..734B}.\footnote{In other words, galaxy colours $g-r \equiv M_{{\rm mod},g}-M_{{\rm mod},r}$, where $M_{{\rm mod},g}$ and $M_{{\rm mod},r}$ are the $g$ and $r$-band model absolute magnitudes corrected for both Galactic and internal extinction and $k$-corrected to $z=0$, respectively.} 
We denote absolute magnitudes and colours that have not been corrected for internal extinction with the superscript ``NC'', e.g., $M_r^{\rm NC}$, $M_i^{\rm NC}$, $(g-r)^{\rm NC}$. We reserve symbols without superscripts for quantities that have been corrected for internal extinction. 

In addition to these photometric quantities from the SDSS photometric pipeline, we also use \citet{1968adga.book.....S} indices $\ns$ provided with the NYU-VAGC, determined from radial profile fits performed by \citet{2005ApJ...629..143B}. 

\subsection{P07 data}
\label{subsec:data_piz}

P07 selected 234 target galaxies from the SDSS DR2 Main galaxy sample with redshifts $z<0.05$, absolute magnitudes $-22<M_r^{\rm NC}<-18.5$, and $r$-band isophotal axis ratios smaller than $0.6$. In addition, they imposed an absolute magnitude-dependent upper limit on redshift, so that the galaxies in their sample have apparent half-light radii larger than 2\arcsec. In this work, we use a different set of selection criteria to define a disk galaxy sample (see \S\ref{subsec:samp_parent}). 

P07 observations were carried out at the Calar Alto Observatory using the TWIN spectrograph mounted on the 3.5 m telescope with 1200 lines mm$^{-1}$ grating, 6200--7300\AA\ spectral coverage, and 1.5\arcsec\ slit width, and at the MDM Observatory using the CCDS spectrograph mounted on the 2.4 m Hiltner telescope with 600 lines mm$^{-1}$ grating, 6500--6994\AA\ spectral coverage, and 2\arcsec\ slit width. Integration times per galaxy vary from 1200 s for bright galaxies to three exposures of 1200 s for faint galaxies. We refer the reader to the original paper for further details about the long-slit spectroscopy observations and data reduction.

P07 obtained usable $\ha$ rotation curves for 162 out of their 234 target galaxies (69 per cent); the other 72 targets had insufficient $\ha$ line emission for obtaining rotation curves. Out of these 162 galaxies, we found that four were flagged for redshift measurement warnings, 
and one is not included in the NYU-VAGC. Out of the remaining 157 galaxies (hereafter referred to as the ``P07 galaxy sample''), 99 pass our sample selection criteria (described in \S\ref{subsec:samp_parent}) and are included in our analysis. For these galaxies, we re-processed the long-slit spectroscopy data obtained by P07 with the same pipeline used for data from our new observations.

\section{Sample Selection}
\label{sec:samp_sele}

As motivated in \S\ref{sec:intr}, we define two disk galaxy samples-- (i) a \textit{parent} disk sample of 169~563 galaxies, large enough to allow stacked weak lensing analysis, as well as detailed statistical analysis of galaxy properties, and (ii) a \textit{child} disk sample of 189 galaxies, a subsample of the parent disk sample with measured rotation curves for kinematic analysis. 

Insofar as this child disk sample is a fair subsample of the parent disk sample, results derived from it can be applied to the larger sample. Thus, we have made an effort to construct a child disk sample that spans the parameter space of galaxy properties--- colour, physical size, and galaxy type--- covered by the parent disk sample. Our child disk sample is composed of 99 galaxies from the P07 galaxy sample (see \S\ref{subsec:data_piz}) and 90 galaxies for which we acquired new long-slit spectra, totalling 189 galaxies with $\ha$ rotation curves. 

We describe the selection criteria used to define the parent disk sample in \S\ref{subsec:samp_parent}. We describe the construction of the child disk sample in \S\ref{subsec:samp_child}. Finally, we show that the distributions of galaxy properties are the same for the two samples in \S\ref{subsec:samp_comp}.

\subsection{Parent disk sample}
\label{subsec:samp_parent}

We define the parent disk sample to be a large sample of nearby, star-forming, disk galaxies from the SDSS, appropriate for weak lensing and detailed statistical analysis. Galaxy colour and S{\'e}rsic index are often used in the literature to select disk galaxies, but applying such cuts needlessly excludes a significant population of disk galaxies \citep{2009ApJ...691..394M}. Thus, we do not apply any cuts on galaxy colour, and apply only a conservative cut on S{\'e}rsic index. 

To select star-forming galaxies, we primarily rely on the $\ha$ emission-line strength observed through the SDSS fibre as a marker of star formation.\footnote{The 3\arcsec\ SDSS fibres probe the integrated light from the central regions of nearby galaxies (at redshifts of 0.02, 0.07, and 0.10, the fibre covers the central 1.2, 4.0, and 5.6 kpc, respectively).} 
Thus, we explicitly select galaxies that are likely to have sufficient $\ha$ emission for obtaining rotation curves. As a consequence, gas-poor S0s will be excluded automatically. This selection enabled us to efficiently use our observing time and obtain usable rotation curves for all of our targets (compared to only 69 per cent of targets for P07). 

We select the parent disk sample from the SDSS DR7 NYU-VAGC described in \S\ref{subsec:data_sdss}. To clean the sample, we exclude candidate objects that have spectroscopic redshift measurement warnings (0.8 per cent) and those that are not spectroscopically confirmed as having class ``GALAXY'' (2.5 per cent). This selection yields a total of 686~656 galaxies. 

We then apply the following selection criteria:
\begin{enumerate}
\item $0.02 < z < 0.10$ \\
\item $-22.5 < M_r^{\rm NC} < -18.0$\\ 
\item $\fobs > 2 \times 10^{-16} \mbox{ erg s}^{-1} {\rm cm}^{-2}$ \\ 
\item $0.5<\ns<5.9 \mbox{  \textbf{AND}  } \ns < 1.7 - (M_r + 18.0) $\\
\item $\log\left(\frac{{\rm [OIII]}5007}{\hb}\right) <\frac{0.61}{\log( {\rm [NII]6583}/\ha    )-0.47}+1.19$, \\ 
\end{enumerate}
where $z$ is the galaxy redshift, $M_r^{\rm NC}$ is the $r$-band absolute magnitude corrected for Galactic extinction and $k$-corrected to $z=0$ (but not corrected for internal extinction), $\fobs$ is the observed $\ha$ emission-line flux through the SDSS spectroscopic fibre, and $\ns$ is the $i$-band S{\'e}rsic index determined by \citet{2005ApJ...629..143B}. The final cut removes galaxies with active nuclei (AGN) and is based on the ratio of high- to low-ionization emission-line fluxes, as measured through the SDSS fibre. We describe each of these cuts in detail below.

{\bf (i) Redshift cut}. We apply a lower limit on redshift that ensures that the uncertainty in the absolute magnitude due to peculiar velocities ($\approx 300\ \kms$) is no larger than $5\log(1+300/6000)=0.11$ mag. We also note that galaxies with $z<0.02$ are not very useful for weak lensing because they would have negligible tangential shear compared to lens galaxies at higher redshifts. 

The maximum redshift of $0.1$ is a practical choice given the angular resolution required to obtain resolved rotation curves, as well as the need to obtain long-slit spectroscopic observations for the child disk sample within a reasonable time. After applying this cut, a total of 327~027 candidates remain. 

{\bf (ii) Absolute magnitude cut}. Following P07, the bright-end cut at $M_r^{\rm NC}=-22.5$ mag excludes many non-starforming (i.e., elliptical) galaxies that may contaminate the sample. This cut removes 10~852 candidates or 3.3 per cent of those from (i).

{\bf (iii) $\ha$ flux cut}. We apply a lower limit on the $\ha$ emission-line flux observed through the SDSS spectroscopic fibre. We measure the $\ha$ flux, $\fobs$, by simultaneously fitting a set of three Gaussians plus a linear continuum to the $\ha$ and [NII]6548,6583 lines (with the ratio of [NII]6548 to [NII]6583 flux fixed to 1:3 and all line widths required to be equal). 

We choose the lower limit to be the value that roughly divides the bimodal distribution in $\ha$ flux formed by the non-starforming and star-forming galaxy populations. This cut removes 37 per cent of candidates from (ii), leaving a total of 198~769 galaxies. 

We note that this cut introduces some redshift dependence. Since for nearby galaxies, the SDSS fibre is only sensitive to the central regions, some star-forming galaxies with extended $\ha$ emission but suppressed central $\ha$ emission would be rejected. 

{\bf (iv) S{\'e}rsic index cut}. We apply a conservative upper limit on the $i$-band S{\'e}rsic index that depends on absolute magnitude. This cut closely approximates the 90th percentile $\ns$ values at each $M_r^{\rm NC}$ (increasing linearly from 2.2 at $-18.5$ mag to 5.7 at $-22.0$ mag). We also restrict to galaxies with $0.5 < \ns < 5.9$ to ensure that the fits are not at the limits of the the allowed parameter space, $\ns \in [0.5,6.0]$. This cut removes 10~019 candidates or 5 per cent of those from (iii).

{\bf (v) AGN removal cut}. It is important to exclude active galaxies (AGN) from our sample to ensure that most or all of the observed $\ha$ emission comes from the disk, rather than the central engine, and can thus reliably trace the rotation of the disk. We use a standard diagnostic emission-line ratio diagram \citep{1981PASP...93....5B} and apply the theoretical emission-line ratio cut of \citet{2001ApJ...556..121K} to classify AGN. This cut is more conservative (removes fewer galaxies) than the empirical cut of \citet{2003MNRAS.346.1055K}. In line with being conservative, we only exclude galaxies that have $S/N>1$ on all six emission lines. This final cut removes 10 per cent of candidates from (iv),
\footnote{This is a reasonable fraction of AGN for a sample of emission-line galaxies. Out of $\sim 10^5$ SDSS galaxies, \citet{2004MNRAS.351.1151B} classified 68.4 per cent as emission-line galaxies, and 10.4 per cent of those as AGN.}
and yields our parent disk sample with 169~563 galaxies.

To measure the emission-line fluxes, we first subtract a best-fit stellar continuum from the SDSS spectrum, following the procedure of \citet{2005AJ....129.1783H}. We then perform a Gaussian fit to the continuum-subtracted spectrum similar to the procedure described above to measure the $\ha$ flux. We repeat the process to measure the $\hb$ and [OIII]4959,5007 lines (with the ratio of [OIII]4959 to [OIII]5007 flux fixed to 1:3). 

We have not applied an axis ratio cut to the parent disk sample, but we require $r$-band isophotal axis ratios $\qiso < 0.6$ on the child disk sample, following P07, to allow accurate inclination corrections to the galaxy rotation velocities. This cut is satisfied by 45 per cent of the parent disk sample, or 75~668 galaxies. 

\begin{figure*}
\includegraphics[width=6in]{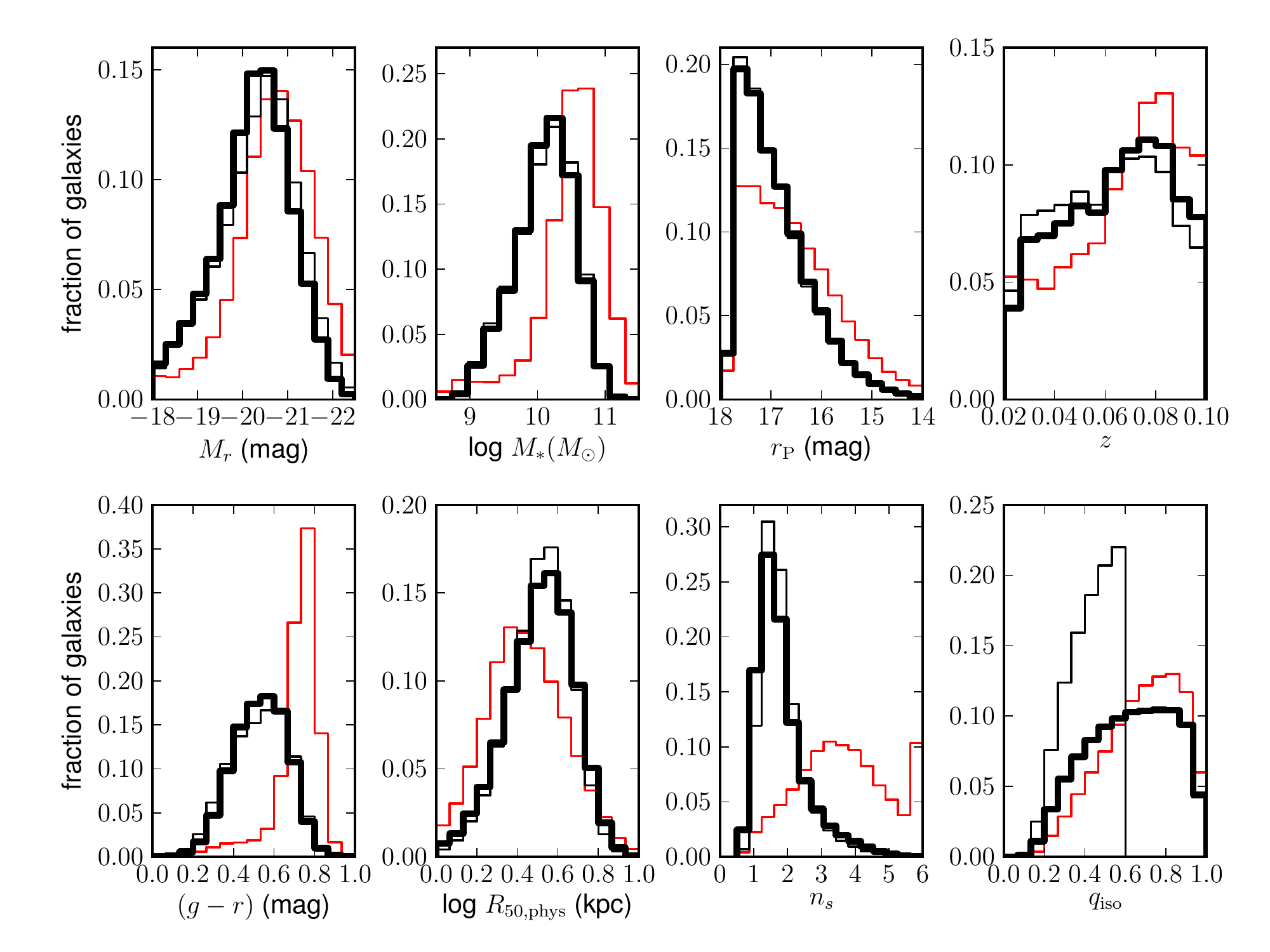}
\caption{Distribution of basic properties of galaxies in the parent disk sample (thick black histograms; 169~563 galaxies), those in the parent disk sample with isophotal axis ratios $\qiso<0.6$ (thin black histograms; 75~668 galaxies), and those galaxies that satisfy the redshift cut (i) but were excluded from the parent disk sample by the rest of the selection criteria (red histograms; 157~464 galaxies). Upper panels show absolute magnitudes $M_r$, stellar masses $M_*$, Petrosian apparent magnitudes $r_{\rm P}$, and redshifts $z$. Lower panels show $g-r$ colours, physical Petrosian half-light radii $\rphys$, S{\'e}rsic indices $\ns$, and isophotal axis ratios $\qiso$.}
\label{fig:samp_hist_parentdisk}
\end{figure*}

Figure~\ref{fig:samp_hist_parentdisk} shows the distributions of galaxy properties for the parent disk sample galaxies (all galaxies and those with $\qiso<0.6$ are shown in thick and thin black histograms, respectively). Also shown are the distributions for galaxies that satisfy the redshift cut (i) but were excluded by the rest of the selection criteria (red histograms). Panels show the following galaxy properties: absolute magnitude $M_r$, stellar mass $M_*$, Petrosian apparent magnitude $r_{\rm P}$, redshift $z$, $g-r$ colour, physical $i$-band Petrosian half-light radius $\rphys$, $i$-band S{\'e}rsic index $\ns$, and $r$-band isophotal axis ratio $\qiso$. Stellar masses $M_*$ correspond to the \citet{2002Sci...295...82K} IMF and were determined from $i$-band absolute magnitudes and $g-r$ colours, using stellar mass-to-light ratio estimates from \citet{2003ApJS..149..289B}, as described in \S\ref{subsubsec:bell_mstr}. The other photometric quantities were defined in \S\ref{subsec:data_sdss}.

The distributions in $g-r$ and $\ns$ indicate that our selection criteria have successfully excluded the population of non-starforming (elliptical) galaxies, which constitute the red peak in $g-r$ colour at $\sim 0.75$ mag and the broad bump of high S{\'e}rsic indices around $\ns \sim 3.5$ (see red histograms in Fig.~\ref{fig:samp_hist_parentdisk} for the excluded galaxies). Recall that our selection relies primarily on the presence of the $\ha$ line emission and does not impose any strict cuts in either colour or S{\'e}rsic index. 

\subsection{Child disk sample}
\label{subsec:samp_child}

We aim to construct a child disk sample that is a fair subsample of the parent disk sample for use in our kinematic analysis. By combining existing data from P07 with newly-acquired observations, we have constructed a child disk sample of 189 disk galaxies with usable $\ha$ rotation curves. 

Of the 157 galaxies in the P07 galaxy sample, 99 pass our selection criteria (in \S\ref{subsec:samp_parent}) and are included in the child disk sample. Of the 58 rejected galaxies, 3 failed due to the absolute magnitude cut, 17 due to the $\ha$ flux cut, 11 due to the S{\'e}rsic index cut, and 27 due to the AGN cut. The AGN fraction of 17 per cent is larger than the overall fraction of 10 per cent (c.f. criterion (v) in \S\ref{subsec:samp_parent}). This may be attributed to the tendency for galaxies in the P07 sample to be of earlier type, and therefore more likely to host AGN \citep[][but also see \citealt{2010ApJ...720..368X}]{2008ApJ...681..931B}.

\begin{figure*}
\bec $
\begin{array}{cc} 
\includegraphics[width=3in]{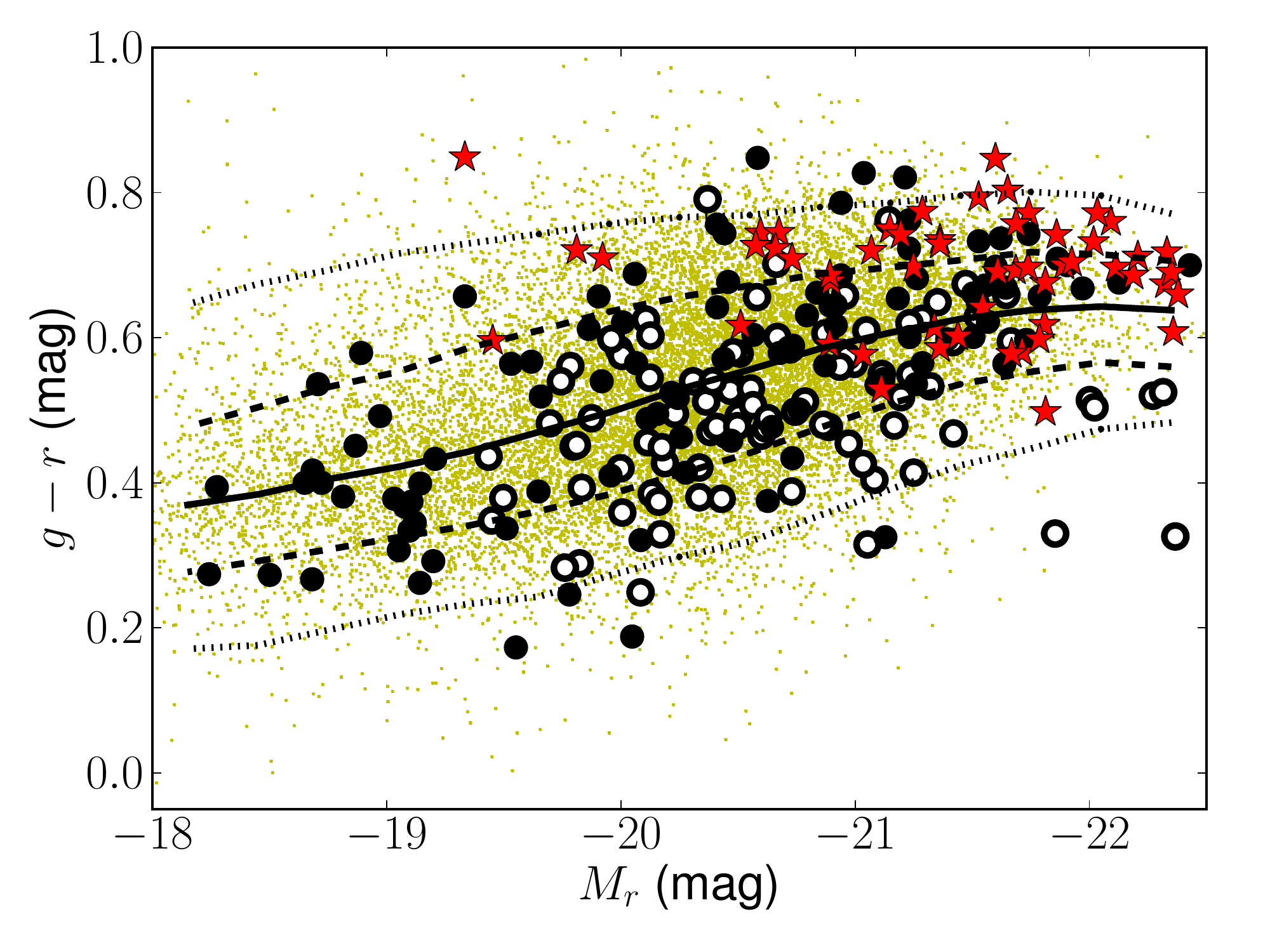} &
\includegraphics[width=3in]{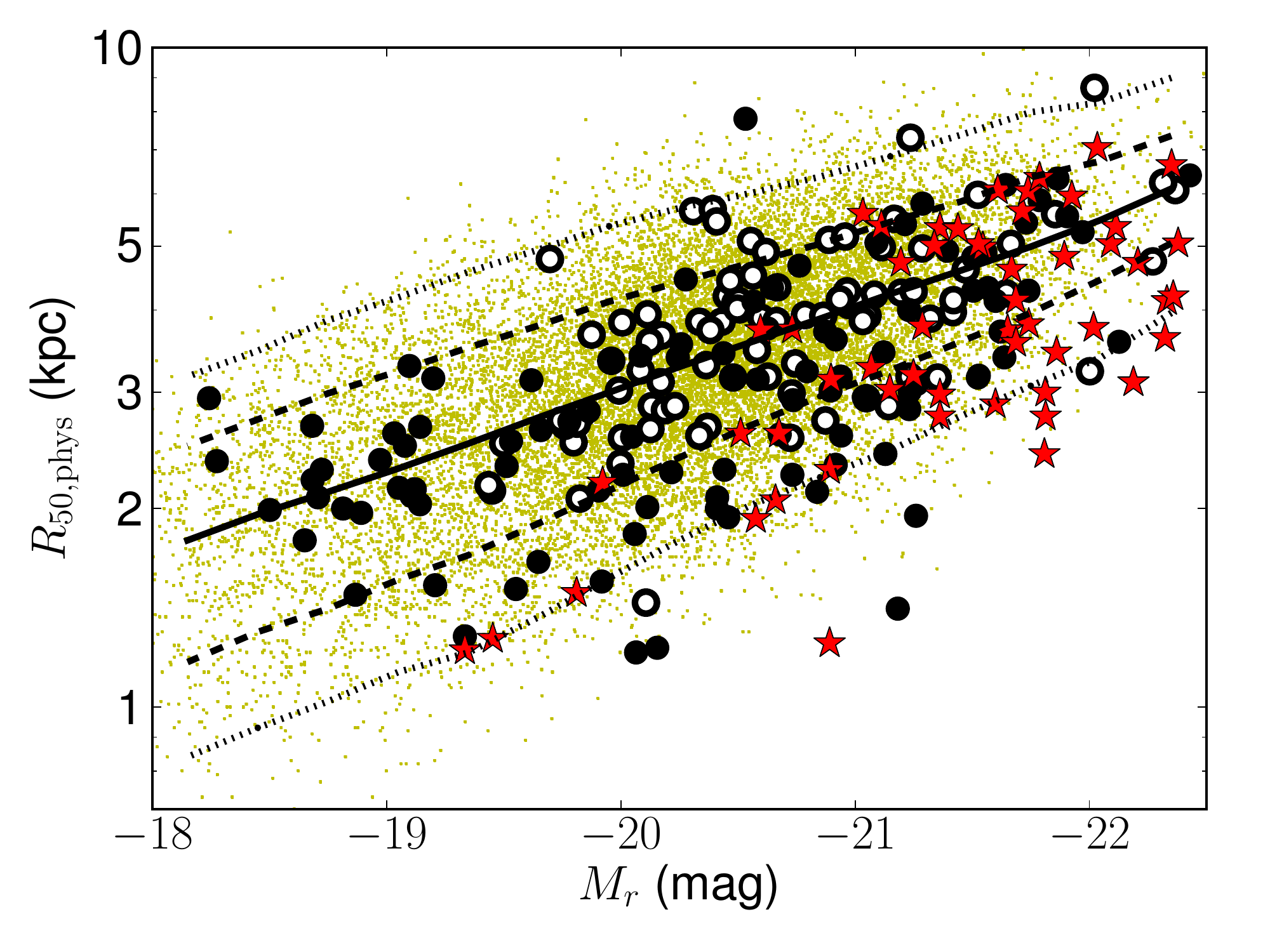}
\end{array} $
\eec
\caption{Distribution in the parameter space of absolute magnitude $M_r$, $g-r$ colour (left panel) and physical half-light radius $\rphys$ (right panel) of 189 galaxies in the child disk sample. Newly-observed galaxies are shown as open circles (90 galaxies), and previously-observed galaxies from P07 are shown as filled circles (99 galaxies). Galaxies in the parent disk sample (169~563 galaxies) are shown as yellow dots, and their median relations and 1- and 2-$\sigma$ percentile bounds are shown in solid, dashed, and dotted lines, respectively. For comparison, galaxies from the P07 galaxy sample that failed our selection criteria (58 galaxies) are shown as red stars.}
\label{fig:samp_param}
\end{figure*}

The other 90 galaxies in the child disk sample have been selected from the parent disk sample to fill in regions of the parameter space in luminosity, colour, and size that were not well covered by the existing dataset (see Fig.~\ref{fig:samp_param}). We require these galaxies to have $r$-band isophotal axis ratios $\qiso<0.6$, following P07, to avoid large inclination corrections to the rotation velocities. We have targeted and acquired long-slit observations of these galaxies, as described in \S\ref{subsec:long_obs}. 

Table~\ref{tab:samp_child} lists the 189 galaxies in the child disk sample, together with their basic properties and the instrument used in their observation. For the five galaxies with repeat observations, instruments from both observations are listed.

\begin{table*}
\begin{center}
\caption{Basic properties of the 189 galaxies in the child disk sample.}
\label{tab:samp_child}
\begin{tabular}{lrrrrrrrl}
\hline \\
 \multicolumn{1}{c}{Galaxy name} & 
 \multicolumn{1}{c}{Spec. ID} & 
 \multicolumn{1}{c}{$z$} & 
 \multicolumn{1}{c}{$M_r$} & 
 \multicolumn{1}{c}{$g-r$} & 
 \multicolumn{1}{c}{$\rphys$} & 
 \multicolumn{1}{c}{$\ns$} & 
 \multicolumn{1}{c}{$\qiso$} & 
 \multicolumn{1}{c}{Instr.} \\
  \multicolumn{3}{c}{ } &
  \multicolumn{1}{c}{(mag)} &
  \multicolumn{1}{c}{(mag)} &
  \multicolumn{1}{c}{(kpc)} &
  \multicolumn{3}{c}{ } \\
 \multicolumn{1}{c}{(1)} & 
 \multicolumn{1}{c}{(2)} & 
 \multicolumn{1}{c}{(3)} & 
 \multicolumn{1}{c}{(4)} & 
 \multicolumn{1}{c}{(5)} & 
 \multicolumn{1}{c}{(6)} & 
 \multicolumn{1}{c}{(7)} & 
 \multicolumn{1}{c}{(8)} &
  \multicolumn{1}{c}{(9)}\\ 
\hline \\
 SDSS J001006.61$-$002609.7 & 0389-303-51794 &  0.0321 & $ -21.13$\,($   0.10$) & $   0.32$\,($   0.03$) & $   2.42$\,($   0.08$) &                $ 1.22$ &                $ 0.42$ &      TWIN\\
 SDSS J001708.75$-$005728.9 & 0390-300-51900 &  0.0189 & $ -19.62$\,($   0.13$) & $   0.57$\,($   0.04$) & $   3.13$\,($   0.17$) &                $ 1.45$ &                $ 0.57$ &      TWIN\\
   SDSS J002844.82+160058.8 & 0417-329-51821 &  0.0947 & $ -22.37$\,($   0.23$) & $   0.33$\,($   0.06$) & $   6.08$\,($   0.06$) &                $ 1.71$ &                $ 0.17$ &       DIS\\
 SDSS J003112.09$-$002426.4 & 0391-063-51782 &  0.0194 & $ -20.41$\,($   0.12$) & $   0.76$\,($   0.03$) & $   2.00$\,($   0.10$) &                $ 2.95$ &                $ 0.53$ &      TWIN\\
   SDSS J004916.23+154821.0 & 0419-602-51879 &  0.0846 & $ -20.31$\,($   0.07$) & $   0.54$\,($   0.04$) & $   5.65$\,($   0.07$) &                $ 1.37$ &                $ 0.53$ &       DIS\\
   SDSS J004935.71+010655.2 & 0394-380-51913 &  0.0176 & $ -21.21$\,($   0.26$) & $   0.82$\,($   0.03$) & $   5.40$\,($   0.31$) &                $ 2.42$ &                $ 0.17$ &      TWIN\\
   SDSS J011750.26+133026.3 & 0423-044-51821 &  0.0326 & $ -20.01$\,($   0.07$) & $   0.36$\,($   0.03$) & $   3.83$\,($   0.12$) &                $ 1.44$ &                $ 0.53$ &       DIS\\
 SDSS J012317.00$-$005421.6 & 0399-254-51817 &  0.0259 & $ -20.92$\,($   0.10$) & $   0.62$\,($   0.03$) & $   2.33$\,($   0.09$) &                $ 2.02$ &                $ 0.39$ &      TWIN\\
   SDSS J012340.12+004056.4 & 0399-478-51817 &  0.0334 & $ -21.23$\,($   0.14$) & $   0.60$\,($   0.03$) & $   3.01$\,($   0.09$) &                $ 2.35$ &                $ 0.32$ &      TWIN\\
 SDSS J012438.08$-$000346.5 & 0399-178-51817 &  0.0277 & $ -20.92$\,($   0.14$) & $   0.64$\,($   0.03$) & $   3.61$\,($   0.13$) &                $ 1.95$ &                $ 0.35$ &      TWIN\\
\hline
\end{tabular}
\end{center}
\begin{flushleft}
Notes. --- {
Col. (1): SDSS name of galaxy. 
Col. (2): Spectroscopic ID: plate, fiber ID, and MJD of SDSS fibre spectrum. 
Col. (3): Redshift. 
Col. (4): $r$-band Petrosian absolute magnitude, $k$-corrected to $z=0$, and corrected for Galactic and internal extinction. 
Col. (5): $g-r$ model colour, $k$-corrected to $z=0$, and corrected for Galactic and internal extinction. 
Col. (6): physical $i$-band Petrosian half-light radius.
Col. (7): $i$-band S{\'e}rsic index. 
Col. (8): $r$-band isophotal axis ratio. 
Col. (9): Instrument(s) used for observation-- TWIN/CCDS (P07), DIS (this work). 
Quantities in parentheses are 1-$\sigma$ uncertainties. The uncertainty in $\rphys$ includes a distance uncertainty assuming a peculiar velocity of 300 $\kms$.}
\end{flushleft}
\end{table*}%

Figure~\ref{fig:samp_param} shows the distribution of child disk sample galaxies in the parameter space of absolute magnitude $M_r$, $g-r$ colour, and physical half-light radius $\rphys$ (open and filled circles), compared with that of parent disk sample galaxies (yellow dots). Also shown are the median relations (solid lines) and the 1- and 2-$\sigma$ percentile limits (dashed and dotted lines, respectively) for the parent disk sample. As desired, the child disk sample spans the parameter space covered by the parent disk sample fairly well over most of the luminosity range covered. Moreover, 68 and 93 per cent of galaxies in the child disk sample lie within the 1- and 2-$\sigma$ percentile bounds of the parent disk sample, close to the expected values of 68 and 95 per cent, respectively.

In Fig.~\ref{fig:samp_param}, previously-observed and newly-observed galaxies are shown as filled and open circles, respectively. The former cover the bright and faint ends, while the latter fill in the region of intermediate absolute magnitudes $-21 \la M_r \la -19.5$ mag, by construction. Galaxies from the P07 galaxy sample that failed our selection criteria are shown as red stars. These excluded galaxies tend to be more luminous and redder than the mean population (i.e., earlier-type disks). 

\begin{figure*}
\includegraphics[width=6in]{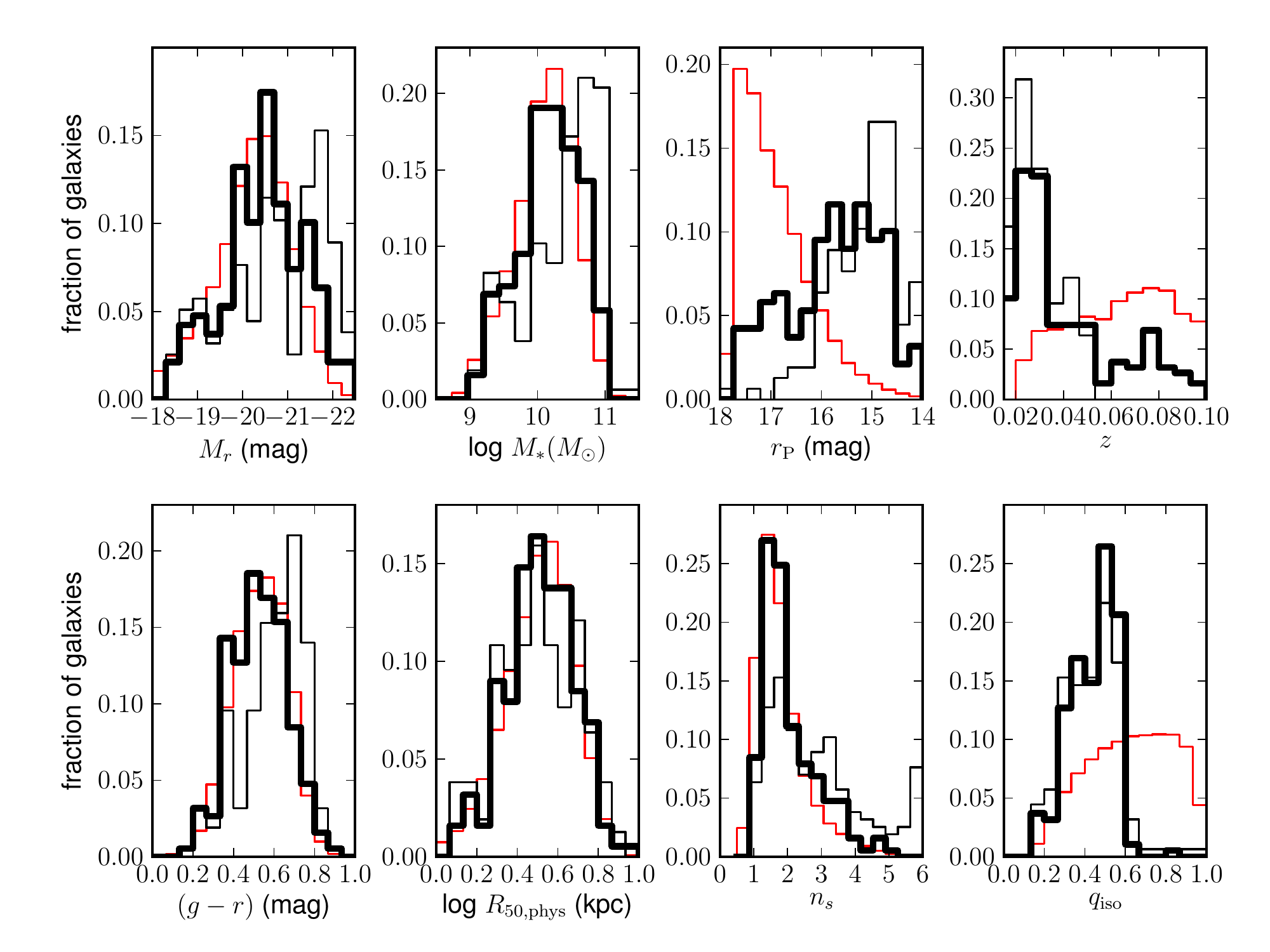}
\caption{Distribution of basic properties for galaxies in the child disk sample (thick black histograms; 189 galaxies), P07 galaxy sample (thin black histograms; 157 galaxies), and parent disk sample (red histograms; 169~563 galaxies). As in Fig.~\ref{fig:samp_hist_parentdisk}, upper panels show absolute magnitudes $M_r$, stellar masses $M_*$, Petrosian apparent magnitudes $r_{\rm P}$, and redshifts $z$. Lower panels show $g-r$ colours, physical Petrosian half-light radii $\rphys$, S{\'e}rsic indices $\ns$, and isophotal axis ratios $\qiso$.}
\label{fig:samp_hist_child}
\end{figure*}

\begin{table}
\begin{center}
\caption{Mean galaxy properties for the parent disk, child disk, and P07 galaxy samples.}
\label{tab:samp_means}
\begin{tabular}{lrrrrr}
\hline \\
\multicolumn{1}{c}{Sample} &
\multicolumn{1}{c}{$N$} &
\multicolumn{1}{c}{$\langle z\rangle$} &
\multicolumn{1}{c}{$\langle M_r\rangle$} &
\multicolumn{1}{c}{$\langle g-r\rangle$} &
\multicolumn{1}{c}{$\langle R_{50}\rangle$}\\
\multicolumn{1}{c}{} &
\multicolumn{1}{c}{} &
\multicolumn{1}{c}{} &
\multicolumn{1}{c}{(mag)} &
\multicolumn{1}{c}{(mag)} &
\multicolumn{1}{c}{(kpc)} \\
\hline \\
Parent & $169~563$ &  $  0.0637$ &   $-20.23$ &  $   0.53$ &   $   3.35$ \\
Child &  $   189$ &  $  0.0407$ &   $-20.50$ &  $   0.53$ &   $   3.49$ \\
P07 &  $   157$ &  $  0.0289$ &   $-20.70$ &  $   0.61$ &   $   3.43$ \\ 
\hline
\end{tabular}
\end{center}
\end{table}

Figure~\ref{fig:samp_hist_child} compares the distribution of galaxy properties for the child, P07, and parent disk galaxy samples (thick black, thin black, and red histograms, respectively). Table~\ref{tab:samp_means} compares mean values of some basic galaxy properties for the three samples. 

Figures~\ref{fig:samp_param} and \ref{fig:samp_hist_child} show that the P07 galaxy sample is skewed toward brighter luminosities, higher stellar masses, redder $g-r$ colours, and higher S{\'e}rsic indices, compared to the parent and child disk samples. By construction, the child disk sample follows the distribution of the parent disk sample more closely. However, because the child disk sample includes many galaxies from the P07 galaxy sample, it retains a slight bias towards brighter luminosities, higher stellar masses, and higher S{\'e}rsic indices (but not redder colours). 

\subsection{Child vs. parent disk sample}
\label{subsec:samp_comp}


We check that the child disk sample is a fair subsample of the parent disk sample once the small difference in their luminosity distributions are taken into account. To do this, we construct ten random subsamples of the parent disk sample with the same number of galaxies and the same $M_r$ distribution (in 0.3 mag wide bins) as the child disk sample. We find that the distributions in $g-r$ colour, S{\'e}rsic index $\ns$, and physical half light radius $\rphys$ for the child disk sample are consistent with the mean distributions for the ten random subsamples of the parent disk sample within the sampling variance (as can be gleaned from Fig.~\ref{fig:samp_hist_child}).

We also perform a Kolmogorov-Smirnov (KS) test to check whether each pair of distributions is consistent with being drawn from the same underlying population. We find mean D-statistic values equal to 0.12, 0.11, and 0.14 for the distributions in $g-r$, $\ns$, and $\rphys$, respectively ($D=0$ if the two distributions are identical). The large values of the corresponding probabilities $P(D>\mbox{observed})$--- 15, 26, and 9 per cent, respectively--- indicate that we cannot reject the null hypothesis that the two distributions were drawn from the same population. 

In contrast, the P07 galaxy sample is highly inconsistent with being drawn from the same distribution as the parent disk sample. Performing the same matching in absolute magnitude distributions and Kolmogorov-Smirnov tests, we find D-statistic values equal to 0.20, 0.27, and 0.24, all with probabilities $P(D>\mbox{observed})$ much less than 1 per cent--- $7\times 10^{-3}$, $9\times 10^{-5}$, and $8\times 10^{-4}$. 

\section{Bulge-disk decompositions}
\label{sec:bdfit}

First, we describe our motivation for performing bulge-disk decomposition (B+D) fits and how these results are used in this work (\S\ref{subsec:bdfit_motivation}). Then, we describe the B+D fitting procedure (\S\ref{subsec:bdfit_methodology}). Finally, we show example fits and discuss some results (\S\ref{subsec:bdfit_results}).

\subsection{Motivation and use}
\label{subsec:bdfit_motivation}

We perform two-dimensional B+D fits of galaxies in the child disk galaxy sample to determine their disk sizes and axis ratios, and enable accurate rotation velocity estimates. The fits also provide disk-to-total light ratios, $D/T$, and surface brightness profiles, which we will present and analyze in future work. 

In this work, we use only three basic parameters from the B+D fits:
\begin{itemize}
\item $\qd$: disk axis ratio
\item $\rd$: disk scale length
\item $R_{80}$: radius containing 80 per cent of the total galaxy light 
\end{itemize}
We fit to the SDSS $i$-band galaxy images because this band is less affected by dust than the $g$ and $r$ bands (while the $z$ band has noisier photometry). 

Disk axis ratios $\qd$ from the B+D fits are more accurate than the isophotal axis ratios $\qiso$ from the SDSS photometric pipeline (used in our sample selection), because the latter is affected by the presence of a bulge and variations in seeing conditions. We therefore use $\qd$ to determine internal extinction corrections to absolute magnitudes and colours (as described in \S\ref{subsec:int_ext_corr}). More importantly, we use $\qd$ to determine inclination corrections to the observed rotation velocities (c.f. Eq.~\ref{eq:sin_theta}). 

\subsection{Fitting procedure}
\label{subsec:bdfit_methodology}

Numerous studies have performed two-dimensional bulge-disk decomposition of galaxy light
profiles, in order to quantify their morphological
properties \citep{2009MNRAS.393.1531G,2007MNRAS.379..841B,2006MNRAS.371....2A,2002AJ....124..266P}. 
However, a bulge-disk decomposition into an exponential disk plus a S{\'e}rsic bulge requires a fit with a large number (12) of
free parameters, many of which are degenerate, in particular, the bulge
S{\'e}rsic index, the bulge effective radius, and the bulge central
surface brightness.  

Fortunately, the surface density profiles of nearby disk galaxies are adequately
described, on average, by a double-exponential profile \citep{2003ApJ...582..689M,2008MNRAS.388.1708G,2011arXiv1105.0002M}.\footnote{From their B+D fits of nearby disk galaxies, \citet{2003ApJ...582..689M} found bulge S{\'e}rsic indices ranging from $0.1$ to 2, with a mean very close to 1.} 
Since we are not interested in the detailed morphology of galaxy bulges in this work, we have 
chosen to fix the bulge S{\'e}rsic index to 1, yielding more robust fits to the bulge and disk.\footnote{The degeneracies are not completely removed, but fixing the bulge S{\'e}rsic index decreases the covariance between bulge flux normalization and bulge scale length.} 

We fit the $i$-band galaxy images with a double-exponential distribution:
\beq
I(R) = I_{\rm d} \exp(-R/\rd) + I_{\rm b} \exp(-1.68 R/R_{\rm eff}),
\eeq
where $\rd$ is the disk scale length, $R_{\rm eff}$ is the effective
radius of the bulge (equal to 1.68 times its scale length), and $I_{\rm
d}$, $I_{\rm b}$ are the central surface brightness of the disk and
bulge components, respectively. The radial coordinate $R$, is given by
an ellipse centered at $(x_0,y_0)$
\beq
R = \left( (x-x_0)^2 + ((y-y_0)/q_{(b,d)})^2 \right)^{1/2}\ ,
\eeq
where $q_{(b,d)}$ denotes the axis ratios of the bulge and disk,
respectively. As with the SDSS exponential models, the surface brightness of each component 
is truncated at 4 times the effective radius, with a smooth fall-off to zero between 3 and 4 effective radii.

We use the SDSS DR7 $i$-band atlas images from the most recent
reductions (\verb+v5_6+) \citep{2011ApJS..193...29A}. 
These are sky-subtracted images\footnote{Improper sky subtraction will affect the B+D fits. The general trend in SDSS DR7 (and DR8) is for the sky to be overestimated near bright, large galaxies, which leads to underestimation of the galaxies' sizes and luminosities. These sky subtraction issues are not expected to affect the typical galaxy in our sample.} 
of typically less than $200\times200$ pixels. The fits are done by minimizing the weighted
difference between the image and the model convolved with the PSF at the galaxy position. 
The minimizer used is the Levenberg-Marquardt minimizer \verb+mpfit2dfun+ in IDL
\citep{2009ASPC..411..251M}. Each pixel is weighted by its inverse variance, which is computed using the locally measured
sky background counts from SDSS for each galaxy. 

The initial conditions for the fit are taken from the SDSS measurements of \verb+R_EXP+, the
exponential scale length, and \verb+AB_EXP+, the exponential fit axis
ratio, and the total flux of the atlas image. The initial axis ratio and position angle of the bulge 
are chosen to be the same as those of the disk. The initial scale length of the bulge is set to 40\% of the disk,
and the initial $D/T$ is set to 0.6. The fits are robust against changes in the initial conditions, 
specifically changes in the initial bulge size and $D/T$.

In order to compare the model to the atlas image during fitting, the model is 
convolved with the locally measured point-spread-function from the
SDSS pipeline. For each galaxy, the fitter returns the best-fit
parameters, the covariance matrix of these parameters, and
the reduced $\chi^2$ value for the fit. Since each fit consists of two
exponential profiles, we chose the exponential with the smaller scale
length to be the bulge.

Our model consists of two axisymmetric exponential profiles, so we are 
unable to fit any non-axisymmetric features such as spiral arms, bars, and dust lanes. 
For galaxies in which these features
are very prominent, the model is inadequate. Additionally, our model is inadequate
for highly-inclined disks, where the semi-minor axis of the galaxy
image ($\rd\times q_d$) is comparable to the disk scale height. At
inclinations of $\sim 70^\circ$, the errors in the surface brightness
are between $5\%$ and $10\%$, for reasonable disk scale heights. 
If the bulge of such a disk galaxy is disk-like (i.e., a pseudo-bulge) but is not highly inclined, it is still
possible to fit it accurately. 

\subsection{Results}
\label{subsec:bdfit_results}

\begin{figure} 
\bec $
\begin{array}{c} 
\includegraphics[width=3in]{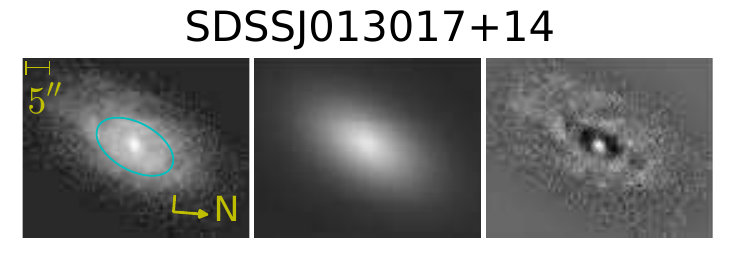} \\
\includegraphics[width=3in]{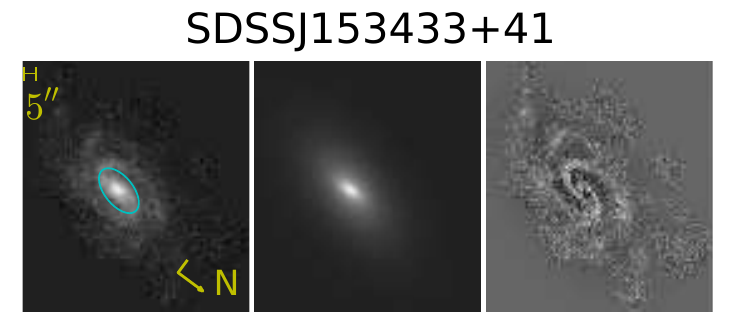} \\
\includegraphics[width=3in]{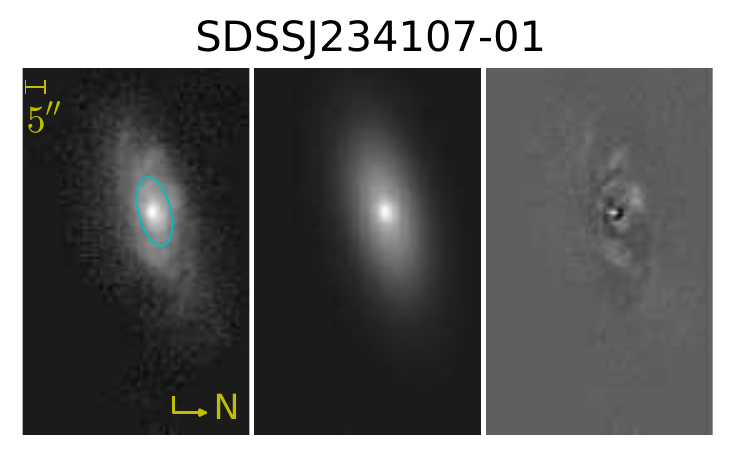} \\
\includegraphics[width=3in]{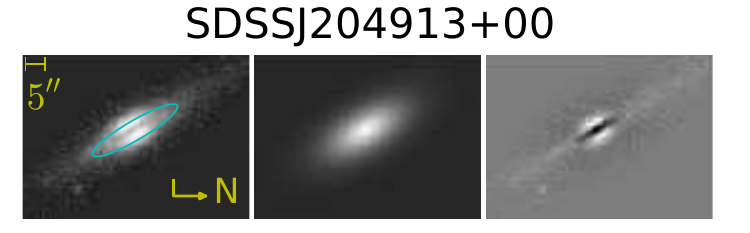}
\end{array} $
\eec
\caption{Example two-dimensional bulge-disk decomposition fits. Left to right panels: SDSS $i$-band image, model image (exponential disk plus exponential bulge, convolved with the PSF), and residual image (data$-$model). The blue ellipse on the SDSS image marks the ellipse containing 80 per cent of the flux, with semi-major axis $R_{80}$ and axis ratio equal to that of the model disk, $\qd$. Both SDSS and model images are logarithmically scaled (with the same stretch), while the residual image is linearly scaled. Reading down the figure, the maximum (minimum) residual equals 36\% (-25\%), 18\% (-11\%), 15\% (-9\%), and 47\% (-46\%) of the central model brightness for each galaxy.}
\label{fig:bdfit_images}
\end{figure}

\begin{figure} 
\includegraphics[width=3in]{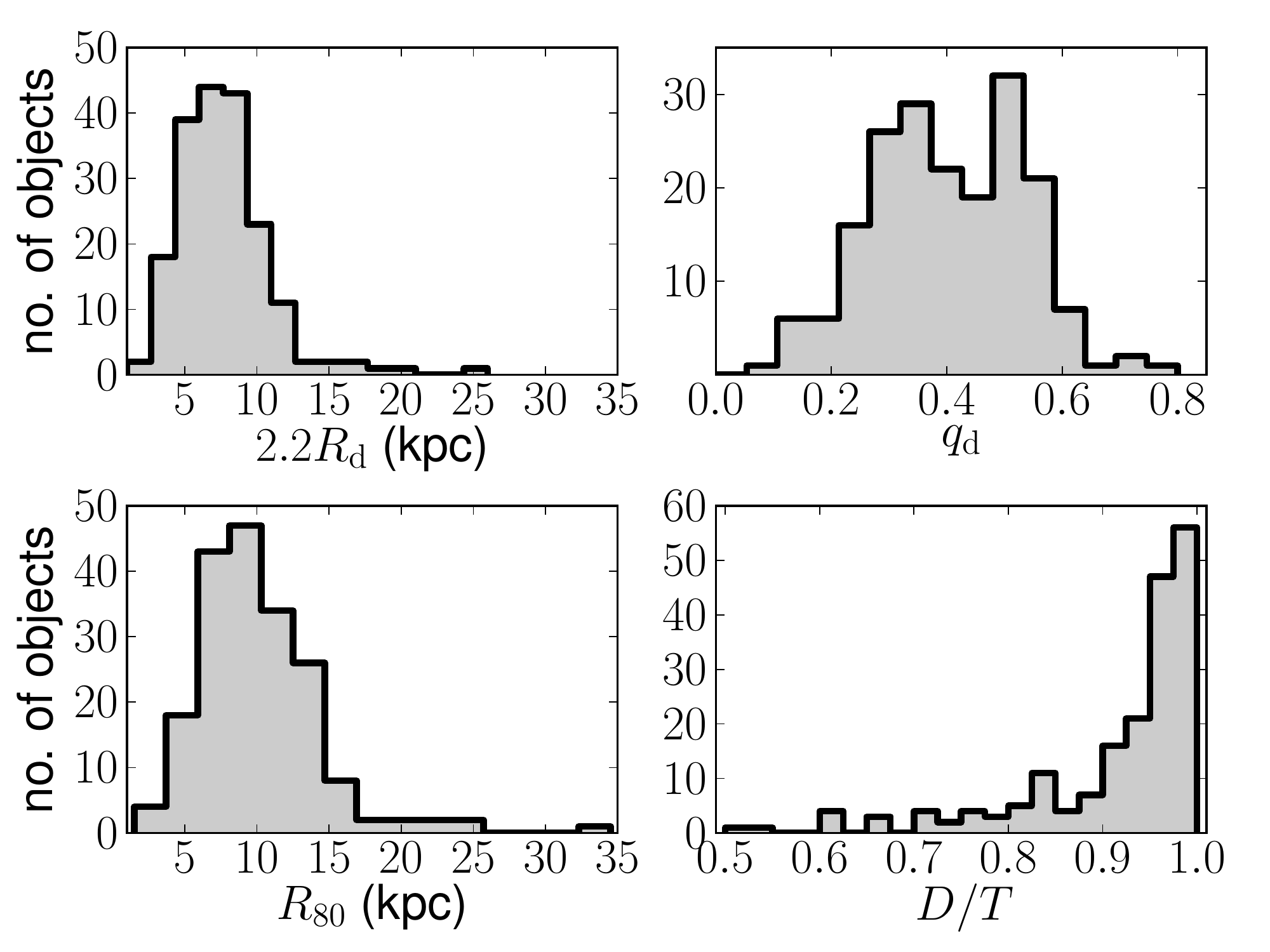}
\caption{Distribution of $i$-band B+D fit parameters-- disk sizes, $2.2\rd$ and $R_{80}$, disk axis ratios, $\qd$, and disk-to-total ratios, $D/T$--- for 189 galaxies in the child disk sample.} 
\label{fig:hist_bdpar}
\end{figure}

Panels in Figure~\ref{fig:bdfit_images} shows the observed, model, and residual images (from left to right) from B+D fits of four galaxies (from top to bottom), chosen for illustration: SDSSJ013017+14 and SDSSJ153433+41 are disk-dominated galaxies, with $D/T=0.99$ and 1.00, respectively; SDSSJ234107$-$01 has a more substantial bulge, with $D/T$=0.81; SDSSJ204913+00 is an extreme case--- it has an edge-on disk and a substantial bulge; the disk light is obscured by dust, resulting in an extremely low reported $D/T$ of 0.54.

We use both $2.2\rd$ and $R_{80}$ to set the radius at which to define the disk rotation velocity amplitude (see \S\ref{subsec:vrot_definition}). We note that $R_{80}$ is less sensitive to any degeneracies in the B+D fit. The ratio $R_{80}/(2.2\rd)$ clearly depends on the relative size of the bulge and disk. Its value is 1.4 for a pure exponential disk ($D/T=1$), and decreases smoothly down to 1.2, 1.1, and 1.0 for $D/T= 0.8$, 0.7, and 0.6, respectively.

Table~\ref{tab:child_bdpar} lists the best-fit parameters from the $i$-band B+D fits of galaxies in the child disk sample, together with their reduced $\chi^2$ values. Figure~\ref{fig:hist_bdpar} shows distributions of $2.2\rd$, $R_{80}$, $\qd$, and $D/T$. The 5th, 50th (median) and 95th percentile values for $2.2\rd$ are 3.4, 7.4, and 12.2 kpc, respectively; for $R_{80}$, they are 4.3, 9.8, and 16.5 kpc, respectively. 

We find that disk axis ratios $\qd$ from the B+D fits correlate well with the isophotal axis ratios $\qiso$ from SDSS, and are lower by 8 per cent on average. As noted earlier, the isophotal axis ratios are affected by seeing and bulge light, and both effects tend to increase the inferred axis ratios. In \S\ref{subsec:int_ext_corr}, we describe corrections to $\qiso$ to estimate disk axis ratios for the parent disk sample (to be used for internal extinction corrections).
 
Out of the 189 galaxies in the child disk sample, 86 (46\%), 135 (71\%), and 165 (87\%) have $D/T$ greater than 0.95, 0.9, and 0.8, respectively. We find that the majority of galaxies with $D/T$ greater than 0.95 have $\ns<2.3$, but there is a large scatter in $\ns$ for any given $D/T$. Two galaxies--- SDSSJ124545+52 and SDSSJ204913+00 (shown in bottom panels of Fig.~\ref{fig:bdfit_images})--- have reported $D/T$ lower than 0.5. Both galaxies have edge-on disks ($\qd<0.2$), which led to an underestimation of the disk light (due to large amounts of dust extinction), and a large overestimation of $R_{80}$. For both galaxies, we do not use the overestimated values for $R_{80}$. Instead, we assign $R_{80}=2.3\rd$, based on the relation between these two disk sizes for B+D model galaxies with $D/T\approx 0.5$. 
 
\begin{table*}
\begin{center}
\caption{Best-fit parameters from $i$-band B+D fits for 189 galaxies in the child disk sample.}
\label{tab:child_bdpar}
\begin{tabular}{lrrrrrr}
\hline 
\multicolumn{1}{c}{Galaxy} &
\multicolumn{1}{c}{$D_{\rm A}(1\arcsec)$} &
\multicolumn{1}{c}{$\qd$} &
\multicolumn{1}{c}{$D/T$} &
\multicolumn{1}{c}{$\rd$} &
\multicolumn{1}{c}{$R_{80}$} &
\multicolumn{1}{c}{$\chi^2/\nu$} \\
\multicolumn{1}{c}{} &
\multicolumn{1}{c}{(kpc)} &
\multicolumn{2}{c}{} &
\multicolumn{1}{c}{(arcsec)} &
\multicolumn{1}{c}{(arcsec)} &
\multicolumn{1}{c}{} \\
 \multicolumn{1}{c}{(1)} & 
 \multicolumn{1}{c}{(2)} & 
 \multicolumn{1}{c}{(3)} & 
 \multicolumn{1}{c}{(4)} & 
 \multicolumn{1}{c}{(5)} & 
 \multicolumn{1}{c}{(6)} & 
 \multicolumn{1}{c}{(7)} \\
\hline 
 SDSS J001006.61$-$002609.7 &  0.6420 & $ 0.284$ & $ 1.00$ &      $ 3.56$\,$(0.01)$ &                $10.67$ &$ 3.91$ \\
 SDSS J001708.75$-$005728.9 &  0.3833 & $ 0.575$ & $ 1.00$ &      $ 7.56$\,$(0.04)$ &                $22.62$ &$ 2.10$ \\
   SDSS J002844.82+160058.8 &  1.7575 & $ 0.144$ & $ 0.83$ &      $ 5.09$\,$(0.14)$ &                $14.17$ &$ 1.23$ \\
 SDSS J003112.09$-$002426.4 &  0.3943 & $ 0.534$ & $ 0.79$ &      $ 5.38$\,$(0.01)$ &                $14.34$ &$ 1.71$ \\
   SDSS J004916.23+154821.0 &  1.5879 & $ 0.568$ & $ 0.95$ &      $ 2.83$\,$(0.12)$ &                $ 8.28$ &$ 1.27$ \\
   SDSS J004935.71+010655.2 &  0.3570 & $ 0.107$ & $ 0.96$ &      $21.93$\,$(0.06)$ &                $64.63$ &$ 3.00$ \\
   SDSS J011750.26+133026.3 &  0.6513 & $ 0.430$ & $ 0.95$ &      $ 5.98$\,$(0.08)$ &                $17.53$ &$ 1.44$ \\
 SDSS J012317.00$-$005421.6 &  0.5211 & $ 0.299$ & $ 0.81$ &      $ 5.28$\,$(0.03)$ &                $14.25$ &$ 2.75$ \\
   SDSS J012340.12+004056.4 &  0.6654 & $ 0.213$ & $ 0.84$ &      $ 5.55$\,$(0.04)$ &                $15.39$ &$ 1.82$ \\
 SDSS J012438.08$-$000346.5 &  0.5561 & $ 0.221$ & $ 0.94$ &      $ 6.75$\,$(0.03)$ &                $19.76$ &$ 1.57$ \\
\hline
\end{tabular}
\end{center}
\begin{flushleft}
Notes. --- {
Col. (1): SDSS name of galaxy. 
Col. (2): Angular diameter distance corresponding to 1\arcsec\ at the redshift of the galaxy, in units of kpc. 
Col. (3): Disk axis ratio, $\qd$.
Col. (4): Disk-to-total light ratio, $D/T$.
Col. (5): Disk scale length, $\rd$, and its 1-$\sigma$ uncertainty (in parenthesis).
Col. (6): Radius containing 80 per cent of the total galaxy light, $R_{80}$.
Col. (7): Reduced $\chi^2$ of the B+D fit ($\nu$ is the number of degrees of freedom).
} 
\end{flushleft}
\end{table*}

\section{Derivation of photometric quantities}
\label{sec:phot}

Our aim is to construct photometric estimators of disk rotation velocities by calibrating their respective TFRs. We begin by describing how we apply internal extinction corrections to galaxy absolute magnitudes and colours (\S\ref{subsec:int_ext_corr}).  We do not apply any internal extinction corrections to the disk scale lengths, but note that their effect on the rotation velocity amplitudes used in the TFRs is small (see \S\ref{subsec:vrot_definition} for details).

Next, we define the different photometric quantities, namely: absolute magnitudes, synthetic magnitudes, stellar masses, and baryonic masses (\S\ref{subsec:phot_absmag}--\S\ref{subsec:phot_mbar}). We also define and characterize the associated observational errors, which are important for estimating the intrinsic scatter in the TFRs. In light of future applications, we have focused on estimators that are readily available for the full parent disk sample (i.e., do not require B+D fits).

\subsection{Internal extinction corrections}
\label{subsec:int_ext_corr}
Disk galaxies are affected by dust obscuration; an inclined disk galaxy appears redder and fainter than if it were face-on \citep[e.g.,][]{1991Natur.353..515B}. Therefore, we apply internal extinction corrections to the absolute magnitudes and colours of galaxies in both the child and parent disk samples. If these corrections are not applied, extinction effects can induce a spurious correlation between disk inclination and residuals from the TFR. We demonstrate that our corrections effectively remove this correlation in \S\ref{subsec:correl_axisratio}.

Traditionally, internal extinction corrections are applied relative to the face-on orientation, but any inclination (or, equivalently, axis ratio) can be used as a reference point. In fact, correcting to a reference inclination that is typical of the galaxy sample has the advantage of minimizing the amount of the correction, as well as the corresponding uncertainty. For this reason, we use extinction corrections to a reference inclination $\theta^*$ (corresponding to an axis ratio $q^*$), which are related to face-on corrections by
\beq \label{eq:rel_int_ext}
A_\lambda^{\theta-\theta^*} = A_\lambda^{\theta-0} - A_\lambda^{\theta-0}(q^*). 
\eeq
Here, the superscript `$\theta-\theta^*$' denotes corrections to a reference inclination $\theta^*$, and `$\theta-0$' denotes corrections to face-on orientation. Note that both corrections are relative, and do not account for extinction in a face-on system. We choose $q^*=\langle \qd \rangle =0.40$, the mean disk axis ratio of the child disk sample.\footnote{For comparison, the mean axis ratio of a sample of randomly-oriented disks with an intrinsic axis ratio (ratio of the vertical and radial scale lengths) of 0.19 is 0.67. Recall that the child disk sample is restricted to low-axis ratio galaxies, with $q_{\rm iso}<0.6$.}

Following the empirical prescription of \citet{1998AJ....115.2264T}, which is based on a study of 87 spiral galaxies in the Ursa Major and Pisces clusters with photometry in the Johnson $B$, Cousins $RI$, and an infrared $K$ band, we calculate the extinction $A_\lambda$, in the passband $\lambda$, to be
\beq \label{eq:int_ext}
A_\lambda^{\theta-0} = - \gamma_\lambda(M_\lambda^{\rm NC}) \times \log(\qd),
\eeq
where $\qd$ is the disk axis ratio, and $M_\lambda^{\rm NC}$ is the Petrosian absolute magnitude without internal extinction correction.\footnote{A more recent work, \citet{2009ApJ...691..394M} explored the dependence of $\gamma_\lambda$ on both $M_K$ and $\ns$. For $\ns=1.0$, their values of $\gamma_\lambda$ are comparable with those of Tully et al. 1998, although the inferred dependence on $M_K$ differs (c.f. their Fig.~10). We have not adopted their corrections because they require infrared data.}

The $\gamma_\lambda$ in Eq.~\ref{eq:int_ext} are linear functions of $M_\lambda$, given in Eqs. (3)--(5) of \citet{1998AJ....115.2264T} for the $BRI$ bands (at effective wavelengths 4448, 6581, and 8059\AA, respectively). To determine $\gamma_\lambda$ for the SDSS bands $ugriz$, we first use conversion formulae from Table 7 of \citet{2002AJ....123.2121S} to estimate $BRI$ absolute magnitudes from the SDSS $g$-band absolute magnitudes, $g-r$ and $r-i$ colours. Then, we fit a quadratic function to $\gamma_{\rm B}$, $\gamma_{\rm R}$, and $\gamma_{\rm I}$ as a function of $\lambda$ and interpolate (or extrapolate) $\gamma_\lambda$ for the SDSS $ugriz$ bands, with effective wavelengths 3557, 4825, 6261, 7672, and 9097\AA, respectively (close to the $BRI$ passbands).

For the child disk sample, disk axis ratios $\qd$ were derived from B+D fits (\S\ref{sec:bdfit}). For the parent disk sample, for which we have not performed B+D fits, we approximate the disk axis ratios by applying corrections $C_q$ to the SDSS isophotal axis ratios $\qiso$
\beq
\qd/\qiso \equiv C_q(f_{\rm PSF}, \qiso) 
\eeq
where $f_{\rm PSF}=\theta_{\rm PSF}/R_{50}$, the ratio of the seeing PSF FWHM to the galaxy's Petrosian half-light radius. 
Motivated by available data from the child disk sample, we adopt the fitting formula
\beq
C_q =
1.21 + m_1(f_{\rm PSF}-0.22) + m_2(\qiso-0.20),
\eeq
for galaxies with $\qiso\le 0.6$. From a least-squares fit to the child disk sample dataset, we find $m_1=0.22$ and $m_2=-0.32$.
For galaxies with $\qiso>0.6$, we set $C_q=1$. The mean value of $C_q$ for the parent disk sample is 1.07.

By construction, the mean internal extinction corrections $A_\lambda^{\theta-\theta^*}$ are small for the child disk sample: 0.03 mag for the $u$, $g$, and $r$ bands and 0.02 mag for the $i$ and $z$ bands. For the $i$-band, the 5th, 50th, and 95th percentile values are $-0.22$, $0.01$, and $0.28$ mag, respectively. The mean internal extinction corrections to face-on orientation $A_\lambda^{\theta-0}$ are larger: 0.67, 0.60, 0.51, 0.41, and 0.31 mag for the $ugriz$ bands, respectively. If we were to correct to face-on orientation instead, the main effect would be a shift in the derived zero-points of the $M_\lambda$ TFRs by an amount given by the relative internal extinction corrections (as we show explicitly in \S\ref{subsec:alt_extcorr}). 

\subsection{Absolute magnitudes}
\label{subsec:phot_absmag}

Internal extinction-corrected absolute magnitudes $M_\lambda$, as described in \S\ref{subsec:data_sdss}, are calculated as
\beq \label{eq:absmag}
M_\lambda=m_{{\rm P,}\lambda} - 5 \log(D_{\rm L}/10 {\rm pc}) - K_\lambda -A_\lambda^{\rm MW} - A_\lambda^{\theta-\theta^*}.
\eeq
Here, $m_{{\rm P,}\lambda}$ is the apparent SDSS Petrosian magnitude in the band $\lambda$, $D_{\rm L}$ is the luminosity distance calculated from the SDSS redshift (with cosmology $\Omega_{\rm m}=0.3$, $\Omega_\Lambda=0.7$, and $h=0.7$), $K_\lambda$ is the $k$-correction to $z=0$ calculated using the {\verb kcorrect } product version {\verb v4_1_4 } of \citet{2007AJ....133..734B}, $A_\lambda^{\rm MW}$ is the correction for Galactic extinction based on dust maps of \citet{1998ApJ...500..525S}, and $A_\lambda^{\theta-\theta^*}$ is the internal-extinction correction to the mean inclination of the child disk sample (Eqs.~\ref{eq:rel_int_ext} \& \ref{eq:int_ext}). $M_\lambda^{\rm NC}$ is defined similarly, but without the internal extinction correction term. 

We estimate the uncertainty in $M_\lambda$ to be
\beqa \label{eq:err_absmag} \nonumber
\left(\delta M_\lambda\right)^2 &=& (\delta m_{{\rm P,}\lambda})^2 + \left(\frac{5}{\ln 10} \frac{\delta V_{\rm pec}}{cz}\right)^2 + (\delta A_\lambda^{\theta-\theta^*})^2,
\eeqa
with terms representing the uncertainty in $m_{{\rm P,}\lambda}$, $D_{\rm L}$, and $A_\lambda^{\theta-\theta^*}$, respectively. We neglect the smaller uncertainties from the redshift, $k$-correction, and Galactic extinction correction measurements. We use uncertainties $\delta m_{{\rm P,}\lambda}$ derived from the {\verb kcorrect } routines, which have imposed a minimum uncertainty of $[0.05, 0.02, 0.02, 0.02, 0.03]$ mag in the $ugriz$ bands (to account for uncertainties in the absolute calibration for each galaxy). For the distance uncertainty, we adopt $\delta V_{\rm pec}=300\,\kms$, the typical amplitude of small-scale peculiar velocities \citep{1995PhR...261..271S}. The second term has 5th, 50th, and 95th percentile values of 0.03, 0.07, and 0.12, respectively. For the internal extinction correction uncertainty, we follow other authors (e.g., P07) and arbitrarily adopt $\delta A_\lambda^{\theta-\theta^*}=A_\lambda^{\theta-\theta^*}/3$. However, since we correct to the mean inclination (instead of face-on orientation), our extinction corrections, as well as the associated uncertainties, are much smaller. The median value of $\delta A_\lambda^{\theta-\theta^*}$ is less than 0.01 mag in all bands. For the $i$ band, the 5th and 95th percentile values are $-0.07$ and $0.09$ mag, respectively. 

The uncertainty in the absolute magnitude is dominated by the uncertainty due to peculiar velocities. The median uncertainty in $M_\lambda$ is $0.09$ mag for the $g$, $r$, and $i$ bands, and 0.14 mag for the $z$ band. For the $i$-band, the 5th and 95th percentile values are 0.04 and 0.15 mag, respectively.

\subsection{Synthetic magnitudes}
\label{subsec:phot_synmag}

We define a synthetic magnitude to be a linear combination of the absolute magnitude in some band $\lambda$ and a chosen colour,
\beq \label{eq:synmag}
\msyn(\alpha; \lambda, {\rm colour}) = M_\lambda + \alpha \times ({\rm colour}),
\eeq
where $\alpha$ is some coefficient that we are free to choose. Using this quantity generalizes the method of applying colour-based $M_*/L$ corrections to luminosities to estimate stellar masses (Bell et~al. 2003; see \S\ref{subsubsec:bell_mstr}) 

In this work, we aim to define optimal photometric estimators of rotation velocity. Therefore, we choose the coefficient $\alpha$ to be the value that minimizes the scatter in the inverse TFR, in which $\msyn$ is the independent variable, and rotation velocity is the dependent variable (c.f. Eq.~\ref{eq:tf_synmag} in \S\ref{subsec:tfr_model}).

Internal extinction-corrected galaxy colours, as described in \S\ref{subsec:data_sdss}, are calculated as
\beqa \label{eq:gminr} \nonumber
g-r &=& m_{{\rm model},g} - m_{{\rm model},r} - (K_{{\rm model},g} - K_{{\rm model},r}) \\
	&-& (A_g^{\rm MW} - A_r^{\rm MW}) - (A_g^{\theta-\theta^*}-A_r^{\theta-\theta^*}),
\eeqa
where $m_{{\rm model},\lambda}$ are apparent SDSS model magnitudes, $K_{{\rm model},\lambda}$ are $k$-corrections to $z=0$ for the model magnitudes, $A_\lambda^{\rm MW}$ are corrections for Galactic extinction, and $A_\lambda^{\theta-\theta^*}$ are corrections for internal extinction (these last three quantities are calculated as in Eq.~\ref{eq:absmag}). $(g-r)^{\rm NC}$ is defined similarly, but without the internal extinction correction terms.

We estimate observational errors in the $g-r$ colour to be
\beqa \label{eq:err_gminr} \nonumber
[\delta(g-r)]^2 &=& (\delta m_{{\rm model},g})^2 + (\delta m_{{\rm model},r})^2 \\ 
	&+& [(A_g^{\theta-\theta^*}-A_r^{\theta-\theta^*})/3]^2.
\eeqa
Other colours are defined similarly. The median uncertainty in colours that do not involve the $u$ band range from 0.03 to 0.05 mag. Colours involving the $u$ band have slightly higher median uncertainties, ranging from 0.06 to 0.08 mag. 

\subsection{Stellar masses}
\label{subsec:phot_mstr}

We consider two kinds of stellar mass estimates: Bell et al. (2003) and MPA/JHU stellar masses, defined in \S\ref{subsubsec:bell_mstr} and \S\ref{subsubsec:mpa_mstr}, respectively. We convert both estimates to correspond to the same IMF normalization. We adopt the \citet{2002Sci...295...82K} IMF, which adequately represents direct observational estimates of the IMF for late-type galaxies. (The normalization of this IMF is 0.3 dex lower than that for a Salpeter IMF with a lower mass cut-off of 0.1 $M_\odot$.) We compare the two different stellar mass estimates in \S\ref{subsubsec:comp_mstr}. 

\subsubsection{Bell et al. (2003) stellar masses}
\label{subsubsec:bell_mstr}
Our primary stellar mass estimate, denoted by $M_*$ or $\mbell$, are calculated following the prescription of \citet{2003ApJS..149..289B}. First, we calculate $i$-band luminosities
\beqa
\log(L_i/L_\odot)&=&-0.4 (M_i^{\rm NC}-M_{\odot,i}+1.1z),  
\eeqa
where $M_{\odot,i}=4.56$ mag, and $1.1z$ is the mean evolution correction (in their stellar population models). Then, we calculate $g-r$ colour-based stellar mass-to-light ratios
\beqa
\log(M_*/L_i) &=& a^{\rm Bell}_i + b^{\rm Bell}_i \times (g-r)^{\rm NC} - 0.15,
\eeqa
with $a^{\rm Bell}_i=-0.22$ and $b^{\rm Bell}_i=0.86$. The final term, $-0.15$, converts the normalization from the ``diet Salpeter'' IMF adopted by Bell et~al. (2003) to a Kroupa IMF.\footnote{The normalization of the ``diet Salpeter'' IMF is 0.15 dex lower than that for a Salpeter IMF with a lower mass cut-off of 0.1 $M_\odot$, and therefore 0.15 dex higher than for a Kroupa IMF.}
Finally, we calculate stellar masses
\beqa \label{eq:bellmass}
\log (M_*/M_\odot) &=& \log(L_i/L_\odot) + \log(M_*/L_i).
\eeqa

We have used absolute magnitudes and colours uncorrected for internal extinction because \citet{2003ApJS..149..289B} did not apply any such corrections. Moreover, as noted in \citet{2001ApJ...550..212B}, dust affects the stellar mass estimates in two different ways that cancel out to first order: dimming of the galaxy light leads to a lower luminosity and therefore a lower stellar mass estimate, but dust reddening leads to a higher stellar mass-to-light ratio, and consequently, a higher stellar mass estimate. Because of this partial cancellation, corrections for the fact that galaxies in the child disk sample have smaller axis ratios compared to the full galaxy population (used by Bell et~al. 2003), and therefore larger internal extinctions, on average, are tiny, and are not applied for simplicity. 

To estimate the statistical uncertainty in $M_*$, we propagate errors from $M_i$ and $g-r$, as given by Eqs.~\ref{eq:err_absmag} and \ref{eq:err_gminr}, with the internal extinction correction terms excluded. For the child disk sample, the mean statistical uncertainty in $\mbell$ (at fixed Kroupa IMF) is 0.041 dex.

\subsubsection{MPA/JHU stellar masses}
\label{subsubsec:mpa_mstr}

The second set of photometric stellar mass estimates, denoted by $\mmpa$, are based on fits to the $ugriz$ spectral energy distribution (SED) from the MPA/JHU group.\footnote{http://www.mpa-garching.mpg.de/SDSS/DR7/Data/ stellarmass.html. See also \citet{2011ApJS..193...29A}.} 
We define $\mmpa$ to be the reported median of the probability distribution function (PDF), with 0.05 dex subtracted, to convert the normalization from their adopted Chabrier IMF to a Kroupa IMF. We define the uncertainty in $\mmpa$ to be half of the difference between the 84th and 16th percentiles of the PDF. For the child disk sample, the mean statistical uncertainty in $\mmpa$ (at fixed Kroupa IMF) is 0.089 dex.

\subsubsection{Comparison of $\mbell$ and $\mmpa$}
\label{subsubsec:comp_mstr}

Both the Bell et~al. (2003) and MPA/JHU stellar masses are based on SED fits to SDSS photometry, but they employ different methods, including different sets of model spectra (the latter allows for a wider possible range in parameters), different stellar population synthesis codes (\citealt{2003MNRAS.344.1000B} vs. \citealt{1997A&A...326..950F}), and different treatments for extinction (the former does not apply internal extinction corrections, while the latter does). 

\begin{figure} 
\includegraphics[width=3in]{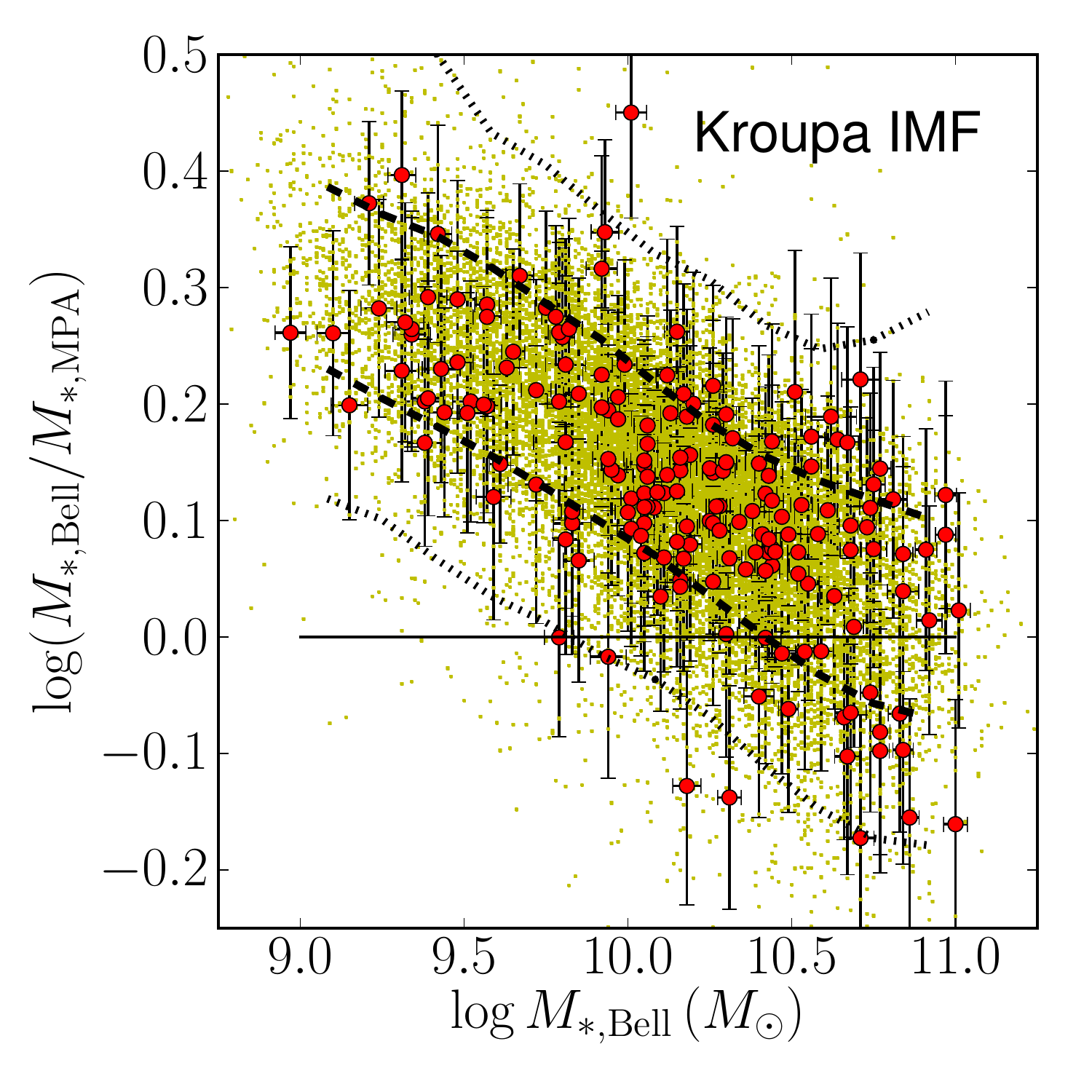} 
\caption{Ratio of the two different stellar mass estimates $\mbell$ and $\mmpa$ for the parent and child disk samples (yellow dots and red circles, respectively). Error bars show 1-$\sigma$ statistical uncertainties in $\mbell$. Dashed and dotted curves show 1-$\sigma$ and 2-$\sigma$ bounds for the parent disk sample, respectively.}
\label{fig:mstr_comp}
\end{figure}

Figure~\ref{fig:mstr_comp} compares the two kinds of stellar mass estimates for the parent and child disk samples (yellow dots and red circles, respectively). There is a clear systematic difference between the two, which increases toward lower stellar masses. It is understandable why the lowest masses are the most problematic; low-mass galaxies have more dust and complicated star formation histories. It is somewhat reassuring that the two estimates approach agreement at the high mass end, but it is unclear which is more correct for lower masses. 

Given these uncertainties in our current knowledge, we will present our main results using both kinds of stellar masses. This will give the reader a sense of the range of allowed scenarios (at fixed IMF), and explicitly show which of our conclusions are robust to systematics associated with the stellar mass estimates.

\subsection{Baryonic masses}
\label{subsec:phot_mbar}

We calculate baryonic mass estimates as a sum of stellar and gas masses: $\mbar=\mbell+\mgas$, with $\mbell$ given by Eq.~\ref{eq:bellmass}. We adopt $u-r$ colour-based gas-to-stellar mass ratio estimates from \citet{2004ApJ...611L..89K}
\beq \label{eq:mgas}
\log(\mgas/\mbell)=1.46-1.06(u-r)^{\rm NC}+0.15,
\eeq
which was derived from 346 galaxies in SDSS DR2 \citep{2004AJ....128..502A} with infrared data from 2MASS \citep{2000AJ....119.2498J} and HI data from HyperLeda \citep{2003A&A...412...57P}.
We have used colours uncorrected for internal extinction because \citet{2004ApJ...611L..89K} did not apply any such corrections. The addition of 0.15 dex converts the normalization of $\mbell$ from their adopted ``diet Salpeter'' IMF to a Kroupa IMF.

To calculate statistical uncertainties in $\mbar$, we propagate errors from $M_*$ and $(u-r)^{\rm NC}$. For the child disk sample, the mean statistical uncertainty in $\mbar$ (at fixed Kroupa IMF) is 0.043 dex.

The above relation, Eq.~\ref{eq:mgas}, has a large scatter, 0.42 dex, which seems to be mostly physical.\footnote{\citet{2004ApJ...611L..89K} obtained a somewhat tighter relation using $u-K$ colour, with a scatter of 0.37 dex. Here, we have adopted the $u-r$ colour-based relation so as not to rely on the availability of $K$ band data.}  
We note that there are substantial systematic uncertainties in this relation as well. However, since we are primarily interested in using $\mbar$ as a photometric estimator of disk rotation velocity, its precision as an estimate of the true baryonic mass of a galaxy is of secondary importance. Regardless, it will be useful to know whether $\mbar$ yields a tighter ITFR than say, $M_*$ or $\msyn$. 

\begin{figure} 
\includegraphics[width=3in]{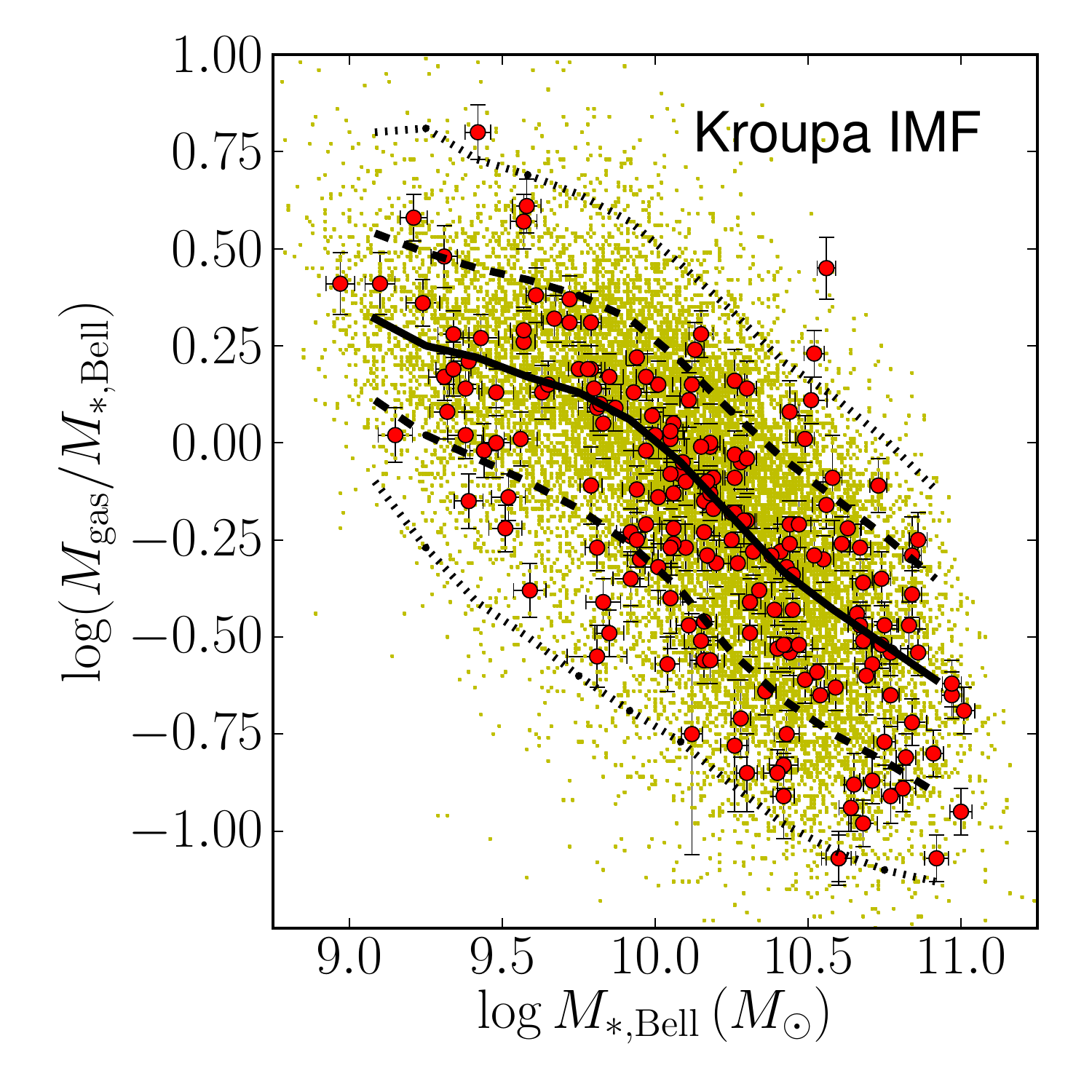} 
\caption{Stellar-to-gas mass ratios derived from $u-r$ colours from \citet{2004ApJ...611L..89K} (Eq.~\ref{eq:mgas}) vs. stellar masses $\mbell$ for the parent and child disk samples (yellow dots and red circles, respectively). Error bars show 1-$\sigma$ uncertainties. Solid, dashed and dotted curves show the median relation and 1- and 2-$\sigma$ limits for the parent disk sample.}
\label{fig:mgas_mstr_ratio}
\end{figure}

Figure~\ref{fig:mgas_mstr_ratio} shows gas-to-stellar mass ratios for the parent and child disk samples (yellow dots and red circles, respectively). This ratio varies from $\sim 30$ per cent at the highest stellar masses, to a factor of $\sim 3$ at the lowest masses. For the child disk sample, the 5th, median, and 95th-percentile values are 0.13, 0.60, and 2.29, respectively.

\section{Derivation of kinematic quantities}
\label{sec:long}

We begin this section by describing our long-slit spectroscopy observations and reductions of the raw data (\S\ref{subsec:long_obs}). Then, we describe the steps in our rotation curve analysis pipeline. First, we extract rotation curves from the two-dimensional long-slit spectra (\S\ref{subsec:rc_extraction}). Then, we perform fits to these rotation curves (\S\ref{subsec:rc_fitting}). Finally, we define the disk rotation velocities that are used in the TFRs (\S\ref{subsec:vrot_definition}). 

In the final subsection, we investigate the effect of systematics on the determination of these rotation velocities. First, we investigate the effect of slit misalignments (\S\ref{subsubsec:syst_pos}). Next, we compare the rotation curves extracted from multiple observations of the same galaxy, for the five galaxies observed by both us and P07 (\S\ref{subsubsec:syst_piz}). Finally, we compare the rotation velocities derived from our analysis pipeline with those reported by P07 (\S\ref{subsubsec:syst_fits}).

\subsection{Long-slit spectroscopy observations and data reduction}
\label{subsec:long_obs}
We carried out long-slit spectroscopy observations with the 3.5 m telescope at Apache Point Observatory (APO) using the Dual Imaging Spectrograph (DIS) with a 1200 lines mm$^{-1}$ grating and 1.5\arcsec\ slit. The spectral range covered is 6433--7615 \AA\,(centered at 7024 \AA). The linear dispersion is 0.58 \AA\, pixel$^{-1}$, and the instrumental resolution is approximately 2 pixels or $50\,\kms$ (at the typical observed wavelength of the H$\alpha$ line). 

We observed a total of 95 galaxies over 25 half-nights between March 2009 and June 2010. In most cases, we took three 1200 s exposures, except for a few bright, nearby galaxies for which we took two 1200 s exposures. All targets were observed with the slit position angle (PA) set on the major axis of the galaxy, determined from SDSS PHOTO exponential fits to the $r$-band images in most cases (see \S\ref{subsubsec:syst_pos} for details). Table~\ref{tab:long_obs} lists the galaxy names, redshifts, slit PA's, exposure times, and observation dates.

To test systematics, we observed two galaxies with the slit PA set on the minor axis, and another two galaxies with the slit PA set at $\pm$10 degrees off the major axis (see \S\ref{subsubsec:syst_pos}). In addition, five galaxies from the P07 sample
were re-observed with DIS (see \S\ref{subsubsec:syst_piz}); for these galaxies, we use the DIS observations in the final analysis. 

Flat-fielding, bias-subtraction, wavelength calibration, linearization, and flux calibration were performed using standard IRAF\footnote{http://iraf.noao.edu} routines. Wavelength calibration and linearization used HeNeAr arc frames. For flux calibration, Feige 34, Feige 56, BD26+2606, and HD 192281 were used as standard stars (selected to match the airmass and time of observation). The use of multiple exposures for each observation allows cosmic ray removal. We combine frames using the IRAF routine {\it imcombine} with option {\verb reject=crreject }. For sky subtraction, we took the average sky spectrum over 21 pixels (12 \AA) on either side of the galaxy's $\ha$ emission. Then, from each area of the CCD on either side of the galaxy's continuum center, we subtracted the average sky spectrum measured from that side. 
 
\begin{table*}
\caption{Observations with DIS at APO 3.5m (95 galaxies).}
\begin{center}
\begin{tabular}{lrrrl}
\hline \\
 \multicolumn{1}{c}{Galaxy name} &
  \multicolumn{1}{c}{$z$} &
 \multicolumn{1}{c}{PA} &
 \multicolumn{1}{c}{Exp. time} &
 \multicolumn{1}{c}{Obs. Date} \\
  \multicolumn{2}{c}{} &
   \multicolumn{1}{c}{(deg)} &
    \multicolumn{1}{c}{(sec)} &
     \multicolumn{1}{c}{} \\
 \multicolumn{1}{c}{(1)} & 
 \multicolumn{1}{c}{(2)} & 
 \multicolumn{1}{c}{(3)} & 
  \multicolumn{1}{c}{(4)} &
 \multicolumn{1}{c}{(5)} \\
 \hline 
    SDSS J002844.82+160058.8 &  0.0947 &$  35.79$ &   $1200 \times 3$ &             2009-09-14\\
   SDSS J004916.23+154821.0 &  0.0846 &$ -73.04$ &   $1200 \times 3$ &             2009-09-14\\
   SDSS J011750.26+133026.3 &  0.0326 &$ -66.45$ &   $1200 \times 2$ &             2009-09-15\\
   SDSS J013017.16+143918.5 &  0.0239 &$ -23.95$ &   $1200 \times 3$ &             2009-09-23\\
   SDSS J020056.00+133116.6 &  0.0312 &$  41.92$ &   $1200 \times 3$ &             2009-09-23\\
   SDSS J020133.02+133126.4 &  0.0978 &$  48.43$ &   $1200 \times 2$ &             2009-09-14\\
   SDSS J020526.65+131938.4 &  0.0251 &$  78.57$ &   $1200 \times 2$ &             2009-09-20\\
 SDSS J020540.31$-$004141.4 &  0.0424 &$ -30.32$ &   $1200 \times 2$ &             2009-09-15\\
   SDSS J020819.43+134944.6 &  0.0798 &$ -22.48$ &   $1200 \times 3$ &             2009-09-23\\
   SDSS J020923.14+125029.8 &  0.0604 &$  64.47$ &   $1200 \times 2$ &             2009-09-15\\
\hline
\end{tabular}
\end{center}
\begin{flushleft}
Notes.---  Col. (1): SDSS name of galaxy. 
Col. (2): Redshift of galaxy.
Col. (3): Major axis position angle in degrees East of North. 
Col. (4): Duration of spectroscopic exposure (sec).
Col. (5): UT date of observation.
\end{flushleft}
\label{tab:long_obs}
\end{table*}%

\subsection{Rotation curve extraction}
\label{subsec:rc_extraction}

We begin by defining the radial bins for the extracted rotation curve. First, we choose the central row to be at the peak of the galaxy continuum flux (defined to be the total flux over rest wavelengths 6590--6630\AA). Then, we bin together rows on either side of the central row with increasing bin sizes of 1, 1, 2, 2, 4, 6, 8, 12, 20, 32, 56, 80, and 100 pixels, for a total of 27 radial bins (1 pixel $=$ 0.577\AA). We take the mean over the rows in each bin to get binned spectra. With this choice of binning, the $S/N$ of the binned spectra remains roughly constant with radius (since the bins cover increasingly larger areas as the galaxy flux falls off with radius). 

For each of the binned spectra, we fit a set of three Gaussians plus a linear continuum to the $\ha$+[NII]6548,6583 emission lines, over the wavelength range 6520--6610\AA. We fix the ratio of [NII]6548 to [NII]6583 flux to the theoretical value 1:3 and require all line widths to be equal. We find the best-fit $\ha$ line center $\lambda_{\ha}$, line width $\sigma_{\ha}$, $\ha$ line flux, and [NII]6548 line flux using a Levenberg-Marquardt least-squares minimization routine in IDL called MPFIT.\footnote{http://purl.com/net/mpfit} We perform a set of 500 Monte-Carlo realizations (varying the flux at each wavelength according to the error in that flux measurement) to estimate errors in the fit parameters. 

For each bin, we also get a flux-weighted average spatial position $R'$, using as weights the total flux over the wavelength range $\lambda_{\ha} \pm 3\sigma_{\ha}$ for each row of that bin. Finally, we convert the rest-frame $\ha$ line centers and errors into circular velocities to obtain the observed rotation curve $V_{\rm obs}(R'_i)$, with uncertainties $\sigma_{V,i}$. 

As desired, $\sigma_{V,i}$ is roughly constant with radius. The 5th, 50th, and 95th percentile values for $\langle \sigma_{V,i}\rangle$ (averaged over all radii) are 0.86, 2.43, and 9.60 $\kms$, respectively. Splitting each rotation curve at its turn-over radius $\rto$ (defined in Eq.~\ref{eq:atan_vprof} below), the $\langle \sigma_{V,i}\rangle$, averaged over the inner and outer regions, have median values of 2.96 and 2.10 $\kms$, respectively (the 5th and 95th percentile values are similar to the full radial average as well). 

\subsection{Rotation curve fitting}
\label{subsec:rc_fitting}

Studies have shown that most observed disk galaxy rotation curves can be modeled by an arctangent model \citep{1997AJ....114.2402C}: 
\beq \label{eq:atan_vprof}
V_{\rm mod}(R') = V_0 + \frac{2}{\pi} V_{\rm c,obs} \arctan \left( \frac{R'-R_0}{\rto}\right).
\eeq
This model has four free parameters: the systemic velocity $V_0$, the asymptotic circular velocity $V_{\rm c,obs}$, the spatial center $R_0$, and the turn-over radius $\rto$, at which the rotation curve starts to flatten out. We denote the radius defined from the centre of the rotation curve as $R = R'-R_0$. Note that all the velocities above are not corrected for inclination.
21
Using the Levenberg-Marquardt routine MPFIT in IDL, we find the best-fit parameters by minimizing $\chi^2$, defined as 
\beq \label{eq:atan_chi2}
\chi^2 = \sum_i \frac{\left[ V_{\rm mod}(R'_i)-V_{\rm obs}(R'_i)\right]^2}{\sigma_{{\rm eff},i}^2}, 
\eeq
where the sum is over radial bins with sufficiently good $S/N$ spectra ($S/N>5$ in the $\ha$ line), and the effective uncertainty in $V_{\rm obs}(R'_i)$ is defined to be $\sigma_{{\rm eff},i}=( \sigma_{V,i}^2+\sigma_{V,{\rm add}}^2)^{1/2}$. Here, $\sigma_{V,i}$ is the formal uncertainty from the Gaussian line fits (described in \S\ref{subsec:rc_extraction}), and $\sigma_{V,{\rm add}}$ is an additional model uncertainty that accounts for non-circular motions in the galaxy. We set $\sigma_{V,{\rm add}}= 10\, \kms$, following P07, but note that the fits are not very sensitive to this choice. Averaging over all radii, $\langle \sigma_{{\rm eff},i}\rangle$ has 5th, 50th, and 95th percentile values of 10.05, 10.37, and 14.89 $\kms$, respectively. 

We define the total uncertainty in a fit parameter $P$ (e.g., $V_{\rm c, obs}$, $\rto$, etc.) as
\beq
\sigma(P) =\left( \sigma_{\rm fit}^2(P) + \sigma_{\rm sys}^2(P) \right)^{1/2},
\eeq
where the fit uncertainty $\sigma_{\rm fit}$ is the formal error from the $\chi^2$ minimization fit 
and the systematic uncertainty $\sigma_{\rm sys}$ is given by
\beq \label{eq:sigma_sys}
\sigma_{\rm sys} = P-P^{\rm (r)},
\eeq
where $P^{\rm (r)}$ is the result of repeating the fit with a radius-weighted scheme that gives greater weight to points at larger radii, i.e., replacing the denominator in Eq.~\ref{eq:atan_chi2} with ${\left(\sigma_{{\rm eff},i}^{\rm (r)}\right)}^2=(1/|R'|)\sigma_{{\rm eff},i}^2$. 

By adopting this definition of total uncertainties, to be used in our fits to the TFRs, we automatically downweight the contribution of those galaxies for which the flat part of the rotation curve is not well-constrained (e.g., irregulars, galaxies in mergers, and other peculiar cases). A large fit uncertainty indicates that $V_{\rm c,obs}$ is not well-constrained because the $S/N$ is low, or because the observed rotation curve does not show a turn-over (i.e., it is still rising at the outermost point). On the other hand, a large systematic uncertainty indicates that $V_{\rm c,obs}$ is not well-defined because the arctangent model is not a good model for the observed rotation curve.

\begin{figure}
\includegraphics[width=3in]{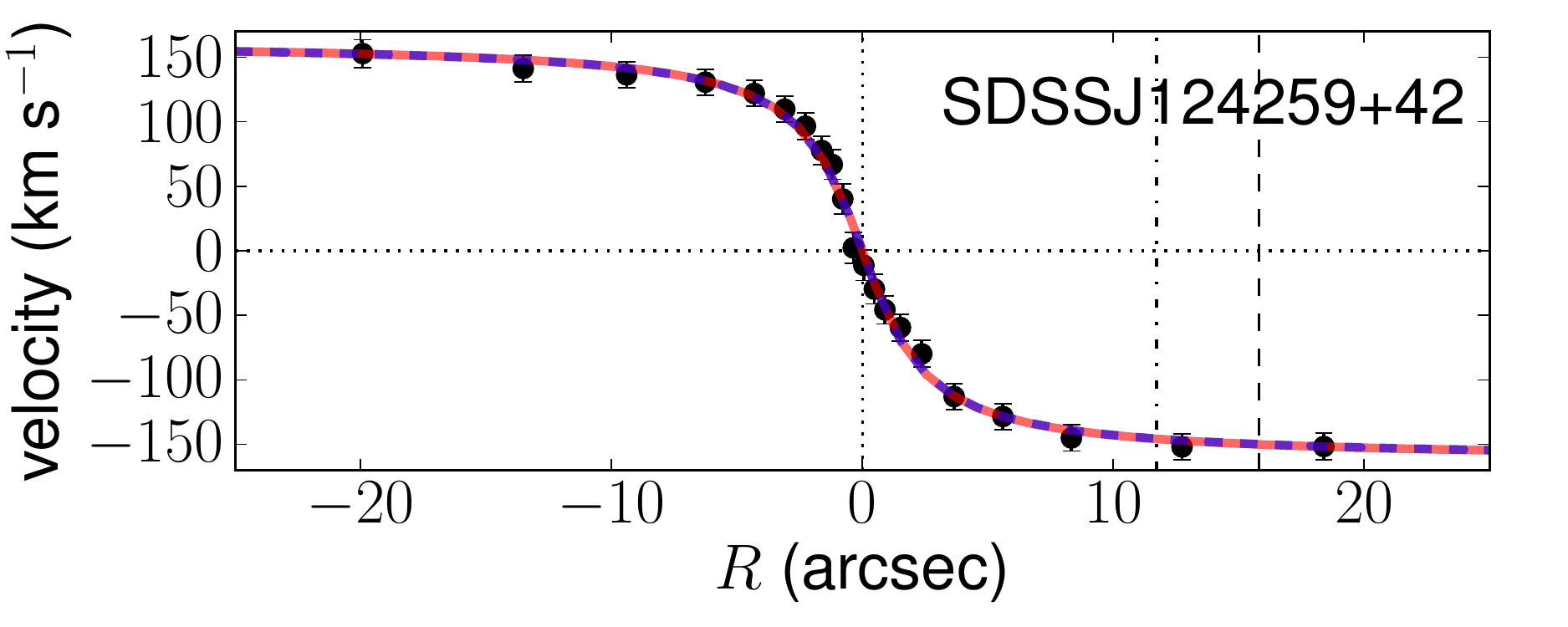} 
\includegraphics[width=3in]{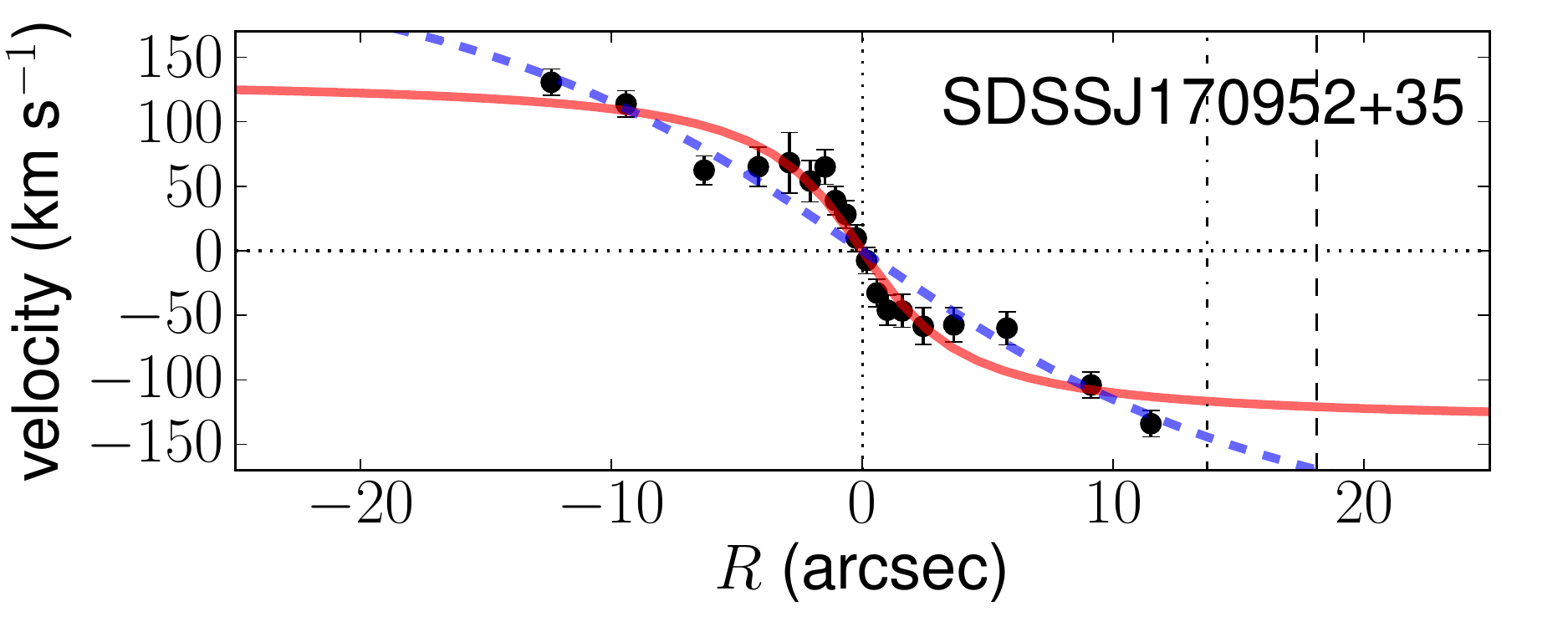}
\caption{Observed rotation curves for SDSSJ124259+42 and SDSSJ170952+35 (top and bottom panels, respectively), shown with the best-fit arctangent models from both unweighted and radius-weighted schemes (solid red and blue dashed curves, respectively). Vertical lines mark characteristic radii, $R=0$, $2.2\rd$, and $R_{80}$ (dotted, dot-dashed, and dashed lines, respectively).}
\label{fig:ex_rcfits}
\end{figure}

For illustration, Figure~\ref{fig:ex_rcfits} shows the unweighted and radius-weighted fits (red solid and blue dashed curves, respectively) for two galaxies. The top panel shows the observed rotation curve of SDSSJ124259+42, a case wherein the two fits coincide, resulting in a small systematic uncertainty in $V_{\rm c,obs}$. The bottom panel shows the peculiar rotation curve of SDSSJ170952+35, one of the few cases wherein the two fits disagree considerably, resulting in a large systematic uncertainty. In this extreme case, $|\sigma_{\rm sys}(V_{\rm c,obs})/V_{\rm c,obs}|=|(-157\ \kms)/(135\ \kms)|=1.16$. By comparison, the 5th, 50th, and 95th percentile values of $|\sigma_{\rm sys}(V_{\rm c,obs})/V_{\rm c,obs}|$ for the child disk sample are 0.0035, 0.033, and 0.14, respectively. 

Table~\ref{tab:child_tfpar} lists parameters from the arctangent model fits for the 189 galaxies in the child disk sample. We list the turn-over radii $\rto$ and inclination-corrected asymptotic circular velocities $V_{\rm c}$ and rotation velocity amplitudes $V_{2.2}$ and $V_{80}$ (defined in \S\ref{subsec:vrot_definition}). $\rto$ is listed with total 1-$\sigma$ uncertainties, while all rotation velocities are listed with both the total and formal fit 1-$\sigma$ uncertainties (the latter are inside parentheses). The table also lists inclination corrections $\sin \theta$ and reduced $\chi^2$ values of the fits. 

\begin{table*}
\caption{Rotation curve fit parameters and rotation velocity amplitudes for the child disk sample (189 galaxies).}
\begin{center}
\begin{tabular}{lrrrrrrr}
\hline \\
\multicolumn{1}{c}{Galaxy name} &
\multicolumn{1}{c}{$\sin \theta$} &
\multicolumn{1}{c}{$\rto$} &
\multicolumn{1}{c}{$V_{\rm c}$} &
\multicolumn{1}{c}{$V_{2.2}$} &
\multicolumn{1}{c}{$V_{80}$} &
\multicolumn{1}{c}{$\chi^2/\nu$} \\
\multicolumn{1}{c}{} &
\multicolumn{1}{c}{} &
\multicolumn{1}{c}{(arcsec)} &
\multicolumn{1}{c}{($\kms$)} &
\multicolumn{1}{c}{($\kms$)} &
\multicolumn{1}{c}{($\kms$)} &
\multicolumn{1}{c}{} \\
 \multicolumn{1}{c}{(1)} & 
 \multicolumn{1}{c}{(2)} & 
 \multicolumn{1}{c}{(3)} & 
 \multicolumn{1}{c}{(4)} &
 \multicolumn{1}{c}{(5)} &
 \multicolumn{1}{c}{(6)} &
  \multicolumn{1}{c}{(7)} \\
 \hline 
 SDSS J001006.61$-$002609.7 &  0.9765 & $ 2.66$\,$( 0.47)$ &     $ 158.2\pm   10.8(5.7)$ &    $ 125.2\pm     5.7(1.0)$ &    $ 133.6\pm     6.9(2.1)$ &    $   2.7$ \\
 SDSS J001708.75$-$005728.9 &  0.8336 & $ 2.81$\,$( 0.98)$ &    $ 137.8\pm   16.9(12.8)$ &    $ 123.1\pm    13.5(6.9)$ &    $ 126.9\pm    14.1(8.4)$ &    $   2.8$ \\
   SDSS J002844.82+160058.8 &  1.0000 & $ 1.45$\,$( 0.68)$ &    $ 144.1\pm   16.7(15.2)$ &    $ 132.3\pm     9.9(8.2)$ &    $ 134.7\pm    11.3(9.6)$ &    $   1.5$ \\
 SDSS J003112.09$-$002426.4 &  0.8611 & $ 0.76$\,$( 0.22)$ &     $ 153.9\pm    9.3(7.3)$ &    $ 147.7\pm     8.6(5.5)$ &    $ 148.8\pm     8.7(5.8)$ &    $   0.8$ \\
   SDSS J004916.23+154821.0 &  0.8386 & $ 1.67$\,$( 0.34)$ &     $ 150.9\pm   12.2(1.9)$ &    $ 125.7\pm    13.4(0.7)$ &    $ 131.8\pm    12.9(1.0)$ &    $   1.7$ \\
   SDSS J004935.71+010655.2 &  1.0000 & $17.70$\,$( 1.52)$ &     $ 267.2\pm   11.0(1.3)$ &    $ 207.4\pm    14.4(0.4)$ &    $ 221.8\pm    13.9(0.6)$ &    $  17.7$ \\
   SDSS J011750.26+133026.3 &  0.9194 & $ 4.06$\,$( 0.92)$ &     $ 108.8\pm   11.5(2.4)$ &    $  88.1\pm     7.9(1.6)$ &    $  93.1\pm     9.0(1.8)$ &    $   4.1$ \\
 SDSS J012317.00$-$005421.6 &  0.9718 & $ 2.23$\,$( 0.57)$ &    $ 197.7\pm   13.1(11.4)$ &    $ 173.8\pm     7.2(4.8)$ &    $ 178.2\pm     8.3(6.0)$ &    $   2.2$ \\
   SDSS J012340.12+004056.4 &  0.9952 & $ 1.76$\,$( 0.24)$ &     $ 206.6\pm    8.6(6.1)$ &    $ 187.8\pm     6.1(3.6)$ &    $ 191.6\pm     6.6(4.1)$ &    $   1.8$ \\
 SDSS J012438.08$-$000346.5 &  0.9933 & $ 4.18$\,$( 0.42)$ &     $ 196.7\pm    7.3(2.3)$ &    $ 162.3\pm     9.4(0.6)$ &    $ 170.6\pm     9.1(1.0)$ &    $   4.2$ \\
\hline
\end{tabular}
\end{center}
\begin{flushleft}
Notes.---  Rotation velocities (cols. 4--6) are listed with their 1-$\sigma$ total (after the $\pm$ symbol) and formal fit uncertainties (in parentheses).
Col. (1): SDSS name of galaxy. 
Col. (2): Inclination correction. 
Col. (3): Turn-over radius, with 1-$\sigma$ total uncertainty. 
Col. (4): Inclination-corrected asymptotic circular velocity. 
Col. (5): Inclination-corrected velocity amplitude at $2.2\rd$. 
Col. (6): Inclination-corrected velocity amplitude at $R_{80}$. 
Col. (7): Reduced $\chi^2$ of the fit.
\end{flushleft}
\label{tab:child_tfpar}
\end{table*}%

The differences between the fit $\chi^2$ values are not being driven by statistical (photon) noise, so their actual values do not carry the expected meaning. Instead, they are driven by systematic deviations from the arctangent model, due to asymmetry or ``hooks'' in the observed rotation curve. So although the typical value of $\chi^2/\nu$ is 0.25, a value larger than unity does not imply that the observed rotation curve is not globally well-described by the arctangent model.

\begin{figure*}
\includegraphics[width=6in]{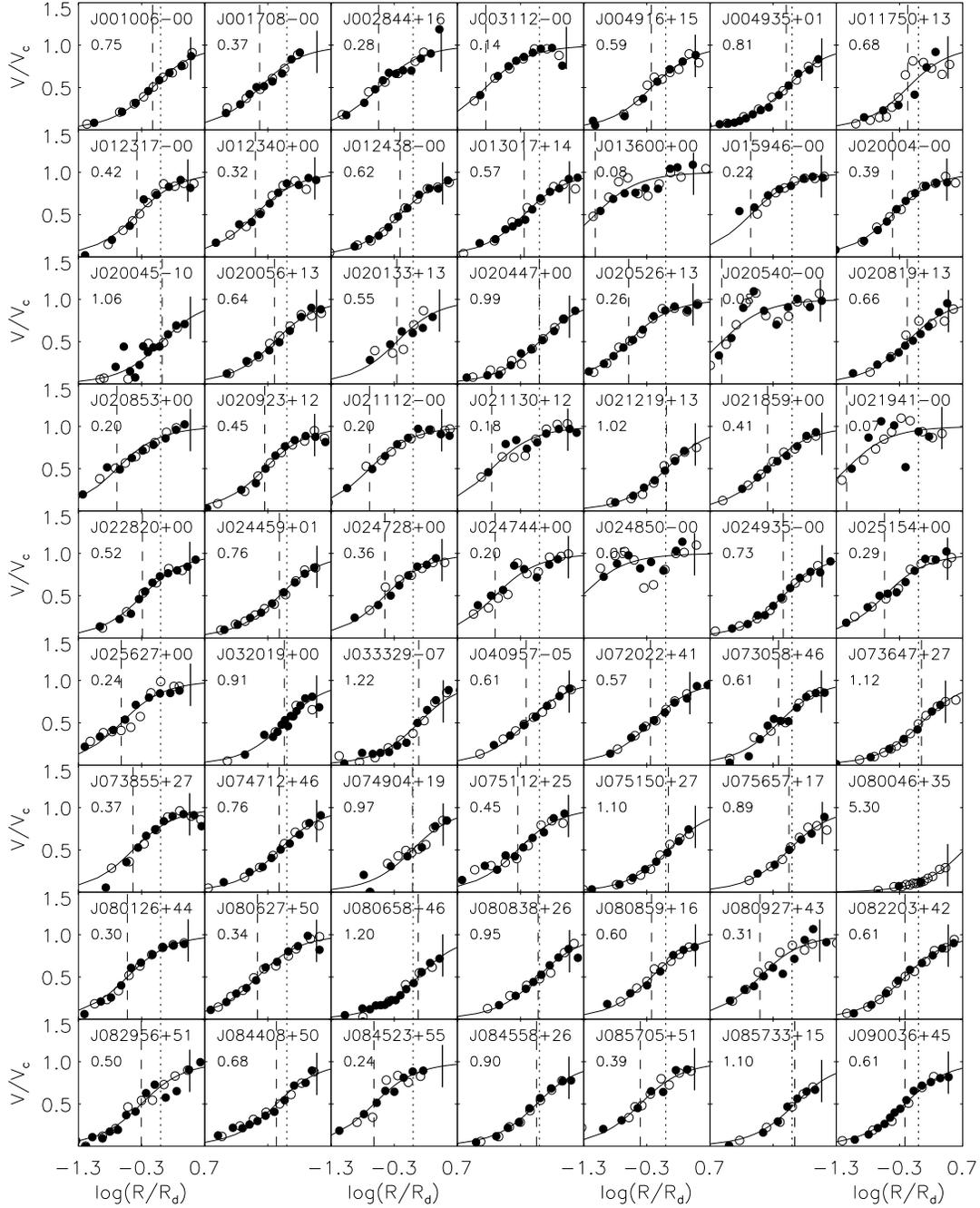}
\caption{Normalized, folded, and scaled rotation curves for 63 (out of the 189) galaxies in the child disk sample, sorted by galaxy name. Velocities are scaled by the best-fit asymptotic circular velocity $V_{\rm c}$; radii are scaled by the disk scale length $\rd$ and shown in a logarithmic scale from 0.05--5$\rd$. Filled and open circles show opposite arms of the rotation curve, folded at the best-fit spatial centre $R_0$. The solid curve shows the best-fit arctangent model. Dashed and dotted lines mark $\rto$ and $\rd$, respectively, and the vertical bar marks $R_{80}$, the radius containing 80 per cent of the $i$-band galaxy light. Note that we define rotation velocity amplitudes at optical radii $2.2\rd$ and $R_{80}$. Panels are labelled with the galaxy name and the ratio of the turn-over radius to the disk scale length, $\rto/\rd$. The rest of the galaxies are shown in Figs.~\ref{fig:rc_panel_b}--\ref{fig:rc_panel_c}.} 
\label{fig:rc_panel_a}
\end{figure*}

\begin{figure*}
\includegraphics[width=6in]{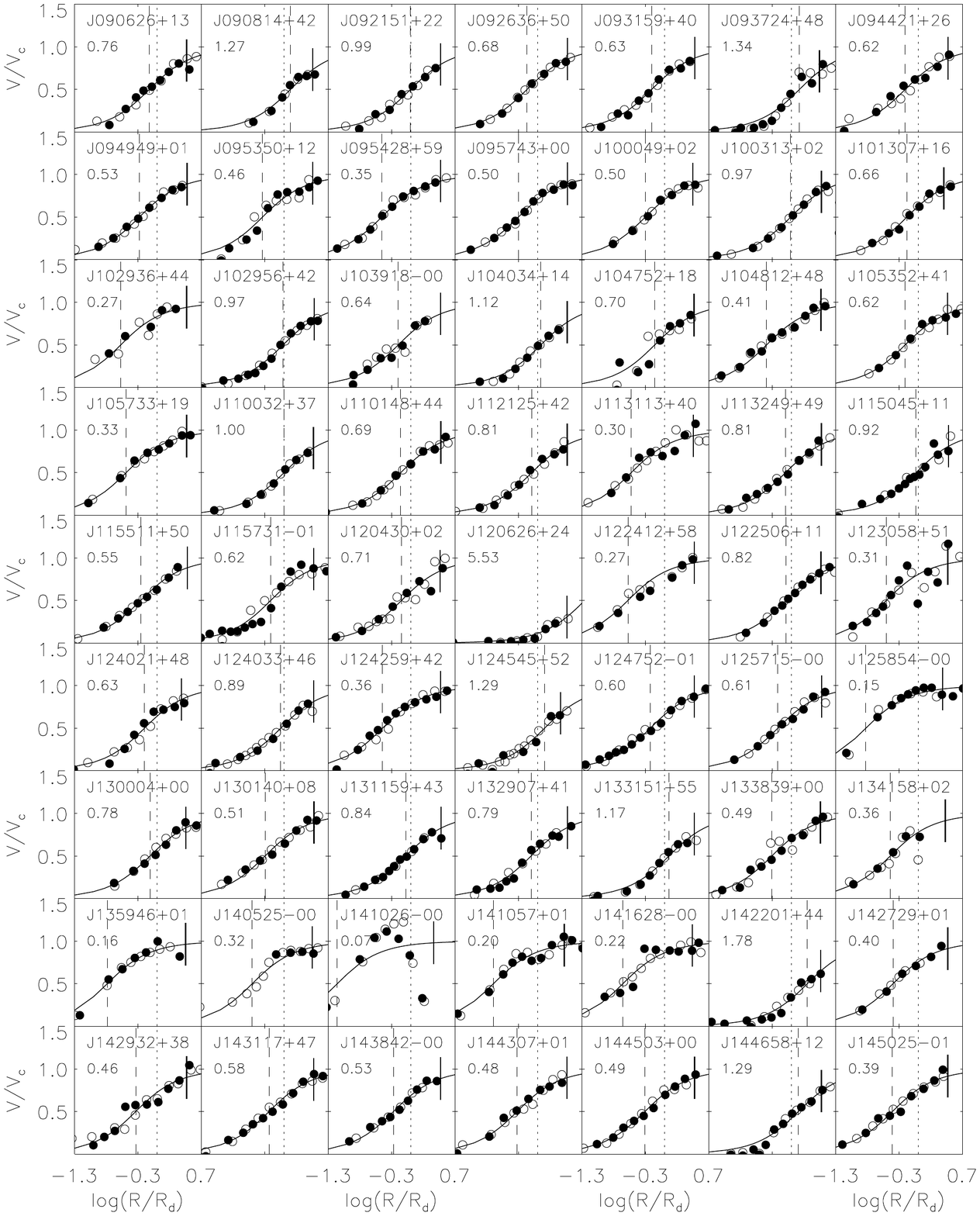}
\caption{Same as Fig.~\ref{fig:rc_panel_a}. Normalized, folded, and scaled rotation curves for 63 galaxies in the child disk sample (2 of 3).}
\label{fig:rc_panel_b}
\end{figure*}

\begin{figure*}
\includegraphics[width=6in]{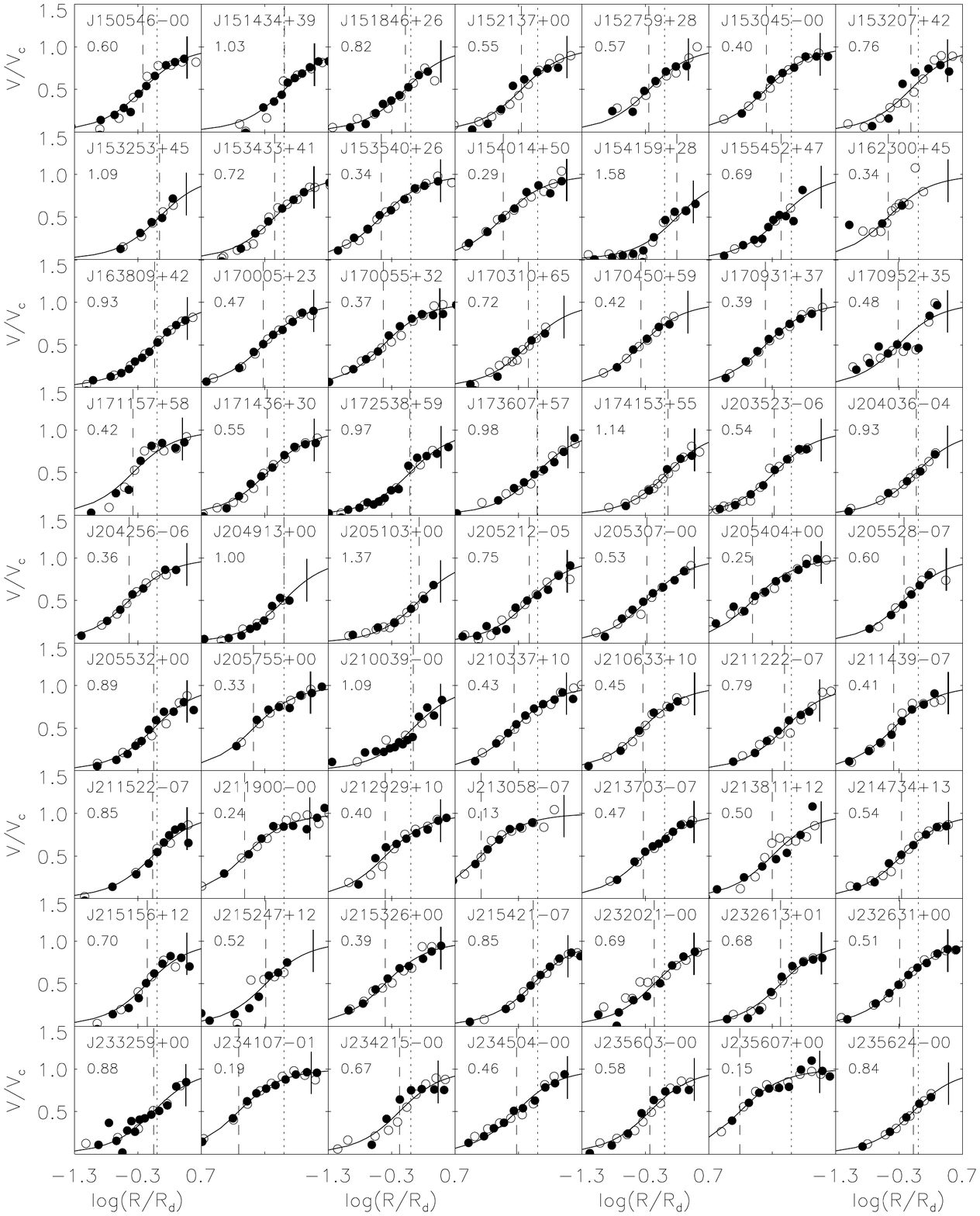}
\caption{Same as Fig.~\ref{fig:rc_panel_a}. Normalized, folded, and scaled rotation curves for 63 galaxies in the child disk sample (3 of 3).}
\label{fig:rc_panel_c}
\end{figure*}

Figures~\ref{fig:rc_panel_a}--\ref{fig:rc_panel_c} show folded, normalized, and scaled rotation curves and best-fit arctangent models for the 189 galaxies in the child disk sample. The rotation curves have been folded at the best-fit spatial centre $R_0$, normalized by the asymptotic circular velocity $V_{\rm c}$, and scaled by the disk scale length $\rd$. Panels are labelled with a shortened galaxy name and $\rto/\rd$. In all but two cases (1 per cent of the sample), the observed rotation curve can be adequately modeled on average by the arctangent model, albeit many galaxies show hooks. 
The only two exceptions are SDSSJ170952+35 (also shown in Fig.~\ref{fig:ex_rcfits}) and SDSSJ141026-00, for which the observed rotation curve turns over toward lower velocities instead of flattening out (see Fig.~\ref{fig:rc_panel_b}). The fits to these two galaxies have extremely poor $\chi^2/\nu$ values of 2.5 and 21.6, respectively. 

For the child disk sample, the median value of the total 1-$\sigma$ uncertainty in $\log V_{\rm c}$ is 0.030 dex. As expected, this uncertainty decreases toward higher stellar masses (i.e., higher $S/N$ data). The median values are 0.045, 0.029, and 0.020 dex for the bottom, intermediate, and top $1/3$ bins in $M_*$, respectively. If the systematic uncertainty is not included, the median value decreases to 0.024 dex for the full sample, and to 0.037, 0.024, and 0.015 dex, for the three $M_*$ bins, respectively. 

\begin{figure}
\includegraphics[width=3in]{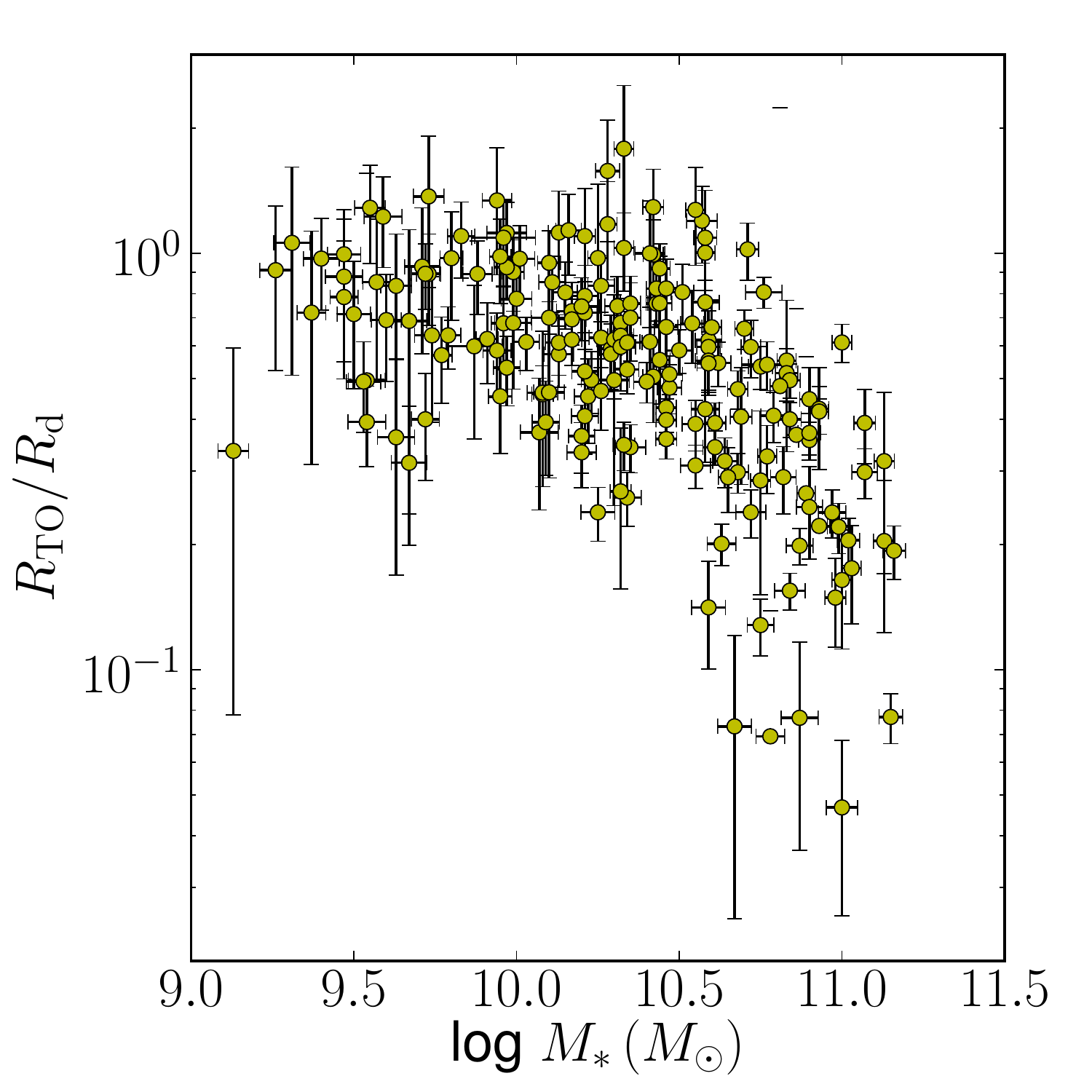}
\caption{Relation between the ratio of turn-over radius to disk scale length $\rto/\rd$ and stellar mass $\mbell$ for 189 galaxies in the child disk sample. Two outlier galaxies--- SDSSJ080046+35 and SDSSJ120626+24--- have $\rto/\rd>5$ and lie beyond this plot.}
\label{fig:mstr_fto}
\end{figure}

Figure~\ref{fig:mstr_fto} shows the relation between stellar mass $M_*$ and the ratio of turn-over radius to disk scale length, $\rto/\rd$. We confirm that the most luminous galaxies tend to have very steeply-rising rotation curves \citep[e.g.,][]{1996MNRAS.281...27P,2006ApJ...640..751C}. At the other extreme, we find two outliers with $\rto/\rd>5$ (SDSSJ080046+35 and SDSSJ120626+24). The rotation curves of both galaxies are observed out to their $R_{80}$, but are still rising at the outermost point.\footnote{SDSSJ120626+24 appears to be a central galaxy of a group, which explains its rising rotation curve beyond its optical radius.} 
Consequently, the asymptotic circular velocity is not well-constrained, resulting in extremely large values of $\sigma(V_{\rm c})$ of 960 $\kms$ and 605 $\kms$, respectively, as well as largely overestimated values of $V_{\rm c}$, equal to 398 $\kms$ and 457 $\kms$, respectively. 

\subsection{Rotation velocity amplitudes}
\label{subsec:vrot_definition}

We now define inclination-corrected rotation velocity amplitudes that we will use in the TFRs. In addition to the asymptotic circular velocity $V_{\rm c}$, we define rotation velocities evaluated at some suitably chosen optical radius $\ropt$. A common choice for $\ropt$ is $2.2$ times the disk scale length $2.2\rd$ \citep[][]{1999ApJ...513..561C}, chosen because this is the radius at which the rotation curve of a pure self-gravitating exponential disk would peak \citep{1970ApJ...160..811F}. We also adopt $R_{80}$, following P07, the radius containing 80\% of the $i$-band flux. As noted in \S\ref{subsec:bdfit_results}, $R_{80}/\rd=3.03$ for a pure exponential disk, but is smaller for galaxies with significant bulges. Both $\rd$ and $R_{80}$ have been corrected for the effect of seeing, but compared to $\rd$, $R_{80}$ has the advantage of being less sensitive to the degeneracies of the B+D fits. Moreover, $R_{80}$ is closer to the peak of the total (disk+DM halo) rotation curve, which is located at $\sim$3$\rd$ for typical haloes of disk galaxies \citep{1998MNRAS.295..319M}. 

We follow convention and correct the disk rotation velocities to the edge-on orientation (assuming circular symmetry)\footnote{Without assuming circular symmetry, the rotation velocity amplitudes would depend not only on inclination, but also on the orientation of the galaxy with respect to the line of sight. When we look down the long axis of gas orbits, the observed velocities are larger than when we look down the short axis. Deviations from circular symmetry and the variation in viewing angles contribute to the observed scatter in the TFRs (see \S\ref{subsec:tfr_interpret} for further discussion).} 
\beqa \label{eq:vcobs}
V_{\rm c} &=& V_{\rm c,obs}/\sin\theta, \\ \label{eq:v22obs}
V_{2.2} &=& V_{\rm 2.2,obs}/\sin\theta =V(R=2.2 \rd)/\sin\theta, \\ \label{eq:v22obs}
V_{80} &=& V_{\rm 80,obs}/\sin\theta = V(R=R_{80})/\sin\theta.
\eeqa
Here, the inclination corrections are given by
\beq \label{eq:sin_theta}
\sin \theta 
= \left(\frac{1-q_{\rm d}^2}{1-q_z^2}\right)^{1/2},
\eeq
where $q_z$ is the intrinsic axis ratio, or the ratio of the vertical and radial scale lengths of the disk. Following P07, we adopt a single value for $q_z=0.19$ \citep{1984AJ.....89..758H}, although it is known to vary slightly from disk to disk.\footnote{From analyses of the distribution of axis ratios of SDSS galaxies, \citet{2004ApJ...601..214R} found $q_z=0.22\pm 0.06$ and \citet{2008MNRAS.388.1321P} found $q_z=0.21\pm 0.02$.}

\begin{figure}
\includegraphics[width=3in]{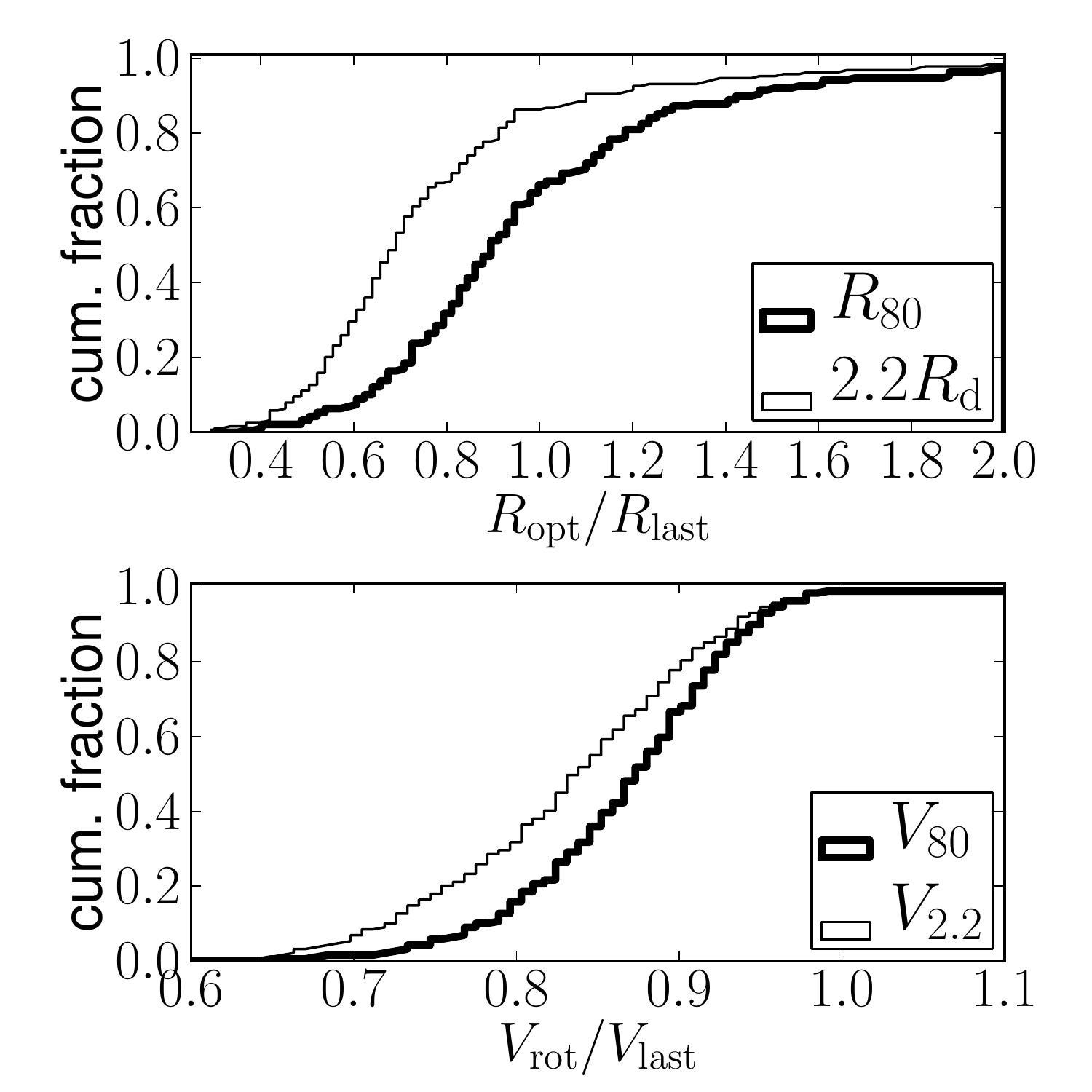} 
\caption{Cumulative distributions of $R_{80}/R_{\rm last}$, $2.2\rd/R_{\rm last}$ (thick and thin curves in the upper panel, respectively) and $V_{80}/V_{\rm last}$ and $V_{2.2}/V_{\rm last}$ (thick and thin curves in the lower panel, respectively) for the child disk sample.
}
\label{fig:chist_panels}
\end{figure}

Figure~\ref{fig:chist_panels} shows cumulative histograms in the ratios $\ropt/R_{\rm last}$ and $\vrot/V_{\rm last}$, where $R_{\rm last}$ is the radius of the outermost point in the observed rotation curve, and $V_{\rm last}$ is the circular velocity at that radius based on the best-fit arctangent model. We find that most of the observed rotation curves extend close to or beyond $R_{80}$, so a substantial extrapolation of the rotation curve is rarely required. Moreover, $R_{80}$ is far enough out to be close to $V_{\rm c}$ in most cases. This is also evident in the panels of Figs.~\ref{fig:rc_panel_a}--\ref{fig:rc_panel_c}, where $R_{80}$ is marked by solid vertical bars. These results support the empirical logic of adopting the rotation velocity amplitude $V_{80}$ (as in P07). 

Moreover, $V_{80}$ is relatively insensitive to changes in the value of $R_{80}$ because it samples the flat portion of the rotation curve in most cases. For example, even though $R_{80}$ varies from 1.2--1.4 $\times (2.2\rd)$, $V_{80}$ is within 10\% of $V_{2.2}$ for most galaxies, and is only 5\% larger on average. Therefore, the uncertainty in $R_{80}$ or $\rd$, including, e.g., the effect of internal extinction,\footnote{Some authors have corrected the observed disk scale lengths for internal extinction. We do not apply such corrections because the magnitude and even the sign of the correction, which depends on the spatial distribution of dust in the disk, is a priori unclear. \citet{2009ApJ...691..394M} found that more highly-inclined late-type galaxies have larger half-light radii than face-on ones, indicating that attenuation of dust is stronger in the center of galaxies. \citet{2007ApJ...671..203C} applied a correction factor of $(1+0.4\log(a/b))^{-1}$, following \citet{1997AJ....113...53G}. This correction is at most 29 per cent, comparable to the difference between $R_{80}$ and $2.2\rd$. The associated systematic bias in $V_{80}$ is thus comparable to the typical measurement uncertainty in $V_{80}$, of around 0.03 dex.} 
has a relatively small effect on the rotation velocity amplitudes. 


To calculate the total error in $V_{80}$, we propagate the formal errors in the best-fit model parameters using the full covariance matrix, then include a systematic uncertainty similar to that defined for the fit parameters (Eq.~\ref{eq:sigma_sys}), as well as a contribution from the uncertainty in the inclination correction
\beqa \label{eq:err_v80}
\left(\delta V_{80}\right)^2 &=& \left. \sum_{ij} \left(\frac{\partial V(R)}{\partial a_i}\right)  \left(\frac{\partial V(R)}{\partial a_j}\right) C_{ij} \right|_{R=R_{80}} \\
&+& \left( V_{80} - V_{80}^{\rm (r)} \right)^2 + \left(\delta(\sin\theta)\right)^2.
\eeqa
Here, $V(R)$ is the arctangent model function defined in Eq.~\ref{eq:atan_vprof}, and $a_i=\{V_0, V_{\rm c}, R_0, \rto\}$ are the fit parameters with formal fit covariance matrix $C_{ij}$, $V_{80}^{\rm (r)}$ is the velocity amplitude at $R_{80}$ for the best-fit model derived from a fit using the radius-weighted scheme (c.f. Eq.~\ref{eq:sigma_sys}), and $\delta(\sin\theta)$ is based on the formal fit error in $\qd$. $\delta V_{2.2}$ is defined similarly. 

The error in $V_{80}$ is largely dominated by the formal fit error; the contribution from the formal inclination correction uncertainty is negligible. The systematic uncertainty in $V_{80}$ is smaller than that in $V_{\rm c}$, by approximately half. For the child disk sample, the 5th, 50th, and 95th percentile values of $\log V_{80}$ are 0.012, 0.027, and 0.10 dex, respectively. The errors in $\log V_{2.2}$ are similar. 

\subsection{Tests of systematics}
\label{subsec:long_syst}

We now investigate potential systematic effects on the measured rotation curves. In \S\ref{subsubsec:syst_pos}, we look at the effect of slit misalignments by comparing observations at different slit PAs (along the minor axis and $\pm 10^\circ$ off the major axis). In \S\ref{subsubsec:syst_piz}, we look for any systematic differences between observations performed by P07 with CCDS/TWIN and those performed with DIS. In \S\ref{subsubsec:syst_fits}, we compare the best-fit parameters from our rotation curve fits with those from P07.

\subsubsection{Off-axis PA observations}
\label{subsubsec:syst_pos}

\begin{figure*}
\bec 
\begin{tabular}{m{1.5in}m{3.5in}}
\includegraphics[width=1.5in]{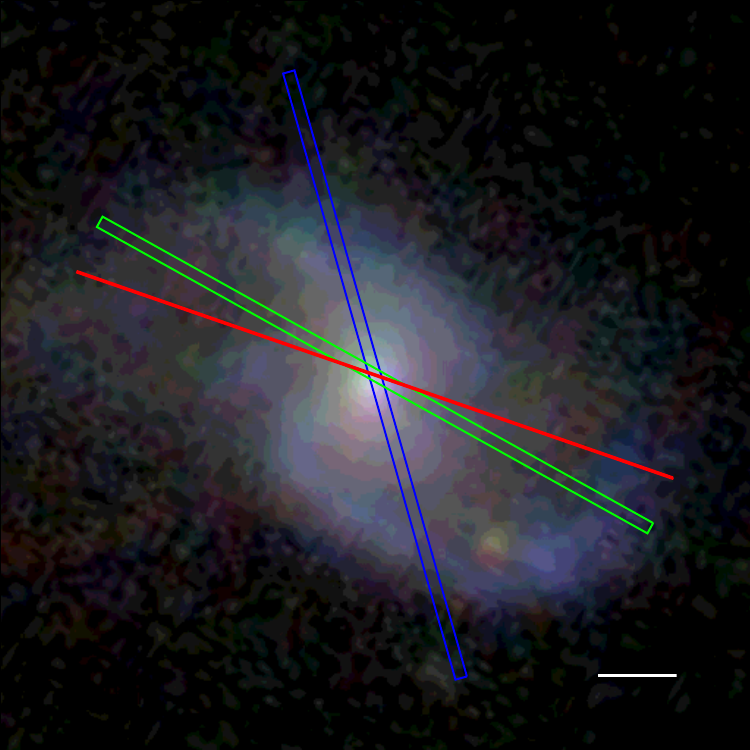} &
\includegraphics[width=3.5in]{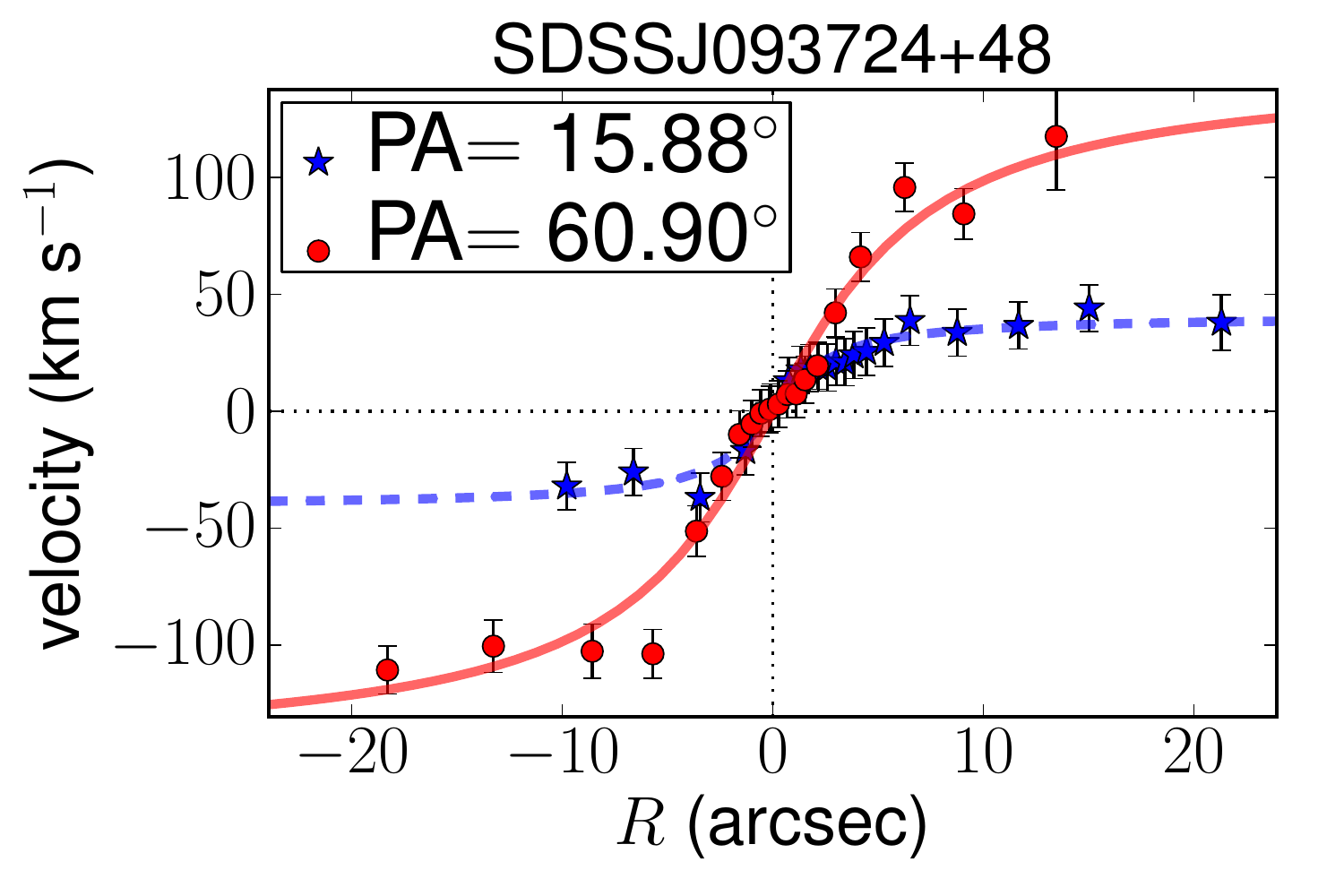} \\
\end{tabular} 
\eec 
\caption{{\it Left panel:} SDSS image of the galaxy SDSSJ093724+48, the only case where the exponential fit and isophotal PAs are significantly different. We observed this galaxy at two slit positions: the $r$-band exponential fit PA (E of N) of $15.88^\circ$ (blue rectangular box) and at an eye estimate of the galaxy major axis PA of $60.9^\circ$ (green rectangular box). The isophotal PA of 70.9$^\circ$ is also shown for reference (thick red line). In this image, North is up and East is toward the left; the horizontal line on the lower-right corner is 5\arcsec\ long; the green box is 1.5\arcsec\ by 40\arcsec\ (the actual slit covers 1.5\arcsec\ by 2$^\prime$, so its length spans the whole extent of the galaxy). {\it Right panel:} Rotation curves observed at the two slit PAs shown in the left panel. The observation at PA $=60.9^\circ$ yields a best-fit asymptotic circular velocity 3.6 times larger than that for the observation at PA $=15.88^\circ$ ((red circles and solid curve vs. blue stars and dashed curve), a strong indication that this PA is closer to the true major axis of the galaxy; therefore, we adopt it in our final analysis.} 
\label{fig:syst_rc_mismatch}
\end{figure*}

\begin{figure*}
\bec $
\begin{array}{cc} 
\includegraphics[width=1.75in]{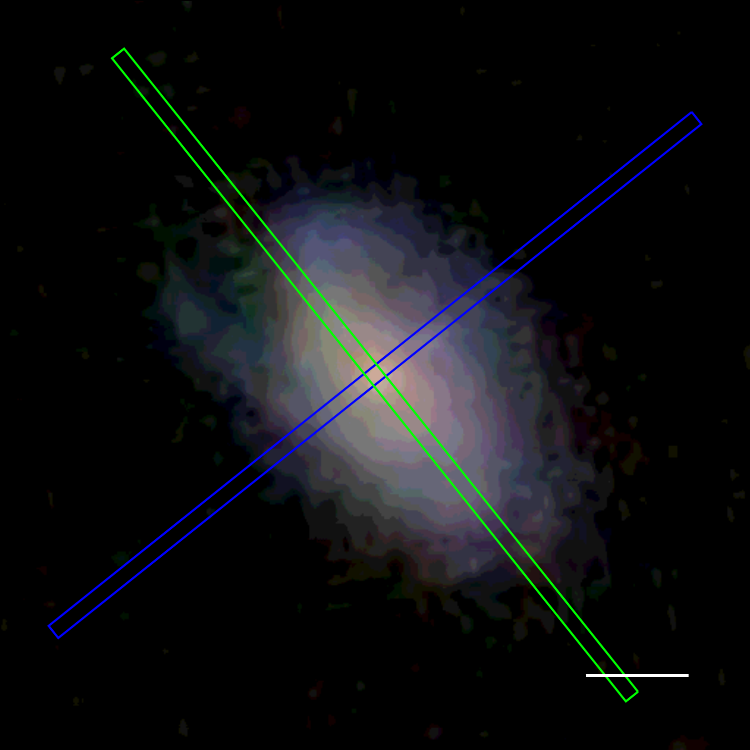} &
\includegraphics[width=1.75in]{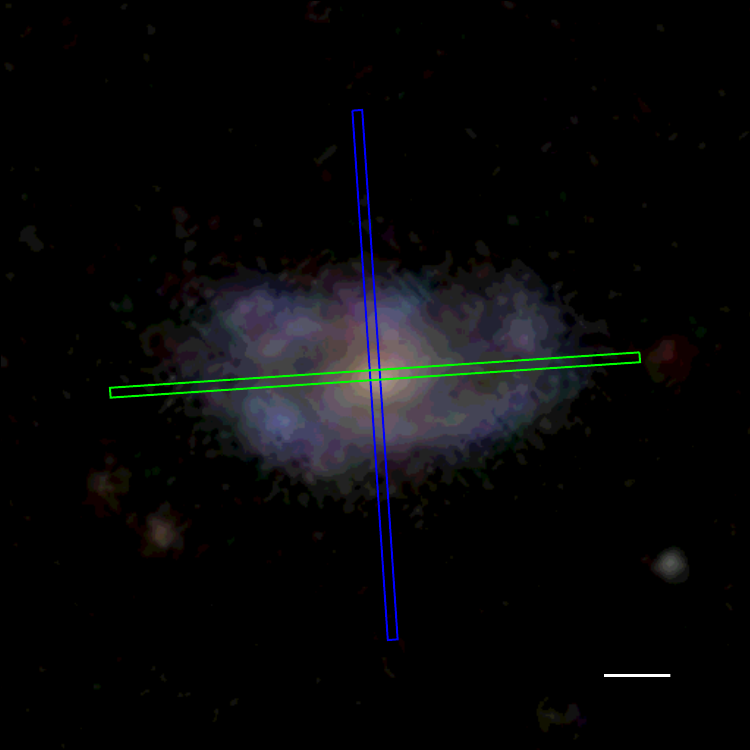} \\
\includegraphics[width=1.75in]{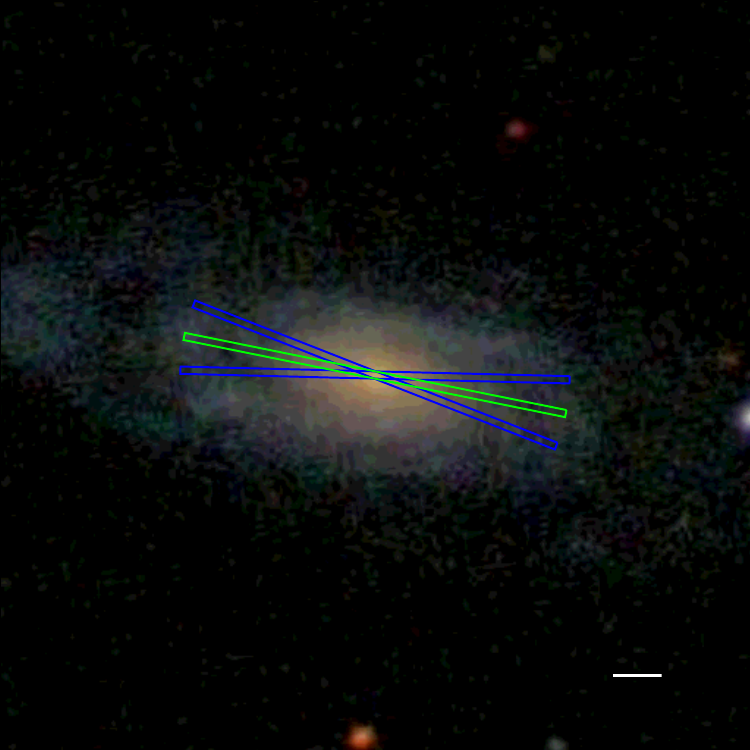} &
\includegraphics[width=1.75in]{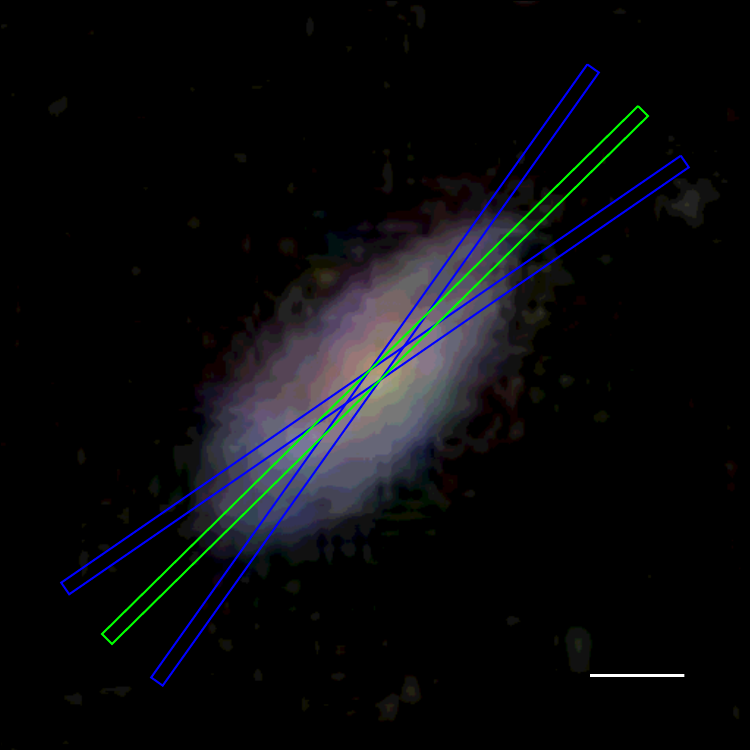} \\
\end{array} $
\eec
\caption{
SDSS images of four galaxies with off-axis slit PA observations. Here, green rectangular boxes indicate the major axis PAs, and blue boxes indicate the off-axis slit PAs. As in Fig.~\ref{fig:syst_rc_mismatch}, North is up and East is toward the left; the horizontal line on the lower-right corner is 5\arcsec\ long; green rectangles are 1.5\arcsec\ by 40\arcsec. %
{\it Upper panels}: SDSSJ144658+12 and SDSSJ154014+50 were observed along the minor axis PA (blue rectangular boxes).
{\it Lower panels}: SDSSJ020526+13 and SDSSJ130140+08 were observed at $\pm$10$^\circ$ off the major axis PA (blue rectangular boxes).
}
\label{fig:syst_pos_img}
\end{figure*}

We do not have a direct measurement of the kinematic major axis PA, and rely on photometry to determine the major axis PA. The SDSS photometric pipeline provides three different measurements of the photometric major axis PA: (i) from the exponential and deVaucoleurs model fits, (ii) from ellipticity measurements at the 25 mag arcsec$^{-2}$ isophote, and (iii) from flux-weighted adaptive second moments. 
The B+D fits performed in \S\ref{sec:bdfit} provide another measurement of the disk major axis PA. The model and B+D fits account for the effect of the seeing, but the other two measurements do not. 

If the velocity field is noncircular, the photometric major axis may be offset from the kinematic major axis. 
For a potential with an ellipticity of 0.1, the mean PA difference is 5$^\circ$ (averaged over all viewing angles), roughly in agreement with observed values \citep{1992ApJ...392L..47F}. In addition, there is some uncertainty in the determination of the photometric major axis PA itself. Comparing the different photometric major axis PA estimates, we find that except for one outlier--- SDSSJ093724+48--- all galaxies in the child disk sample have an isophotal PA within $\pm 11^\circ$, and a B+D fit PA within $\pm 6^\circ$, of its exponential fit PA. Taking into account both effects, the typical level of PA offsets is expected to be $\la 10^\circ$. 

As the default configuration, we observed galaxies at the photometric major axis PA determined from the $r$-band exponential model fit performed by the SDSS photometric pipeline. This was done for 86 out of the 95 targets, including SDSSJ093724+48. Seven galaxies\footnote{
These galaxies are: SDSSJ110148+44, SDSSJ124021+48, SDSSJ131159+43, SDSSJ132907+41, SDSSJ133151+55, SDSSJ143117+47, and SDSSJ171157+58.
} were observed at the $r$-band isophotal PA, and two galaxies\footnote{
These galaxies are SDSSJ142201+44 and SDSSJ151434+39.
} were observed at a PA determined by eye, which turn out to be within 3$^\circ$ of the exponential fit PA (and within 1$^\circ$ of the isophotal PA). All observed slit PAs are listed in Table~\ref{tab:long_obs}.

For the outlier SDSSJ093724+48, a two-armed spiral, the ideal slit orientation is unclear (see left panel of Fig.~\ref{fig:syst_rc_mismatch}). We first observed at the exponential fit PA of $15.88^\circ$ (blue rectangular box in Fig.~\ref{fig:syst_rc_mismatch}), and then at an eye estimate for the galaxy major axis PA of $60.9^\circ$ (green rectangular box). The SDSS isophotal PA is $70.9^\circ$ (thick red line), 10 deg off the observed PA. The right panel of Fig.~\ref{fig:syst_rc_mismatch} shows the rotation curves and best-fit arctangent models for the two observations, obtained using procedures described in \S\ref{subsec:rc_extraction} and \S\ref{subsec:rc_fitting}. The slit orientation with PA$=60.9^\circ$ clearly yielded a larger rotation velocity amplitude, with a best-fit asymptotic circular velocity (before inclination correction) of $\vcobs=148 \pm 16\,\kms$, compared to $41\pm 7\, \kms$ for the other slit orientation. This orientation is therefore closer to the true major axis of the galaxy, and we adopt this observation in our final analysis. 

\begin{figure}
\includegraphics[width=3in]{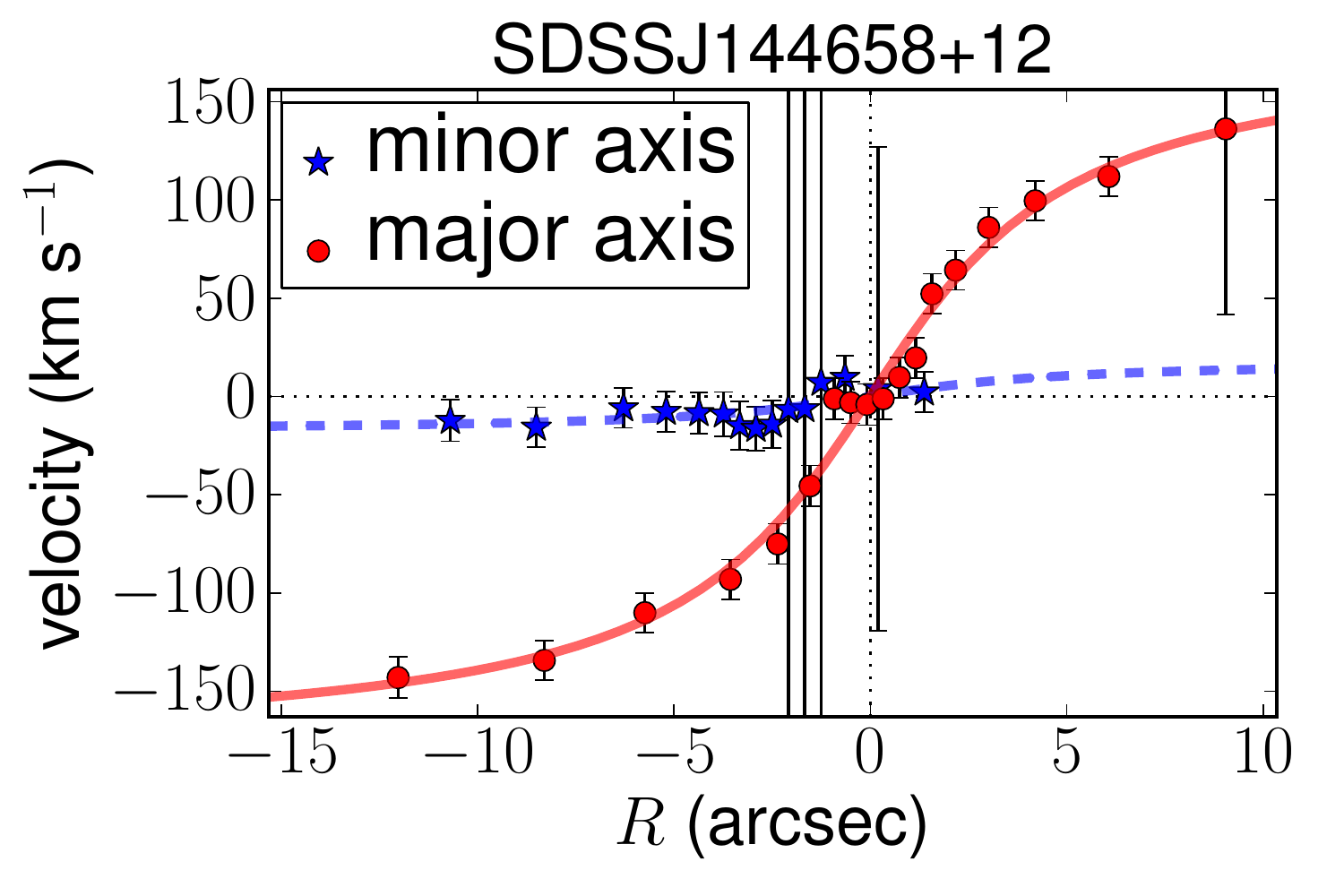} 
\includegraphics[width=3in]{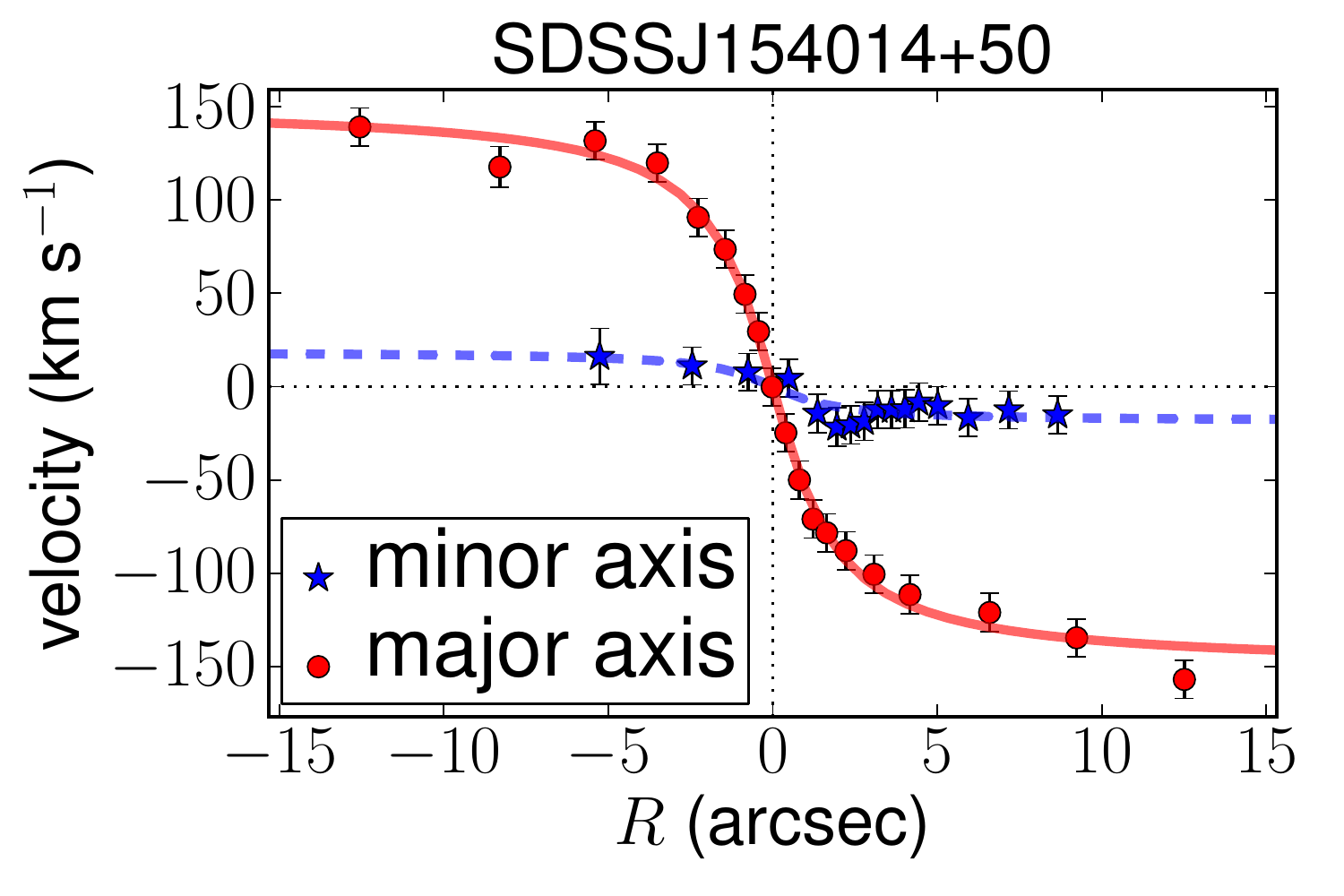} 
\caption{Major and minor axis rotation curves for SDSSJ144658+12 and SDSSJ154014+50 observed along the minor axis, i.e. at 90$^\circ$ off the major axis (blue stars and dashed curves), compared to those observed along the major axis (red circles and solid curves).}
\label{fig:syst_pos_rc_minor}
\end{figure}

To check for a velocity gradient along the minor axis, a clear signature of noncircular motions, we observed two of our targets---SDSSJ144658+12 and SDSSJ154014+50--- along the minor axis, i.e., perpendicular to the observed major axis PA (blue and green rectangular boxes in the upper panels of Fig.~\ref{fig:syst_pos_img} show the minor and major axis PAs, respectively). Figure~\ref{fig:syst_pos_rc_minor} compares the minor and major axis rotation curves and best-fit arctangent models (red circles/solid curves and blue stars/dashed curves, respectively). 

For SDSSJ144658+12, the best-fit parameters are $\vcobs=181 \pm 17\,\kms$\, and $\rto = 3.8 \pm 0.6\arcsec$, with a reduced $\chi^2$ of 1.1. Fitting the minor axis rotation curve with $\rto$ fixed at this value, we find $\vcobs=18 \pm 29\,\kms$, consistent with zero (with a reduced $\chi^2$ of 0.25). Performing a similar procedure for SDSSJ154014+50, we find $\vcobs=151 \pm 6 \mbox{ and } 19 \pm 6\,\kms$, with reduced $\chi^2$ values of 0.57 and 0.32, for the major and minor axis observations, respectively. In this case, there is some evidence for rotation in the minor axis rotation curve, but only at 1/8 of the value of that for the major axis rotation curve.

\begin{figure}
\includegraphics[width=3.in]{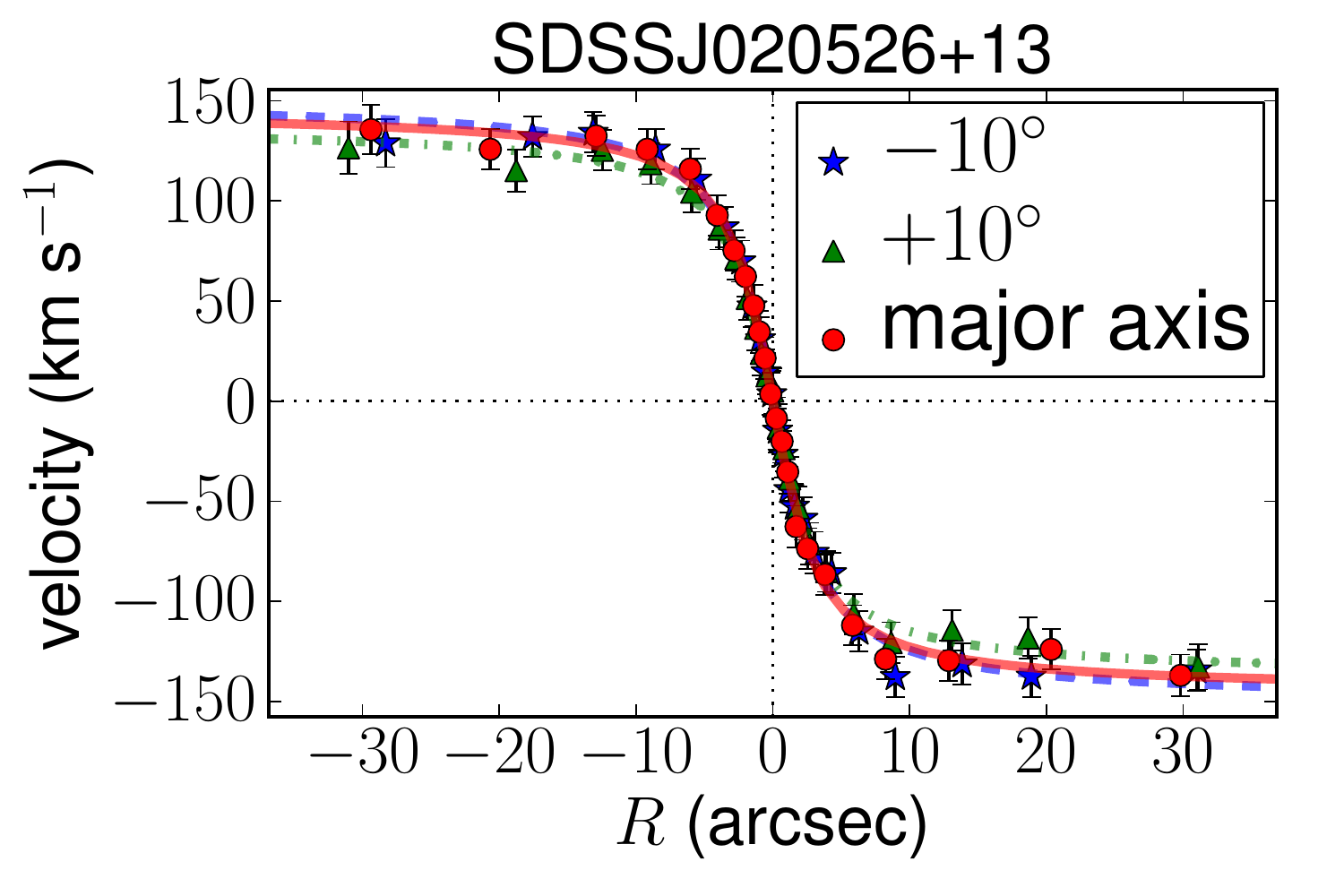} 
\includegraphics[width=3in]{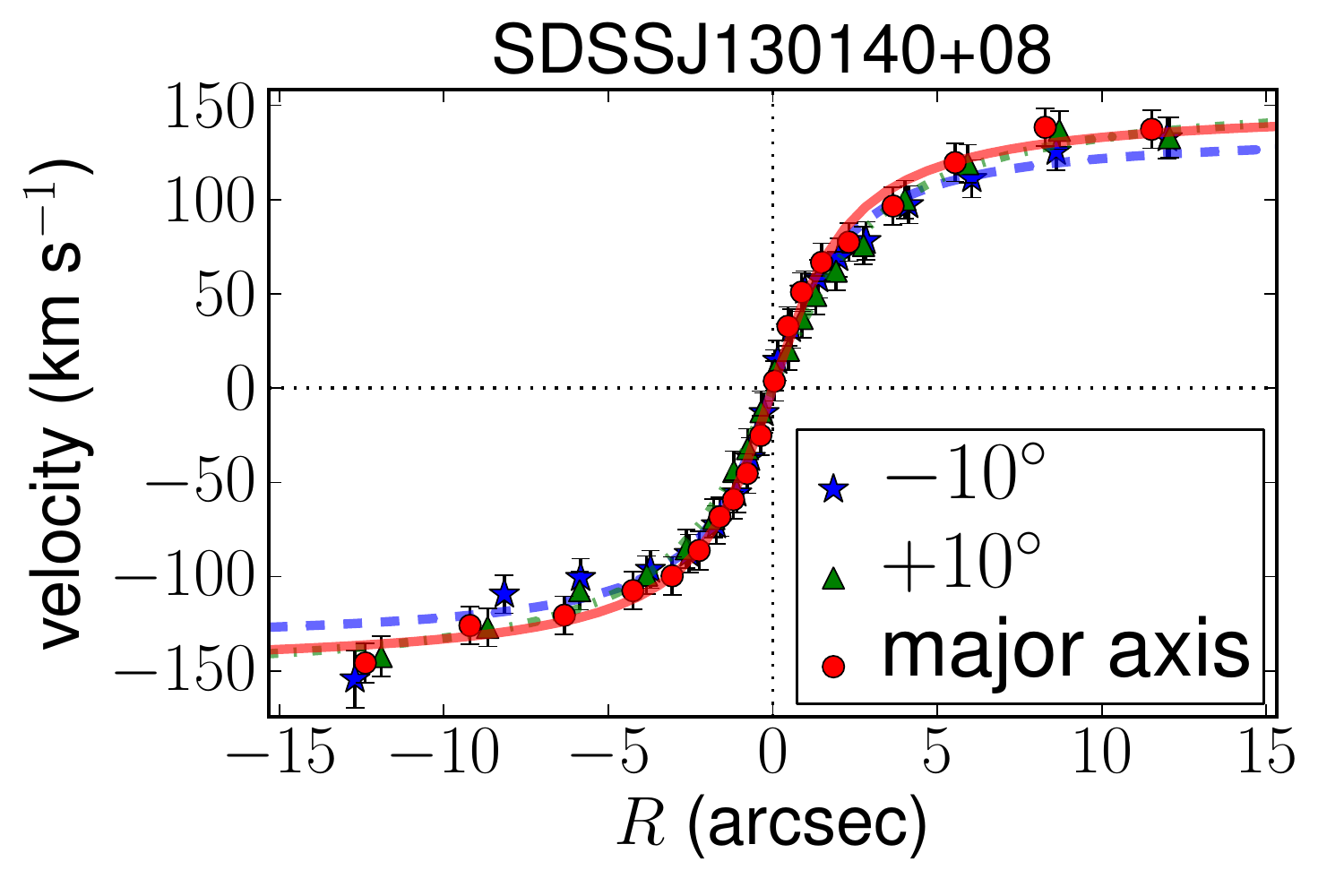} 
\caption{Rotation curves for SDSSJ020526+13 and SDSSJ130140+08 observed at $\pm$10$^\circ$ off the major axis (green triangles and dot-dashed curves, and blue stars and dashed curves, respectively), compared to those observed along the major axis (red circles and solid curves).}
\label{fig:syst_pos_rc_offset}
\end{figure}

To test the effect of slit misalignments on the inferred velocity amplitudes, we observed two of our targets---SDSSJ020526+13 and SDSSJ130140+08--- at $\pm 10^\circ$ offsets from the observed major axis PAs (blue rectangular boxes in the lower panels of Fig.~\ref{fig:syst_pos_img} show the offset slit PAs and green boxes show the major axis PA). Figure~\ref{fig:syst_pos_rc_offset} compares the observed rotation curves for the different slit orientations and the best-fit arctangent models. For both galaxies, one of the off-axis rotation curves has a slightly lower velocity amplitude compared to the other two (as expected if the major axis PA is well determined). Table~\ref{tab:syst_offaxis_rcfit} lists the best-fit arctangent parameters for these observations, together with their formal 1-$\sigma$ fit uncertainties. The rotation velocity amplitudes $V_{\rm 80,obs}$ (col.~6) for the different observations are consistent to within 1$\sigma$, indicating that this quantity is robust to slit PA offsets of order $10^\circ$ (the expected level of slit misalignments, as noted earlier). 

We conclude that slit misalignments are not expected to introduce a significant systematic bias in the TFR toward lower rotation velocity amplitudes. We note, though, that they are expected to contribute to the observed scatter in the relation (as discussed in \S\ref{subsec:tfr_interpret}).


\begin{table*}
\caption{Rotation curve fits for targets with both major axis and off-axis observations (2 galaxies).}
\begin{center}
\begin{tabular}{crrrrrr}
\hline 
 \multicolumn{1}{c}{Galaxy name} &
 \multicolumn{1}{c}{Slit PA} &
 \multicolumn{1}{c}{$\vcobs$} &
 \multicolumn{1}{c}{$\rto$} &
 \multicolumn{1}{c}{$R_{80}$} &
 \multicolumn{1}{c}{$V_{\rm 80,obs}$} &
 \multicolumn{1}{c}{$\chi^2/\nu$} \\
 \multicolumn{1}{c}{} &
 \multicolumn{1}{c}{(deg)} & 
 \multicolumn{1}{c}{($\kms$)} &    
 \multicolumn{1}{c}{(arcsec)} &
 \multicolumn{1}{c}{(arcsec)} &
 \multicolumn{1}{c}{($\kms$)} & 
 \multicolumn{1}{c}{} \\ 
 \multicolumn{1}{c}{(1)} & 
 \multicolumn{1}{c}{(2)} & 
 \multicolumn{1}{c}{(3)} & 
 \multicolumn{1}{c}{(4)} &
 \multicolumn{1}{c}{(5)} &
 \multicolumn{1}{c}{(6)} &
 \multicolumn{1}{c}{(7)} \\
\hline
SDSSJ020526+13 & $ 78.57$ & $144.9 \pm  5.0$ & $2.43 \pm 0.26$ & $ 25.27$ & $136.0 \pm 4.8$ & $ 0.27$ \\
            -- & $ 68.57$ & $149.6 \pm  5.3$ & $2.69 \pm 0.29$ & $     -$ & $139.5 \pm 5.1$ & $ 0.36$ \\
            -- & $ 88.57$ & $137.1 \pm  5.4$ & $2.58 \pm 0.31$ & $     -$ & $128.2 \pm 5.1$ & $ 0.22$ \\
\hline
SDSSJ130140+08 & $-45.43$ & $149.8 \pm  6.4$ & $1.78 \pm 0.21$ & $ 10.49$ & $133.8 \pm 6.0$ & $ 0.41$ \\
            -- & $-55.43$ & $136.6 \pm  6.7$ & $1.72 \pm 0.23$ & $     -$ & $122.4 \pm 6.3$ & $ 0.71$ \\
            -- & $-35.43$ & $157.1 \pm  8.1$ & $2.50 \pm 0.32$ & $     -$ & $133.7 \pm 7.4$ & $ 0.26$ \\
\hline
\end{tabular}
\end{center}
\begin{flushleft}
Notes.---  Col. (1): Galaxy name. Col. (2): Slit position angle in degrees East of North. The first line is the major axis PA; the next two lines are offsets by $\pm 10^\circ$.
Col. (3): Best-fit observed asymptotic circular velocity $\vcobs$.
Col. (4): Best-fit turn-over radius $\rto$.
Col. (5): Radius containing 80 per cent of the $i$-band light.
Col. (6): Velocity amplitude evaluated at $R_{80}$, before inclination correction, $V_{\rm 80,obs}$.
Col. (7): Reduced $\chi^2$ of the fit.
Cols. (3), (4), \& (6) are listed with their formal 1-$\sigma$ fit errors (not including systematic uncertainties). 
\end{flushleft}
\label{tab:syst_offaxis_rcfit}
\end{table*}%

\begin{figure*}
\bec $
\begin{array}{cc} 
\includegraphics[width=3in]{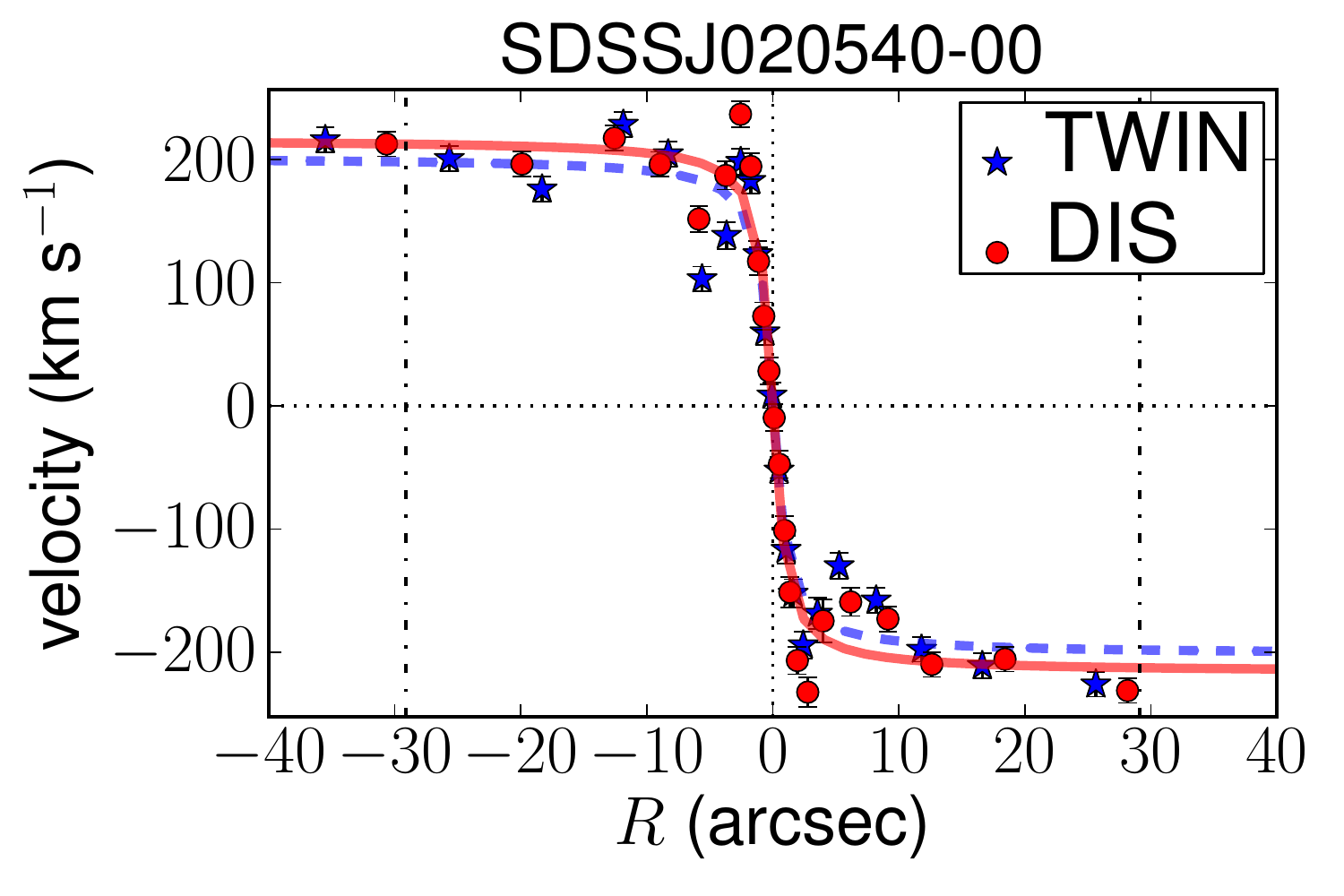} &
\includegraphics[width=3in]{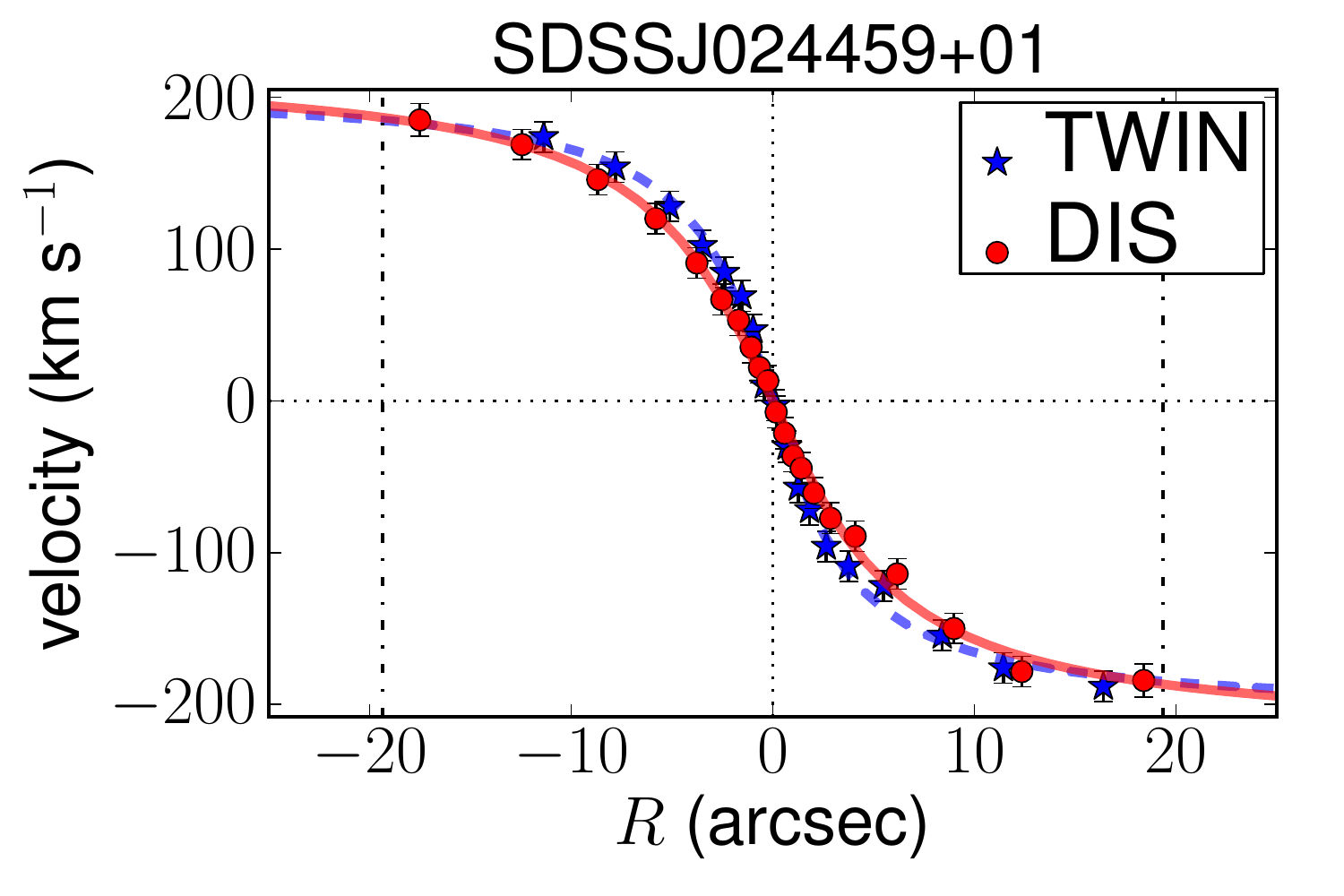} \\
\includegraphics[width=3.in]{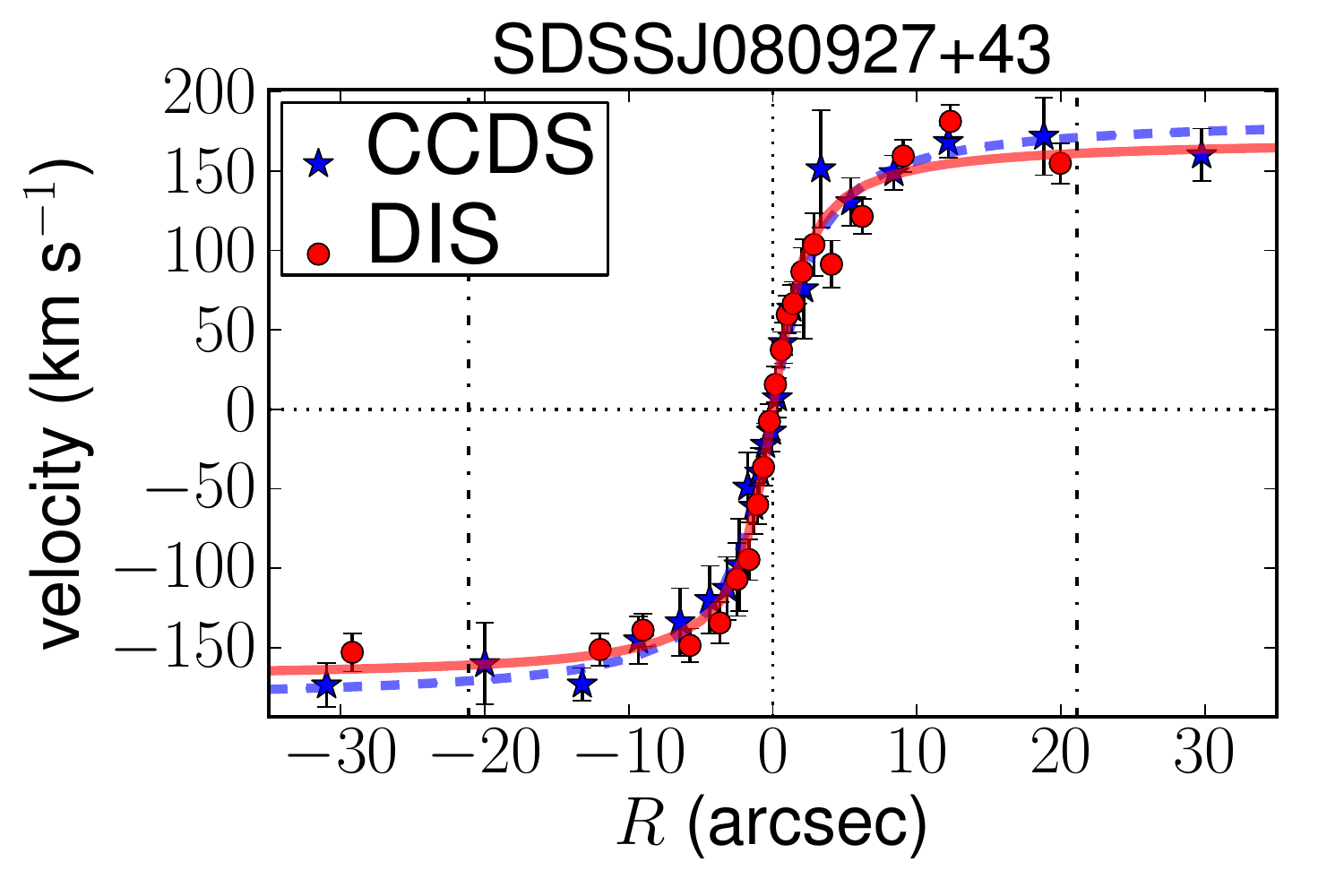} &
\includegraphics[width=3in]{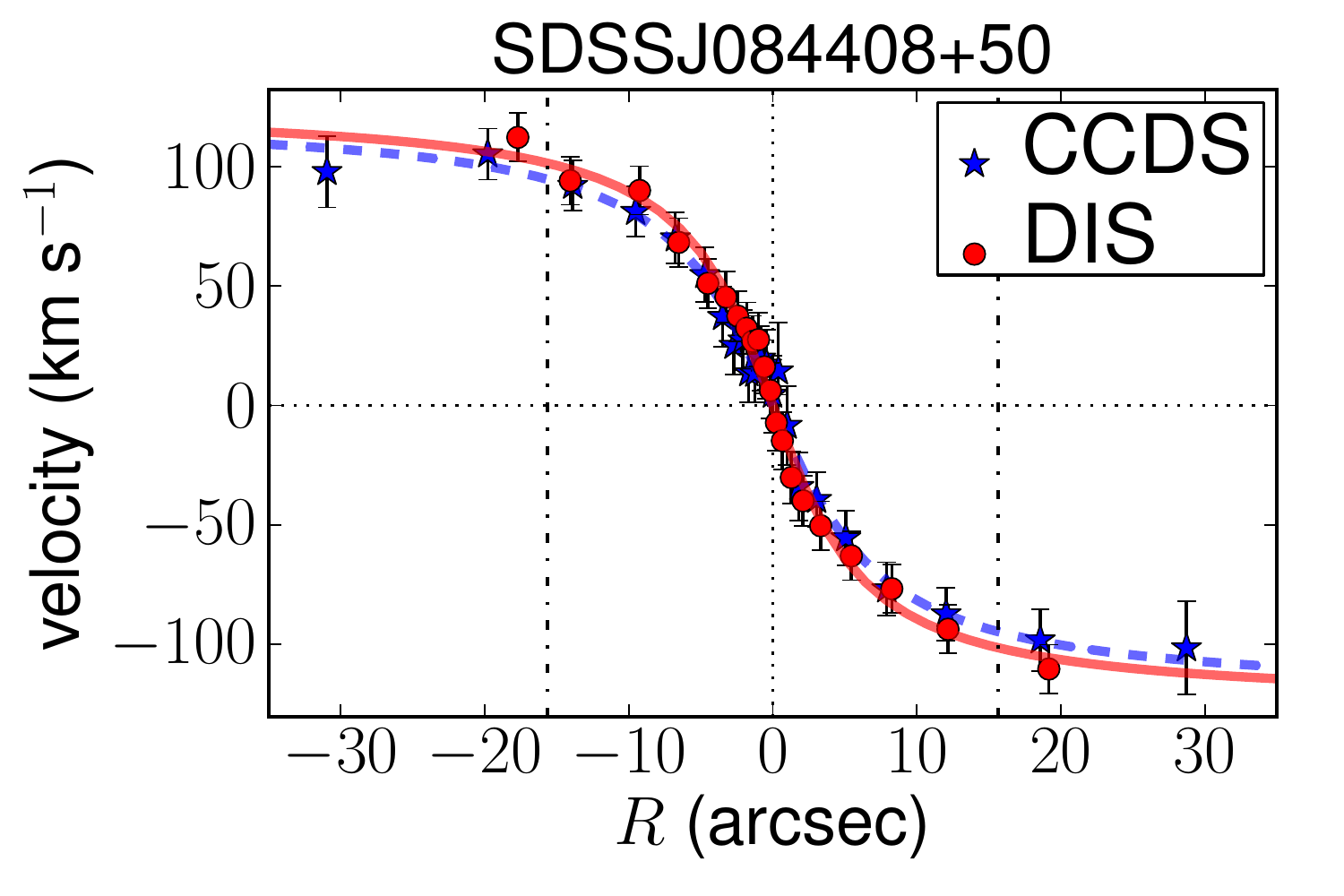} \\
\multicolumn{2}{c}{\includegraphics[width=3in]{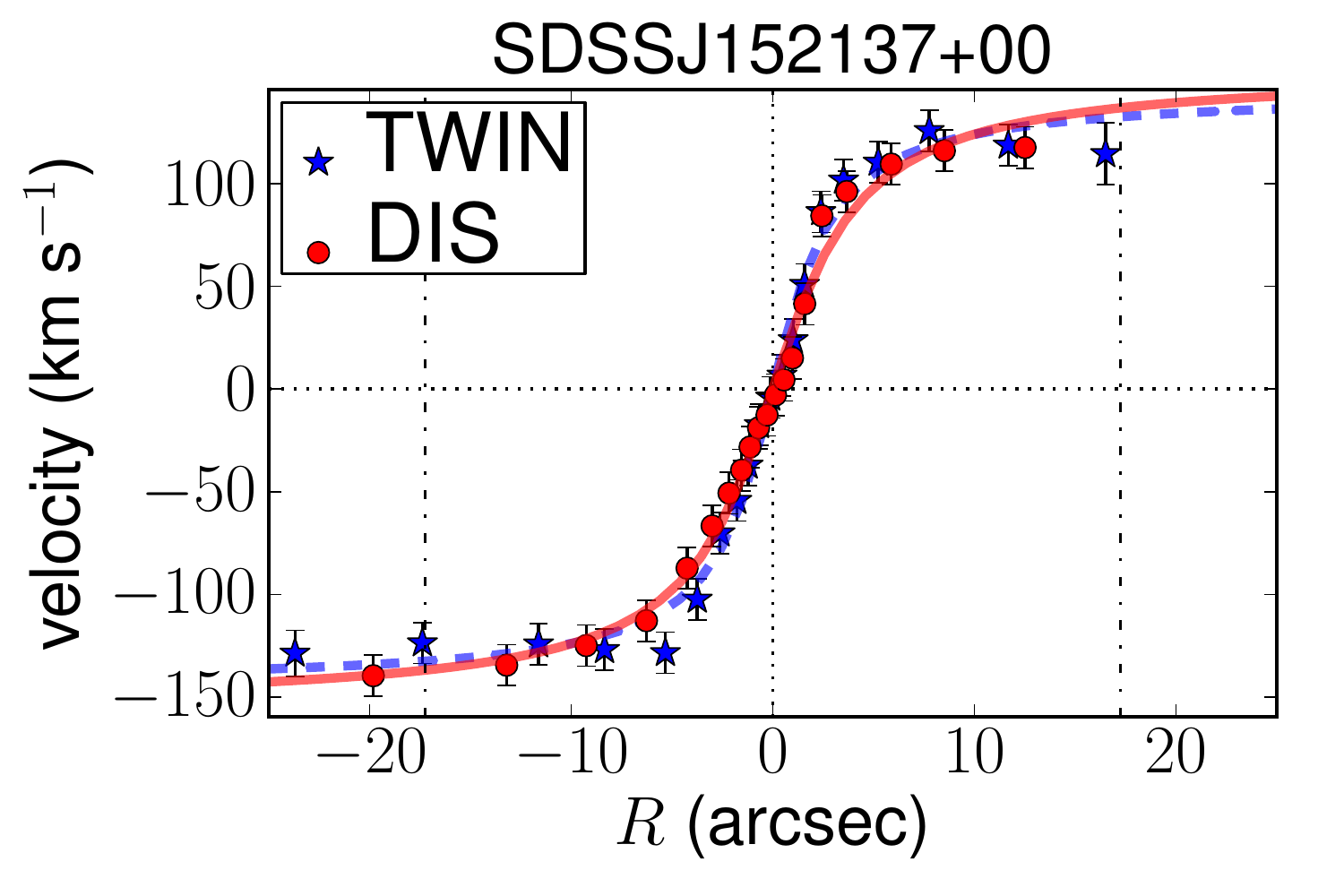}} 
\end{array} $
\eec
\caption{Comparison of rotation curves obtained from DIS (red circles) and P07 TWIN and CCDS observations (blue stars). Best-fit arctangent models are shown in red solid and blue dashed curves, respectively. Dot-dashed vertical lines mark $R_{80}$, the radius enclosing 80 per cent of the $i$-band light, at which the rotation velocity amplitude $V_{80}$ is defined.}
\label{fig:syst_piz_rc}
\end{figure*}

\begin{figure}
\includegraphics[width=3in]{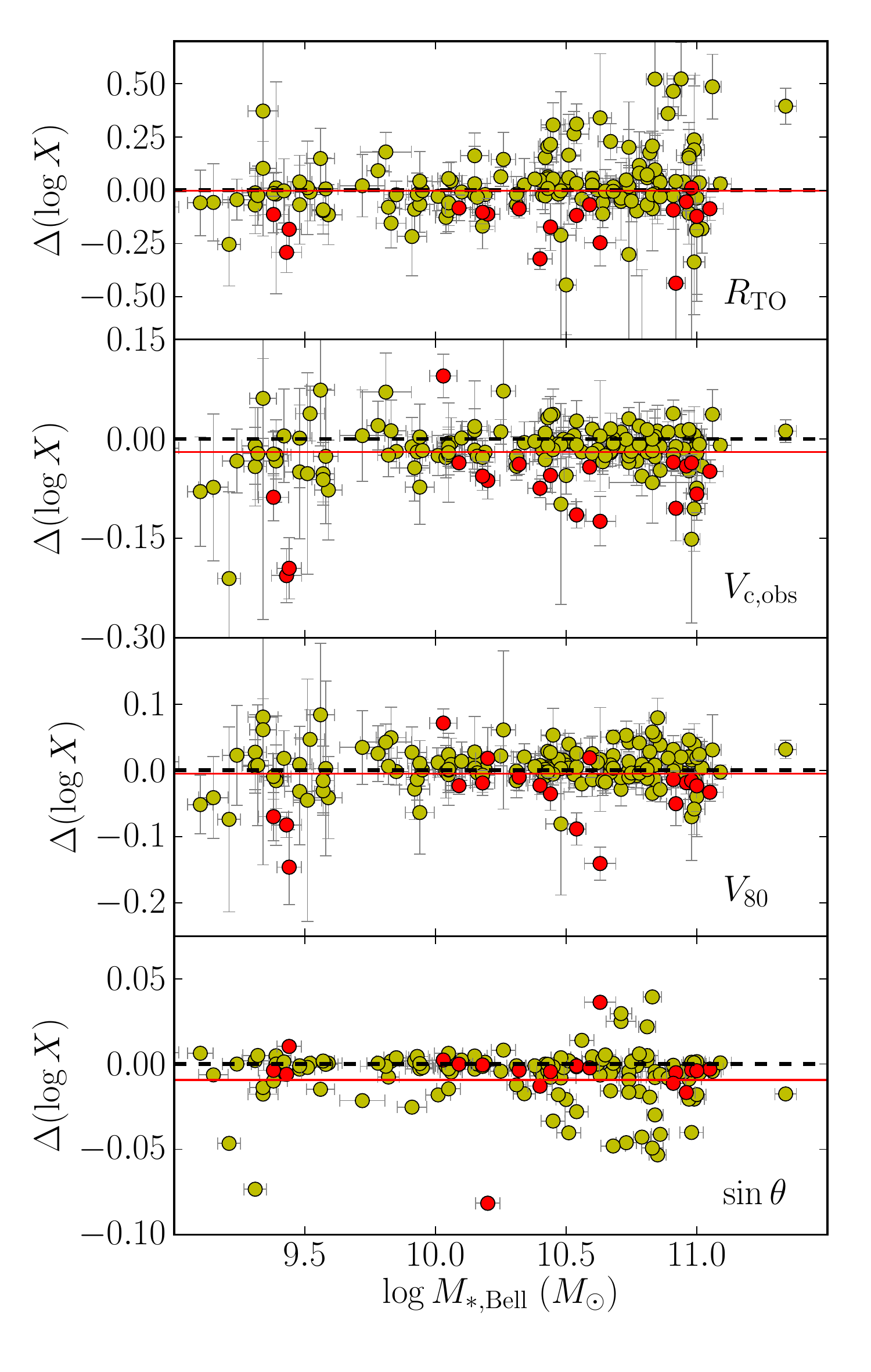}
\caption{Comparison between our measurements and those from P07 for 157 galaxies in the P07 sample (filled circles). Panels show $\Delta(\log X)$, where $X$ is as labelled, from top to bottom: rotation curve fit parameters $\rto$, $\vcobs$, inclination corrected rotation velocity $V_{80}$, and inclination corrections $\sin\theta$. The error bars shown are {\it our} derived 1-$\sigma$ measurement uncertainties. Galaxies for which $|\Delta\vcobs|$ is greater than 2$\sigma(\vcobs)$ are marked by red circles in all panels (19 galaxies, or 12 per cent of the sample). Horizontal red solid lines mark mean values, $\langle \Delta(\log X)\rangle$; black dashed lines mark the zero axis.}
\label{fig:delta_P07param}
\end{figure}

\subsubsection{Comparison with P07 observations}
\label{subsubsec:syst_piz}

Since the child disk sample combines long-slit spectroscopy observations taken using different instruments, it is important to check that this difference does not introduce any systematic bias in the derived kinematic quantities. For this purpose, we have randomly selected five out of the 99 galaxies from the P07 sample that satisfy our selection criteria, and re-observed them with DIS. 

Figure~\ref{fig:syst_piz_rc} compares the rotation curves and best-fit arctangent models for these galaxies. There is a good match between the different observations (blue stars and dashed curve for TWIN or CCDS, and red circles and solid curve for DIS). For each galaxy, the best-fit rotation curve from the DIS observation is consistent, to within a few $\sigma$, with the observed rotation velocities from the TWIN or CCDS observation, and vice versa. 

Table~\ref{tab:syst_piz_rcfit} lists the best-fit arctangent model parameters, together with their 1-$\sigma$ formal fit uncertainties, for the different observations. 
The rotation velocity amplitudes $V_{\rm 80,obs}$ (col.~6) for the different observations are consistent to within 1$\sigma$ for four out of the five galaxies. The exception is SDSSJ020540-00, for which the arctangent fits are affected by hooks in the rotation curve (evident around 5\arcsec). Focusing on the outer arms, it is clear that the rotation curve data themselves are consistent with one other. 

In detail, the shapes of the rotation curves differ from one another, as reflected in the differences in the best-fit $\rto$. This is especially apparent for SDSSJ024459+01 and SDSSJ152137+00, in which the differences in $\rto$ are 50 and 40 per cent, respectively. This difference can be attributed to differences in seeing conditions. 
We conclude that there is no evidence for any systematic bias between the long-slit spectroscopy observations performed by us and P07.


\begin{table*}
\caption{Rotation curve fits for targets with observations from both P07 (CCDS/TWIN) and DIS (5 galaxies).}
\begin{center}
\begin{tabular}{clrrrrr}
\hline 
 \multicolumn{1}{c}{Galaxy name} &
 \multicolumn{1}{c}{Instru.} &
 \multicolumn{1}{c}{$\vcobs$} &
 \multicolumn{1}{c}{$\rto$} &
 \multicolumn{1}{c}{$R_{80}$} &
 \multicolumn{1}{c}{$V_{\rm 80,obs}$} &
 \multicolumn{1}{c}{$\chi^2/\nu$} \\
  \multicolumn{1}{c}{} &
 \multicolumn{1}{c}{(deg)} & 
 \multicolumn{1}{c}{($\kms$)} &    
 \multicolumn{1}{c}{(arcsec)} &
 \multicolumn{1}{c}{(arcsec)} &
 \multicolumn{1}{c}{($\kms$)} & 
 \multicolumn{1}{c}{} \\ 
 \multicolumn{1}{c}{(1)} & 
 \multicolumn{1}{c}{(2)} & 
 \multicolumn{1}{c}{(3)} & 
 \multicolumn{1}{c}{(4)} &
 \multicolumn{1}{c}{(5)} &
 \multicolumn{1}{c}{(6)} &
 \multicolumn{1}{c}{(7)} \\
\hline 
SDSSJ020540-00 &  DIS & $216.2 \pm  3.6$ & $0.80 \pm 0.06$ & $ 29.13$ & $212.4 \pm 3.5$ & $ 8.33$ \\
            -- & TWIN & $201.7 \pm  3.5$ & $0.85 \pm 0.07$ & $     -$ & $198.0 \pm 3.5$ & $10.11$ \\
\hline                        
SDSSJ024459+01 &  DIS & $221.9 \pm 10.2$ & $4.93 \pm 0.49$ & $ 19.35$ & $186.6 \pm 9.2$ & $ 0.36$ \\
            -- & TWIN & $206.3 \pm  8.8$ & $3.20 \pm 0.32$ & $     -$ & $184.8 \pm 8.1$ & $ 0.53$ \\
\hline                        
SDSSJ080927+43 &  DIS & $170.1 \pm  5.6$ & $1.77 \pm 0.21$ & $  15.66$ & $157.9 \pm 5.2$ & $ 1.31$ \\
            -- & CCDS & $184.4 \pm  7.2$ & $2.42 \pm 0.36$ & $     -$ & $166.4 \pm 6.0$ & $ 0.38$ \\
\hline                        
SDSSJ084408+50 &  DIS & $125.3 \pm  9.5$ & $4.83 \pm 0.83$ & $  21.12$ & $ 107.4 \pm 11.1$ & $ 0.36$ \\
            -- & CCDS & $121.9 \pm 10.4$ & $5.72 \pm 1.13$ & $     -$ & $ 101.4 \pm 13.3$ & $ 0.26$ \\
\hline            
SDSSJ152137+00 &  DIS & $155.6 \pm  8.1$ & $3.29 \pm 0.41$ & $ 17.24$ & $136.9 \pm 7.4$ & $ 0.85$ \\
            -- & TWIN & $144.7 \pm  6.0$ & $2.29 \pm 0.27$ & $     -$ & $132.5 \pm 5.7$ & $ 1.14$ \\
\hline
\end{tabular}
\end{center}
\begin{flushleft}
Notes.---  Columns are the same as in Table~\ref{tab:syst_offaxis_rcfit}, except for Col. (2); here, we list the instrument used for the observation.
\end{flushleft}
\label{tab:syst_piz_rcfit} 
\end{table*}%
 
\subsubsection{Comparison with P07 rotation curve fits}
\label{subsubsec:syst_fits}

In this section, we compare our best-fit arctangent model parameters--- $\vcobs$ and $\rto$--- with those published in P07, for 157 galaxies in the P07 sample (99 of which are in our child disk sample). We use the long-slit spectroscopy observations of P07, but perform our own analysis of the data to extract and fit rotation curves. Comparison of our results with theirs provide an estimate of the level of systematic differences that can arise from different analysis methods. 

Figure~\ref{fig:delta_P07param} shows the differences between P07 measurements and ours, $\Delta(\log X)=(\log X)^{\rm (P07)}-\log X$, where $X=\{\rto$, $\vcobs$, $V_{80}$, $\sin\theta\}$ (top to bottom panels). Nineteen out of the 157 galaxies have $|\Delta\vcobs| >2\sigma(\vcobs)$ (marked by red circles, in all panels), and for almost all of them, $\Delta\vcobs<0$. We surmise that the reason for this is that our extracted rotation curves tend to extend out farther than those of P07, which led to lower (i.e., less overestimated) $\vcobs$. As expected, we find that the differences in $V_{80}$ are smaller than those in $\vcobs$, explicitly illustrating the relative robustness (and therefore, advantage) of this choice of velocity amplitude. Finally, we note that differences in the estimated disk axis ratios led to differences in $V_{80}$ as large as $\sim$0.05 dex (see bottom panel). Both we and P07 used two-dimensional bulge-disk decompositions to determine disk axis ratios, but our fits were performed on images from a more recent SDSS reduction. We use our own fitting routine (described in \S\ref{sec:bdfit}), while P07 used the publicly-available code GALFIT \citep{2002AJ....124..266P}.

Later in \S\ref{subsec:alt_P07}, we compare the TFRs derived from our and P07's measurements. We find that the $M_i$--$V_{80}$ relations are consistent within the reported 1-$\sigma$ uncertainties, indicating that our results are robust to systematic differences between the analysis pipelines.


\section{Derivation of TFRs}
\label{sec:tfr_derive}
In this section, we describe our derivation of the scaling relation between the disk rotation velocity $V_{\rm rot}$ and some photometric quantity ${\cal O}$. The different photometric quantities we consider--- absolute magnitudes $M_\lambda$, synthetic magnitudes $M_{\rm syn}$, stellar masses $\mbell$ and $\mmpa$, and baryonic masses $\mbar$--- were defined in \S\ref{sec:phot}. 

We aim to identify the ``optimal'' photometric estimator of disk rotation velocity, calibrate it, and characterize its scatter, using the child disk sample (189 galaxies). The photometric estimates can then be applied to the full (spectroscopic) parent disk sample ($\sim 170~000$ galaxies), and potentially, to an even larger photometric galaxy sample. 

We describe the modeling and fitting of the TFRs in \S\ref{subsec:tfr_model} and \S\ref{subsec:tfr_fit}, respectively. In \S\ref{subsec:tfr_interpret}, we describe how the fits should be interpreted. Results of the fits will be presented in the next two sections, \S\ref{sec:itfr_calib} and \S\ref{sec:alt_fits}.

\subsection{Modelling the TFRs}
\label{subsec:tfr_model}
Since we are mainly interested in calibrating photometric estimators of disk rotation velocities, we will focus on fits to the so-called inverse TFRs (ITFRs), in which the rotation velocity is the dependent variable ($y\equiv \log V_{\rm rot}$) and the photometric quantity is the independent variable ($x\equiv {\cal O}=\{M_\lambda, \msyn, \log \mbell, \log\mmpa, \log\mbar\}$). Fits to the different ITFRs will be presented in \S\ref{sec:itfr_calib}.

For each photometric quantity ${\cal O}$, we model the ITFR as a linear relation with an intrinsic Gaussian scatter of width $\tilde{\sigma}$ in $\log V_{\rm rot}$:
\beqa \label{eq:tf_absmag}
\log V_{\rm rot} &=& a + b \times (M_\lambda - M_{\lambda,{\rm p}}), \\  \label{eq:tf_synmag}
\log V_{\rm rot} &=& a + b \times (M_{\rm syn} - M_{{\rm syn},{\rm p}}). \\ \label{eq:tf_bellmass}
\log V_{\rm rot} &=& a + (-2.5b) \times (\log M_* - \log M_{*,{\rm p}}), 
\eeqa
Here, $a$ is the zero-point, $b$ is the slope, and ${\cal O}_{\rm p}$ is the pivot value of the ITFR. Both $a$ and $\tilde{\sigma}$ have units of $\log (\kms)$ and $b$ has units of $\log (\kms)\ {\rm mag}^{-1}$. We will refer to the slope of the ITFR as steeper (shallower) if $b$ is more (less) negative.
In most cases, we set the pivot value ${\cal O}_{\rm p}$ to be the weighted mean of the fit sample. Doing so makes the covariance between the error in $a$ and the error in $b$ negligible (in other words, changing the value of $b$ by its 1-$\sigma$ uncertainty does not change the best-fit value of $a$). 

For completeness, we also consider forward TFRs, in which the independent variable $y\equiv {\cal O}$ and the dependent variable $x\equiv \log V_{\rm rot}$. Fits to the forward TFRs will be presented in \S\ref{subsec:alt_directions}. 

For each photometric quantity, we model the forward TFR as a linear relation with an intrinsic Gaussian scatter of width $\tilde{\sigma}_{\rm fwd}$ in ${\cal O}$:
\beqa \label{eq:tf_absmag_fwd}
M_\lambda &=& a_{\rm fwd} + b_{\rm fwd} \times \log(V_{\rm rot}/ V_{\rm rot,p}), \\ \label{eq:tf_synmag_fwd}
\msyn &=& a_{\rm fwd} + b_{\rm fwd} \times \log (V_{\rm rot}/V_{\rm rot,p}). \\ \label{eq:tf_bellmass_fwd}
\log M_* &=& a_{\rm fwd} + (-0.4 b_{\rm fwd}) \times \log (V_{\rm rot}/V_{\rm rot,p}), 
\eeqa
Here, both the zero-point $a_{\rm fwd}$ and $\tilde{\sigma}_{\rm fwd}$ have units of mag, and the slope $b_{\rm fwd}$ has units of ${\rm mag}\, (\log(\kms))^{-1}$. 
To directly compare the results of the forward and inverse fits, we convert the forward fit parameters to the equivalent inverse fit parameters: $a_{\rm conv}=\log V_{\rm rot,p}+ ({\cal O}_{\rm p}-a_{\rm fwd})/b_{\rm fwd}$, $b_{\rm conv}=1/b_{\rm fwd}$, and ${\tilde{\sigma}}_{\rm conv}={\tilde{\sigma}}_{\rm fwd}/b_{\rm fwd}$. These converted parameters have the same units as $a$, $b$, and $\tildesig$, as defined in Eqs.~\ref{eq:tf_absmag}--\ref{eq:tf_bellmass}. 

Note that even in the ideal case in which there is zero uncertainty in the data, fits to the forward and inverse TFRs will not necessarily yield the same relation. Moreover, fits to the forward TFRs are subject to Malmquist bias due to our sample selection cuts on absolute magnitude (in contrast, fits to the ITFRs are not affected by this bias). The inverse and forward fit directions assume different underlying models and must be interpreted accordingly. In most classical TFR studies, forward fits were used to obtain distance measurements, based on the luminosities derived from the disk rotation velocities. In this work, we use fits to the ITFRs to obtain photometric estimators of disk rotation velocities. 

\subsection{Fitting the TFRs}
\label{subsec:tfr_fit}
We perform weighted maximum likelihood fits to the inverse and forward TFRs, defined by Eqs.~\ref{eq:tf_absmag}--\ref{eq:tf_bellmass} and \ref{eq:tf_absmag_fwd}--\ref{eq:tf_bellmass_fwd}, respectively. First, we describe how we assign weights to each galaxy. Then, we define the likelihood function used to determine the fit parameters and the bootstrap resampling method used to estimate their uncertainties. Finally, we describe the two-step method for fitting the TFRs for synthetic magnitudes $\msyn$.

We assign to each galaxy a weight according to its stellar mass, so that the effective stellar mass function (SMF) of the fit sample matches that of the parent disk sample. Weights are given by the nearest integer to $100\times w_{\rm samp}(\mbell)$, where
\beq\label{eq:wsamp}
w_{\rm samp}(M_*)=\frac{N_{\rm parent}(M_*)}{N_{\rm parent}} \frac{N_{\rm samp}}{N_{\rm samp}(M_*)}.
\eeq 
Here, $N_{\rm parent}= 169~563$ galaxies, $N_{\rm samp}$ is the total number of galaxies in the fit sample, and $N_{\rm parent}(M_*)$ and $N_{\rm samp}(M_*)$ are the number of galaxies in the stellar mass bin containing $M_*$, for the parent disk sample and the fit sample, respectively. We use logarithmic stellar mass bins of width $\approx$ 0.21 dex. 

We find that the child disk sample SMF deviates from the parent disk sample SMF only at the lowest and highest stellar mass bins, with $w_{\rm samp}\approx 2$ for galaxies with stellar masses below $10^{9.4} M_\odot$ and $w_{\rm samp}\approx 1/2$ for stellar masses above $10^{10.9} M_\odot$. In contrast, the P07 disk sample SMF is significantly different from that of the parent disk sample, especially for the highest stellar mass bins where $w_{\rm samp} \approx 0.1$.


The full dataset, with repeats, consists of a total of $N= \sum_{j=1}^{N_{\rm samp}} {\rm round}(100 w_j)$ pairs of measurements, $\hat{x}_i$ and $\hat{y}_i$, with measurement errors $\sigma_{x,i}$ and $\sigma_{y,i}$. Assuming that these errors are uncorrelated, and that the measurements are normally distributed around their true values, $x$ and $y$, with Gaussian widths given by their respective measurement errors, then the log-likelihood function is given by
\beqa \label{eq:logl} \nonumber
-\ln {\cal L}&=& \frac{1}{2} \sum_{i=1}^N \ln(\tilde{\sigma}^2+\sigma_{y,i}^2+b^2\sigma_{x,i}^2) \\
	&+& \frac{1}{2} \sum_{i=1}^N \frac{ \left[\hat{y}_i-(a+b \hat{x}_i)\right]^2}{(\tilde{\sigma}^2+\sigma_{y,i}^2+b^2\sigma_{x,i}^2)} + K, 
\eeqa
where $K$ is a constant.

We determine the three best-fit parameters, $a$, $b$, and $\tilde{\sigma}$, by maximizing Eq.~\ref{eq:logl} using the Levenberg-Marquardt routine \verb+mpfit2dfun+ in IDL \citep{2009ASPC..411..251M}. Then, we calculate the 1-$\sigma$ uncertainty in each parameter using bootstrap resampling. We generate 500 sample realizations by randomly drawing $N$ galaxies with replacement. We perform fits to each realization, and check that the distributions in the bootstrap parameters are approximately Gaussian.

To fit for the TFRs for synthetic magnitudes $\msyn$, we first have to fix the coefficient $\alpha$ in its definition (c.f. Eq.~\ref{eq:synmag}). To do this, we adopt a four-parameter model
\beq \label{eq:tfr_alpha}
\log V_{\rm rot} = a_2 + b_2 [M_\lambda - M_{\lambda,p_2} + \alpha \times ({\rm colour})],
\eeq
with a Gaussian intrinsic scatter of width $\tilde{\sigma}_2$ in $\log V_{\rm rot}$. The likelihood function for this model is defined analogously to Eq.~\ref{eq:logl}. Minimizing the likelihood function yields the best-fit coefficient $\alpha$, and bootstrap resampling yields its 1-$\sigma$ uncertainty. Finally, the fits to the inverse and forward $\msyn$ TFRs are determined from Eqs.~\ref{eq:tf_synmag} and \ref{eq:tf_synmag_fwd}, respectively, in a similar manner as for $M_\lambda$, as described above.

We present results of weighted TFR fits in \S\ref{sec:itfr_calib} and \S\ref{sec:alt_fits}. In practice, though, we find that unweighted fits yield almost identical results. This is not surprising because, as noted above, the SMFs of the child and parent disk samples are not significantly different. Therefore, for simplicity, we have not used weights in fits for the coefficient $\alpha$ in $\msyn$, nor in the calculations performed in subsequent sections \S\ref{sec:tfr_correl} and \S\ref{sec:mass_ratios}. 


\subsection{Interpreting the TFR scatter}
\label{subsec:tfr_interpret}

Our TFR models include a Gaussian intrinsic scatter of width $\tildesig$. In addition, we define the measured scatter, $\sigmeas$, to be the RMS (root-mean-square) of the TFR residuals $(\Delta y)_i = \hat{y}_i - [a + b \times (\hat{x}_i-x_{\rm p})]$. In the simplest case, in which the measurement errors $\sigma_{x,i}$ and $\sigma_{y,i}$ are equal for all galaxies, and in which our assumptions regarding the Gaussian distributions for $\hat{x}_i$ and $\hat{y}_i$ hold, $\sigmeas^2 = \tildesig^2 + \sigma_{y,i}^2 + (b\sigma_{x,i})^2$. In other words, the measured scatter can be attributed partly to the measurement uncertainties for individual galaxies, and partly to the intrinsic scatter in the ITFR itself. Note, though, that what we have defined to be the ``intrinsic scatter'' in the TFR, $\tildesig$, actually includes contributions from other sources of observational scatter that we have not formally accounted for in the measurement uncertainties, including errors in inclinations, slit PA misalignments, and intrinsic disk ellipticities.

As noted in \S\ref{subsubsec:syst_pos}, ellipticities in the potential in the plane of the disk $\epsilon_\Phi$ lead to noncircular motions and differences between the kinematic and photometric major axis PAs. They also lead to a dependence of the inclination correction on the viewing angle.\footnote{If $\epsilon_\Phi \ll 1$, the true inclination correction is $\sin\theta(1-\epsilon_\Phi\cos 2\phi)$ instead of $\sin\theta$, for a viewing angle $(\theta,\phi)$  \citep{1992ApJ...392L..47F}.} 
Observations indicate that $\epsilon_\Phi \approx 0.1$ \citep[][and references therein]{2006ApJ...641..773R,2008MNRAS.388.1321P}, which in turn translate into an expected scatter in the (forward) TFR of 0.46 mag, 
even if kinematic major axis PAs were used \citep{1992ApJ...392L..47F}. The expected scatter will be larger if photometric major axis PAs were used, as is the case in most TFR studies, including ours. In \S\ref{subsec:disc_syst}, we compare the observed scatter in the TFR with the expected contributions from these systematic effects.


\section{Calibrated ITFRs}
\label{sec:itfr_calib}

In this section, we present weighted fits to the ITFRs for the 189 galaxies in the child disk sample using different photometric quantities ${\cal O}=\{$$M_\lambda$, $\msyn$, $\log \mbell$, $\log \mmpa$, and $\log \mbar$$\}$ (all defined in \S\ref{sec:phot}) and rotation velocity amplitudes $V_{80}$ (defined in \S\ref{subsec:vrot_definition}). These calibrated ITFRs provide photometric estimators of disk rotation velocities applicable to the parent disk sample. The procedure for obtaining these fits were described in the previous section, \S\ref{sec:tfr_derive}. Alternative fits will be presented in the next section, \S\ref{sec:alt_fits}. 

\subsection{$M_\lambda$ ITFRs}
\label{subsec:itfr_mlambda}

Table~\ref{tab:itfr_absmag} lists the best-fit parameters $a$, $b$, and $\tildesig$ for the calibrated $M_\lambda$ ITFR, for five SDSS bands, from weighted fits to the child disk sample (189 galaxies). Also listed are the pivot value $x_0=M_{\rm \lambda,p}$, which has been set equal to the weighted mean of $M_\lambda$, and the measured scatter $\sigmeas$ (defined in \S\ref{subsec:tfr_interpret}).

We confirm previous studies and find that the amount of scatter in the ITFR (both intrinsic and measured) systematically decreases toward redder bands. Among the SDSS bands, we choose the $i$ band to be the optimal one, as it yields a tighter ITFR relation than the bluer bands (in addition to being less affected by dust extinction). The scatter in the $z$ band ITFR is similar, but this band has noisier photometry.

\begin{table*}
\begin{center}
\caption{Calibrated $M_\lambda$ ITFRs for the child disk sample (189 galaxies) with $\vrot=V_{80}$.} 
\begin{tabular}{llrrrrr}
\hline 
 \multicolumn{1}{c}{$y$} &
 \multicolumn{1}{c}{$x$} &
 \multicolumn{1}{c}{$x_0$} &
 \multicolumn{1}{c}{$a$} &
 \multicolumn{1}{c}{$b$} &
 \multicolumn{1}{c}{$\tilde{\sigma}$} &
 \multicolumn{1}{c}{$\sigmeas$} \\
 \multicolumn{1}{c}{(1)} & 
 \multicolumn{1}{c}{(2)} & 
\multicolumn{1}{c}{(3)} & 
\multicolumn{1}{c}{(4)} & 
\multicolumn{1}{c}{(5)} &
\multicolumn{1}{c}{(6)} &
\multicolumn{1}{c}{(7)} \\
\hline
    $\log V_{80}$ &        $M_u$ & $-18.731$ & $  2.188$\,($  0.026$) & $ -0.080$\,($  0.009$) & $  0.114$\,($  0.011$) &  $  0.114$ \\
              --- &        $M_g$ & $-19.903$ & $  2.142$\,($  0.006$) & $ -0.129$\,($  0.008$) & $  0.068$\,($  0.008$) &  $  0.082$ \\
              --- &        $M_r$ & $-20.375$ & $  2.142$\,($  0.005$) & $ -0.130$\,($  0.007$) & $  0.056$\,($  0.007$) &  $  0.071$ \\
              --- &        $M_i$ & $-20.558$ & $  2.142$\,($  0.005$) & $ -0.128$\,($  0.006$) & $  0.049$\,($  0.007$) &  $  0.066$ \\
              --- &        $M_z$ & $-20.649$ & $  2.142$\,($  0.005$) & $ -0.119$\,($  0.005$) & $  0.049$\,($  0.007$) &  $  0.065$ \\
\hline
\end{tabular}
\label{tab:itfr_absmag}
\end{center}
\begin{flushleft}
Notes. --- {
Col. (1): $y$ is the dependent variable of the ITFR, or $\log V_{\rm rot}$. 
Col. (2): $x$ is the independent variable of the ITFR, or ${\cal O}$. 
Col. (3): $x_{\rm p}$ is the pivot value for $x$, equal to the weighted mean of the fit sample.
Cols. (4-6): Best-fit ITFR parameters and their 1-$\sigma$ uncertainties, as defined in Eqs.~\ref{eq:tf_absmag}--\ref{eq:tf_synmag}: $a$ is the zero-point, in units of $\log(\kms)$, $b$ is the slope in units of $\log(\kms)\,{\rm mag}^{-1}$, and $\tilde{\sigma}$ is the intrinsic Gaussian scatter in units of $\log(\kms)$.
Col. (7): $\sigma_{\rm meas}$ is the measured scatter in the ITFR, defined to be the RMS of the ITFR residuals $(\Delta y)_i = y_i - [a + b \times (x-x_{\rm p})]$.
}\end{flushleft}
\end{table*}%

\subsection{$\msyn$ ITFRs}
\label{subsec:itfr_msyn}

Table~\ref{tab:itfr_synmag} lists the best-fit parameters for the calibrated $\msyn$ ITFRs. We find that regardless of the choice of absolute magnitude band, the amount of intrinsic and measured scatter for the $\msyn$ ITFRs is the same (0.035 and 0.056 dex in $\log V_{80}$, respectively). The zero-points and slopes are consistent with one another as well. We find that the $\msyn$ ITFRs have smaller intrinsic scatter than the $M_\lambda$ ITFRS, by at least $2\sigma$; their measured scatter are also smaller. We conclude that synthetic magnitudes are better photometric estimators of disk rotation velocities than single-band optical absolute magnitudes.

Table~\ref{tab:itfr_synmag} also shows the best-fit coefficients $\alpha$ (col.~4) in the definition of synthetic magnitudes  $\msyn=M_\lambda+\alpha \times (g-r)$. These are to be compared with the coefficients based on the colour-based estimates of $M_*/L_\lambda$ from Bell et~al. (2003), $\alpha^{\rm Bell}$ (col.~3). For each band $\lambda$, we find that the best-fit $\alpha$ is within 1$\sigma$ of $\alpha^{\rm Bell}$. These colour corrections extrapolate toward a redder band, so that the synthetic magnitudes approximate stellar masses. We find similar results when using colours other than $g-r$. These results suggest that stellar mass is indeed a more fundamental parameter than luminosity in the TFR. Moreover, they provide independent support for the reliability of these colour-based stellar mass-to-light ratios (\citealt{2002AJ....123.2358K} came to similar conclusions regarding the earlier Bell \& de Jong 2001 stellar mass-to-light ratios).   

\begin{table*}
\begin{center}
\caption{Calibrated $\msyn$ ITFRs for the child disk sample (189 galaxies) with $\vrot=V_{80}$, where $\msyn=M_\lambda+\alpha \times (g-r)$.} 
\label{tab:itfr_synmag}
\begin{tabular}{llrrrrrrr}
\hline 
 \multicolumn{1}{c}{$\lambda$} &
 \multicolumn{1}{c}{colour} &
 \multicolumn{1}{c}{$\alpha^{\rm Bell}$} & 
 \multicolumn{1}{c}{$\alpha$} &
 \multicolumn{1}{c}{$x_{\rm p}$} &
 \multicolumn{1}{c}{$a$} &
 \multicolumn{1}{c}{$b$} &
 \multicolumn{1}{c}{$\tilde{\sigma}$} &
 \multicolumn{1}{c}{$\sigmeas$} \\
 \multicolumn{1}{c}{(1)} & 
 \multicolumn{1}{c}{(2)} & 
\multicolumn{1}{c}{(3)} & 
\multicolumn{1}{c}{(4)} & 
\multicolumn{1}{c}{(5)} &
\multicolumn{1}{c}{(6)} &
\multicolumn{1}{c}{(7)} &
\multicolumn{1}{c}{(8)} &
\multicolumn{1}{c}{(9)} \\
\hline
$g$ & $g-r$ & $-3.80$ & $-4.09$\,($0.91$) & $-22.172$ & $  2.159$\,($  0.004$) & $ -0.105$\,($  0.004$) & $0.034$\,($0.006$) & $0.056$ \\
$r$ & $g-r$ & $-2.74$ & $-3.25$\,($0.72$) & $-22.211$ & $  2.160$\,($  0.004$) & $ -0.105$\,($  0.004$) & $0.035$\,($0.005$) & $0.056$ \\
$i$ & $g-r$ & $-2.16$ & $-2.42$\,($0.86$) & $-21.965$ & $  2.160$\,($  0.004$) & $ -0.107$\,($  0.004$) & $0.035$\,($0.006$) & $0.056$ \\
$z$ & $g-r$ & $-1.72$ & $-2.28$\,($1.04$) & $-21.987$ & $  2.160$\,($  0.004$) & $ -0.101$\,($  0.004$) & $0.036$\,($0.006$) & $0.057$ \\
\hline
\end{tabular}
\end{center}
\begin{flushleft}
Notes. --- {
Col. (1)-(2): Band $\lambda$ and colour in the synthetic magnitude $\msyn=M_\lambda+\alpha \times \mbox{(colour)}$.
Col. (3): $\alpha^{\rm Bell}$ is the coefficient of the colour correction in the Bell et al. (2003) stellar mass-to-light ratio estimates for this band and colour.
Col. (4): Best-fit coefficient $\alpha$ and its 1-$\sigma$ uncertainty.
Col. (5): $x_{\rm p}$ is the pivot value for $x=\msyn$.
Col. (6-8): Best-fit ITFR parameters and their 1-$\sigma$ uncertainties: $a$ is the zero-point, in units of $\log(\kms)$, $b$ is the slope in units of $\log(\kms)\,{\rm mag}^{-1}$, and $\tilde{\sigma}$ is the intrinsic Gaussian scatter in units of $\log(\kms)$.
Col. (9): $\sigma_{\rm meas}$ is the measured scatter in the ITFR, defined to be the RMS of the ITFR residuals $(\Delta y)_i = y_i - [a + b \times (x-x_{\rm p})]$.
}
\end{flushleft}
\end{table*}%

\subsection{$\mbell$, $\mmpa$, and $\mbar$ ITFRs}
\label{subsec:itfr_mstr}

Table~\ref{tab:itfr_masses} lists the best-fit parameters for calibrated ITFRs for stellar mass estimates, $\mbell$ and $\mmpa$, and for baryonic mass estimates, $\mbar$. As expected from the systematic difference between $\mbell$ and $\mmpa$ (see Fig.~\ref{fig:mstr_comp} in \S\ref{subsubsec:comp_mstr}), the former has a pivot value higher by 0.15 dex, corresponding to a lower normalization in $\log \vrot$, of 0.04 dex (at fixed $M_*$). The $\mbell$ ITFR is steeper than the $\mmpa$ ITFR at $>4\sigma$ level. Their intrinsic and measured scatter are similar to each other and to the $\msyn$ ITFR. These results confirm our earlier conclusion (in \S\ref{subsec:itfr_msyn}) that colour-based stellar mass estimates are a good tracer of disk rotation velocity. 

Compared to the $\mbell$ and $\mmpa$ ITFRs, the $\mbar$ ITFR has larger intrinsic and measured scatter. True baryonic mass is expected to trace the disk rotation velocity more faithfully than stellar mass alone, since the both stellar and gas mass contributes to the total disk potential. However, we do not find this to be the case for these baryonic mass estimates. Moreover, based on the increasing trend in gas mass fraction as stellar mass decreases, one would expect the baryonic ITFR to have a shallower slope than the stellar mass ITFR; contrary to this expectation, we find a best-fit slope that is steeper by $\sim$3$\sigma$. We conclude that $\mbar$ is not an optimal photometric estimator for disk rotation velocities. In other words, the relation between $u-r$ colour and gas-to-stellar mass ratios (Eq.~\ref{eq:mgas}), on which our baryonic mass estimates are based, is not tight enough for our purposes (unlike the colour-$M_*/L$ relations on which our stellar mass estimates are based). 


\begin{figure*}
\bec
\includegraphics[width=5.5in]{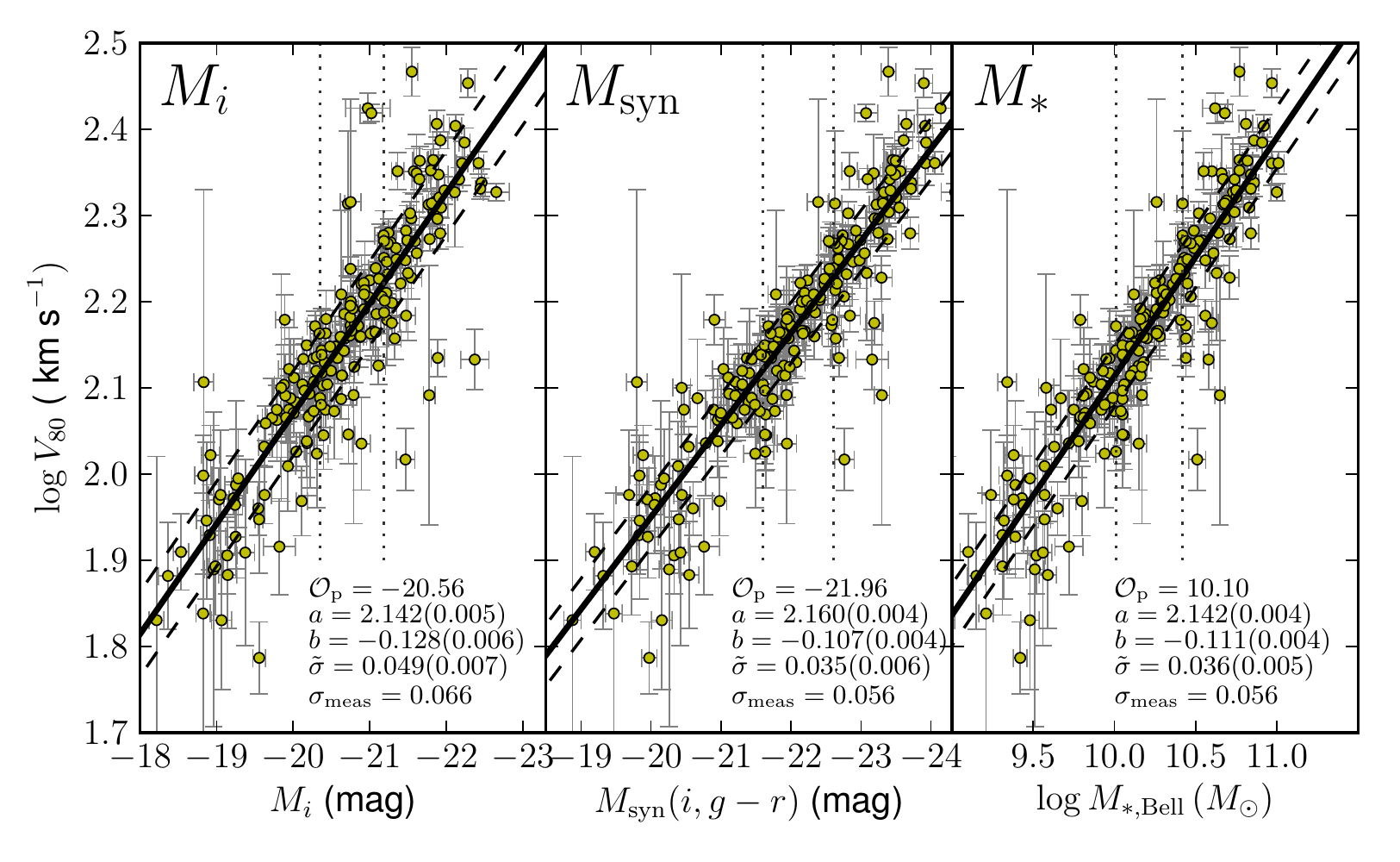}
\eec
\caption{ITFRs between rotation velocities $V_{80}$ and photometric quantities ${\cal O}$=$\{M_i$, $\log M_*$, and $M_{\rm syn}(i,g-r)\}$ (left, middle, and right panels, respectively). Observations for the 189 galaxies in the child disk sample are shown as filled circles with 1-$\sigma$ error bars. 
Thick solid lines show the best-fit ITFRs and dashed lines indicate the amount of intrinsic scatter (they are displaced by $\pm \tildesig$ from the mean relations). Vertical dotted lines divide the sample into three bins of roughly equal number. Labels list the pivot values ${\cal O}_{\rm p}$, best-fit parameters--- zero-point $a$, slope $b$, intrinsic scatter $\tildesig$--- of the ITFR, and measured scatter $\sigmeas$ in $\log V_{80}$.} 
\label{fig:linfits_obs3}
\end{figure*}

\begin{figure} 
\bec
\includegraphics[width=3in]{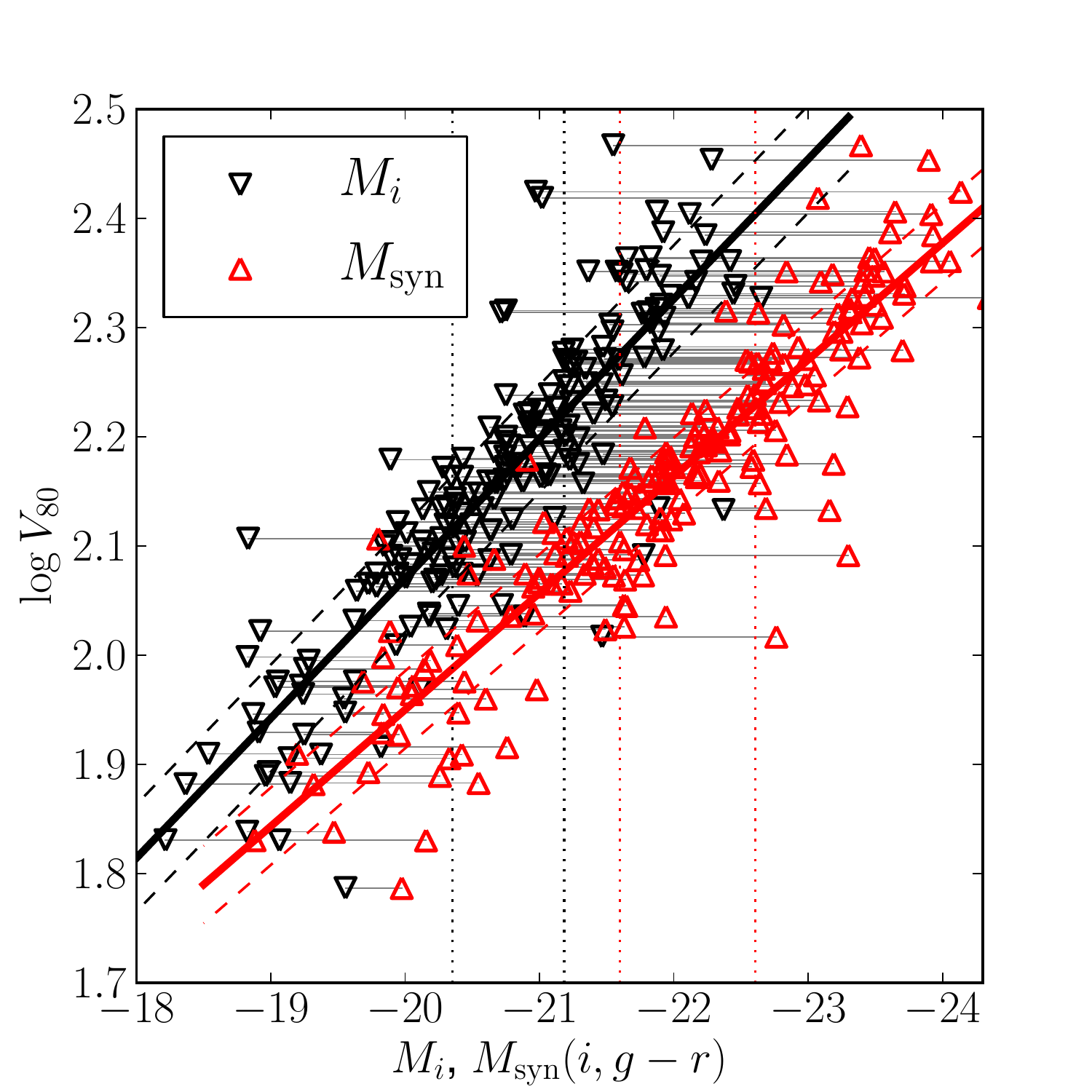}
\eec
\caption{Comparison between $M_i$ and $\msyn$ ITFRs (black inverted triangles and lines vs. red upright triangles and lines, respectively). Pairs of points for every galaxy are connected by gray horizontal lines. The amount of intrinsic scatter in the ITFRs is indicated by the dashed lines, which are displaced by $\pm$ $\tildesig$ from the mean relations. Vertical dotted lines divide the sample into three equal bins in $M_i$ and $\msyn$ (black and red dotted lines, respectively).} 
\label{fig:tfcon_synmag}
\end{figure}

\begin{table*}
\begin{center}
\caption{Calibrated $\mbell$, $\mmpa$, and $\mbar$ ITFRs for the child disk sample (189 galaxies) with $\vrot=V_{80}$.}
\begin{tabular}{llrrrrr}
\hline 
 \multicolumn{1}{c}{$y$} &
 \multicolumn{1}{c}{$x$} &
 \multicolumn{1}{c}{$x_{\rm p}$} &
 \multicolumn{1}{c}{$a$} &
 \multicolumn{1}{c}{$b$} &
 \multicolumn{1}{c}{$\tilde{\sigma}$} &
 \multicolumn{1}{c}{$\sigmeas$} \\
 \multicolumn{1}{c}{(1)} & 
 \multicolumn{1}{c}{(2)} & 
\multicolumn{1}{c}{(3)} & 
\multicolumn{1}{c}{(4)} & 
\multicolumn{1}{c}{(5)} &
\multicolumn{1}{c}{(6)} &
\multicolumn{1}{c}{(7)} \\
\hline
    $\log V_{80}$ &       $M_{\rm *,Bell}$ & $ 10.102$ & $  2.142$\,($  0.004$) & $ -0.111$\,($  0.004$) & $  0.036$\,($  0.005$) &  $  0.056$ \\
              --- &        $M_{\rm *,MPA}$ & $  9.952$ & $  2.140$\,($  0.004$) & $ -0.093$\,($  0.004$) & $  0.034$\,($  0.006$) &  $  0.056$ \\
              --- &          $M_{\rm bar}$ & $ 10.355$ & $  2.144$\,($  0.005$) & $ -0.127$\,($  0.006$) & $  0.053$\,($  0.007$) &  $  0.071$ \\
\hline
\end{tabular}
\label{tab:itfr_masses}
\end{center}
\begin{flushleft}
Notes. --- {Columns are the same as in Table~\ref{tab:itfr_absmag}.}
\end{flushleft}
\end{table*}%

\subsection{Comparison of $M_i$, $\msyn$, and $\mbell$ ITFRs}
\label{subsec:itfr_compare}

In this subsection, we compare three good photometric estimators of disk rotation velocity, choosing one of each type we consider, namely: $M_i$, $\msyn(i,g-r)$, and $\mbell$. Table~\ref{tab:itfr_obs3} puts together the fit results from Tables~\ref{tab:itfr_absmag}--\ref{tab:itfr_masses}, for easier comparisons. Figure~\ref{fig:linfits_obs3} shows the observed relations (filled circles with 1-$\sigma$ error bars), the best-fit ITFRs (thick solid lines), and the amounts of intrinsic scatter in the relations (upper and lower dashed lines are displaced by $\pm 1\tildesig$ from the mean relation). Note that $\msyn(i,g-r)=M_i-2.42(g-r)$ (c.f. Table~\ref{tab:itfr_synmag}).

\begin{table*}
\begin{center}
\caption{Calibrated $M_i$, $\msyn(i,g-r)$, and $\mbell$ ITFRs for the child disk sample (189 galaxies) with $\vrot=V_{80}$ (compiled from Tables~\ref{tab:itfr_absmag}--\ref{tab:itfr_masses}).}
\begin{tabular}{llrrrrr}
\hline 
 \multicolumn{1}{c}{$y$} &
 \multicolumn{1}{c}{$x$} &
 \multicolumn{1}{c}{$x_{\rm p}$} &
 \multicolumn{1}{c}{$a$} &
 \multicolumn{1}{c}{$b$} &
 \multicolumn{1}{c}{$\tilde{\sigma}$} &
 \multicolumn{1}{c}{$\sigmeas$} \\
 \multicolumn{1}{c}{(1)} & 
 \multicolumn{1}{c}{(2)} & 
\multicolumn{1}{c}{(3)} & 
\multicolumn{1}{c}{(4)} & 
\multicolumn{1}{c}{(5)} &
\multicolumn{1}{c}{(6)} &
\multicolumn{1}{c}{(7)} \\
\hline
    $\log V_{80}$ &                  $M_i$ & $-20.558$ & $  2.142$\,($  0.005$) & $ -0.128$\,($  0.006$) & $  0.049$\,($  0.007$) &  $  0.066$ \\
              --- &   $M_{\rm syn}(i,g-r)$ & $-21.965$ & $  2.160$\,($  0.004$) & $ -0.107$\,($  0.004$) & $  0.035$\,($  0.006$) &  $  0.056$ \\
              --- &       $M_{\rm *,Bell}$ & $ 10.102$ & $  2.142$\,($  0.004$) & $ -0.111$\,($  0.004$) & $  0.036$\,($  0.005$) &  $  0.056$ \\
\hline
\end{tabular}
\label{tab:itfr_obs3}
\end{center}
\begin{flushleft}
Notes. --- {Columns are the same as in Table~\ref{tab:itfr_absmag}.}
\end{flushleft}
\end{table*}%

We find that adding colour information to luminosity, in $\msyn$ and $\mbell$, reduces the amount of scatter in the ITFR. In Figure~\ref{fig:tfcon_synmag}, we directly compare the $M_i$ and $\msyn$ ITFRs (black inverted triangles and lines vs. red upright triangles and lines, respectively), with each pair of data points connected galaxy-by-galaxy. The flattening of the slope from one relation to the other is expected from the color-luminosity relation: brighter galaxies are redder and therefore have larger color corrections than fainter galaxies. The reduction in the scatter can also be explained similarly: outliers that lie well above the $M_i$ ITFR relation (in particular, those with $\log V_{80} \ga 2.3$) tend to be redder and have larger color corrections, putting them closer to the mean $M_{\rm syn}$ ITFR.

Among these photometric quantities, we choose $\mbell$ to be the optimal photometric estimator of disk rotation velocity. It yields the one of the tightest relations among those we have considered. It has a natural physical interpretation, as an estimate of stellar mass. Finally, it is straightforward to determine from photometry (as a simple linear combination of absolute magnitude and color).


Before we end this section, we study the distribution of residuals from these ITFRs. Visual inspection of Fig.~\ref{fig:linfits_obs3} indicates that the amount of scatter in the ITFRs varies over the full range in ${\cal O}$. In particular, the relation seems to be tightest for intermediate values of ${\cal O}$. To investigate further, we study the normalized residuals $\Delta(\log V_{\rm rot})_i/\sigma_{T,i}$, where the numerator is the offset of a galaxy's measured rotation velocity from the best-fit ITFR, and the denominator is the ``total scatter'' associated with that galaxy, defined as a combination of its measurement errors and the intrinsic scatter for the full ensemble,
\beq\label{eq:sigma_tot}
\sigma_{T,i}^2 = \tildesig^2 + \sigma_{y,i}^2 + (b\sigma_{x,i})^2.
\eeq
We find that the contribution from the uncertainty in the rotation velocity is the dominant term: for the child disk sample, $\langle\sigma_{y}^2\rangle^{1/2}=0.063$ and $\langle\sigma_T^2\rangle^{1/2}=$ 0.081, 0.073, and 0.073, for the $M_i$, $\msyn(i,g-r)$, and $\mbell$ ITFRs, respectively. The contribution from $\sigma_{x,i}$ is always negligible because it is heavily downweighted by the coefficient $b^2 \approx 0.01$.

\begin{figure*}
\bec
\includegraphics[width=6.5in]{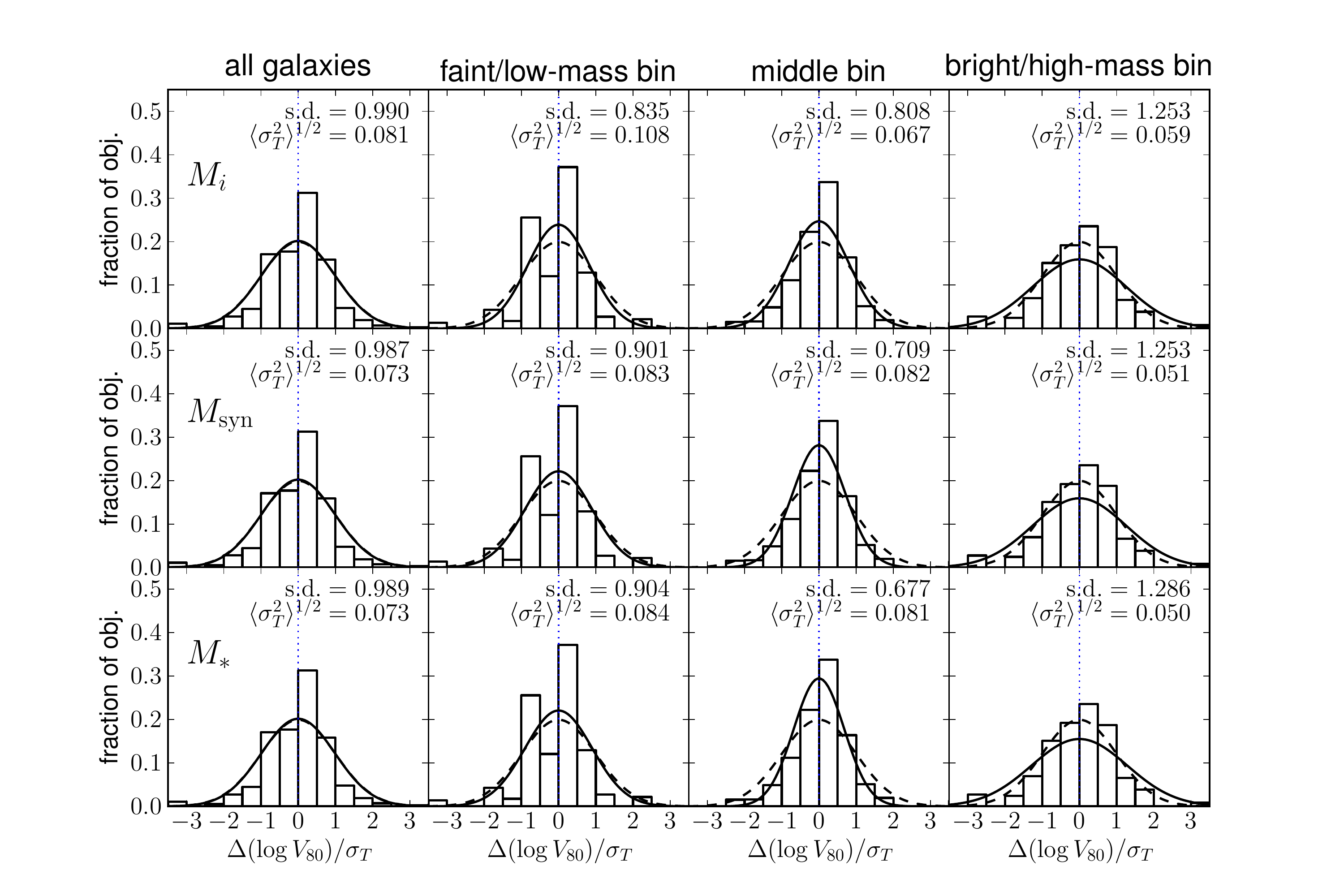}
\eec
\caption{Distribution of normalized residuals $\Delta(\log V_{80})/\sigma_T$ for the $M_i$, $\msyn(i,g-r)$, and $\mbell$ ITFRs (top, middle, and bottom panels, respectively). For each galaxy, $\sigma_{T,i}=(\tildesig^2 + \sigma_{y,i}^2 + b^2\sigma_{x,i}^2)^{1/2}$ is the total scatter, defined to be the root-square-sum of the intrinsic scatter in the ITFR and the galaxy's measurement error. We show distributions for the full child disk sample (left-most panels), as well as for the sample split into three bins in $M_i$, $\msyn$, and $\mbell$ (rest of the panels; from left to right). Each panel is labeled by the standard deviation of the distribution (s.d.) and the RMS value of $\sigma_{T}$. A Gaussian of width equal to the measured s.d. is shown by a solid curve, and one with unit width is shown by a dashed curve, for comparison.}
 \label{fig:resid_obs3}
 \end{figure*}

Figure~\ref{fig:resid_obs3} shows distributions of the normalized residuals from the $M_i$, $\msyn(i,g-r)$, and $\mbell$ ITFRs (histograms in the top, middle, and bottom panels, respectively). For comparison, Gaussian distributions of unit width and of width equal to the standard deviation (s.d.) of the distribution are also shown (dashed and solid curves, respectively). The leftmost panels show distributions for the full child disk sample and the other panels show distributions for the faint (low-$M_*$), intermediate, and bright (high-$M_*$) bins (from left to right).  

We expect that the distributions for the full sample would be Gaussians of unit width. We find that the distribution widths are indeed close to unity, although the distributions are more centrally-peaked than a Gaussian (leftmost panels in Fig.~\ref{fig:resid_obs3}). 
We find that the distribution widths vary with stellar mass and luminosity. The intermediate bins have the narrowest distributions, with s.d. values less than unity. On the other hand, the bright/high-$M_*$ bins have the widest distributions, with s.d. $\approx 1.3$. These galaxies have better $S/N$ measurements, and therefore tend to have smaller measurement errors on average (with $\langle\sigma_T^2\rangle^{1/2}= 0.05$, compared to the sample mean of 0.08). The faint/low-$M_*$ bins have the largest absolute residuals, but those galaxies also have the largest measurement errors.

It is interesting to compare the distributions for the $M_i$ ITFR on the one hand, and the $M_*$ and $\msyn$ ITFRs on the other. We find that the reduction in scatter when going from $M_i$ to $M_*$ (or $\msyn$) can be mostly attributed to galaxies in the middle bin, with the s.d. of the distribution in normalized residuals dropping from 0.8 for $M_i$ to 0.7 for $\msyn$ (or $\mbell$). 

\section{Alternative TFR fits}
\label{sec:alt_fits}

\subsection{Alternative $\vrot$ amplitudes}
\label{subsec:alt_vrot}

In \S\ref{sec:itfr_calib}, we have presented ITFR fits with $\vrot=V_{80}$, our default rotation velocity amplitude definition. Now, we present ITFR fits with alternative rotation velocities $V_{2.2}$ (the rotation velocity evaluated at a radius $2.2\rd$) and $\vc$ (the asymptotic rotation velocity of the best-fit arctangent model). Table~\ref{tab:linfits_alt_vrot} shows the best-fit parameters for the $M_i$, $\mbell$, and $\msyn(i,g-r)$ ITFRs for $\vrot=V_{2.2}$ and $\vc$. We find that $\alpha$ is robust to the choice of rotation velocity definition, and consistent with the Bell et al. (2003) coefficient $\alpha^{\rm Bell}=-2.16$ (as is the case for $\vrot=V_{80}$). 

\begin{figure*} 
\bec
$\begin{array}{cc}
\includegraphics[width=3in]{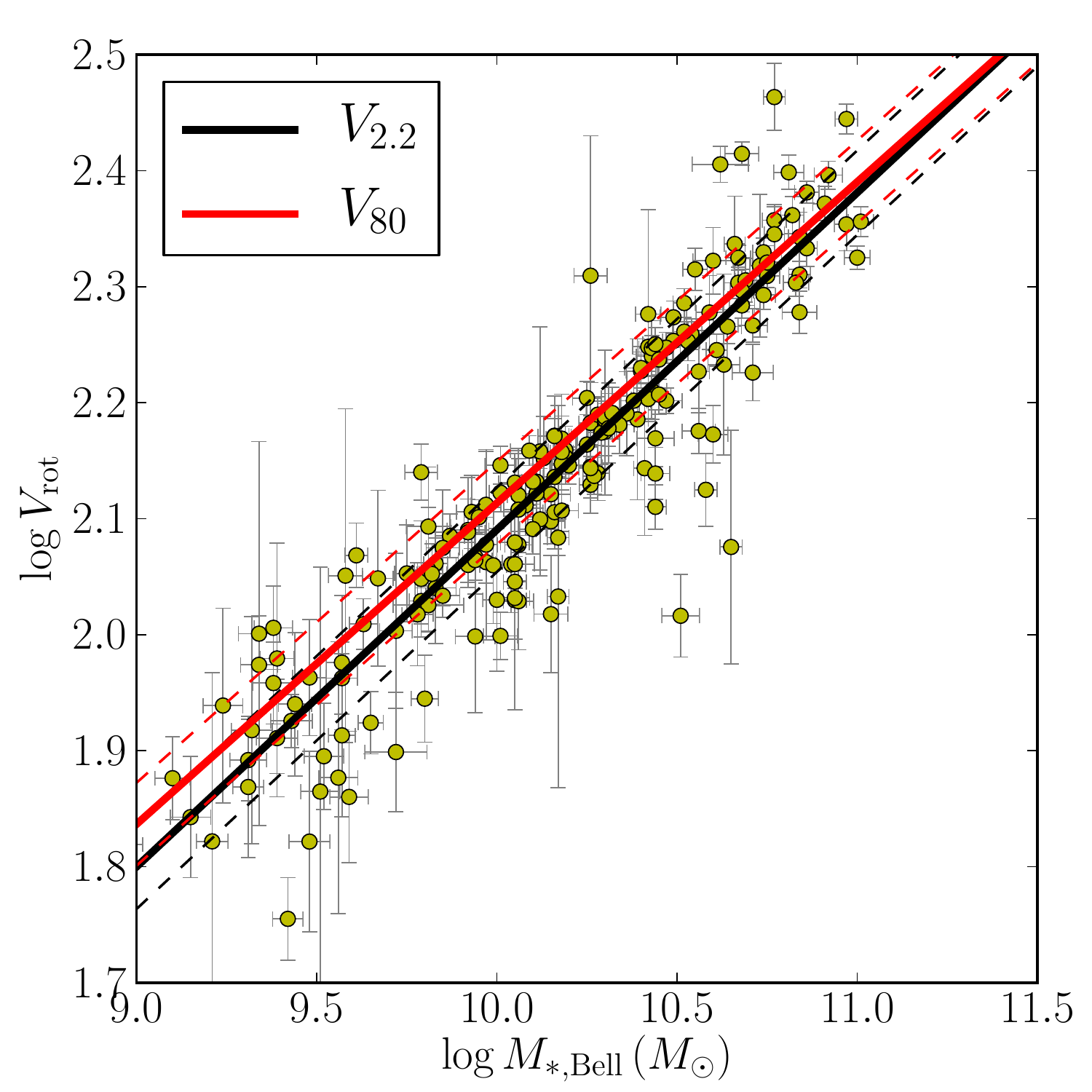} &
\includegraphics[width=3in]{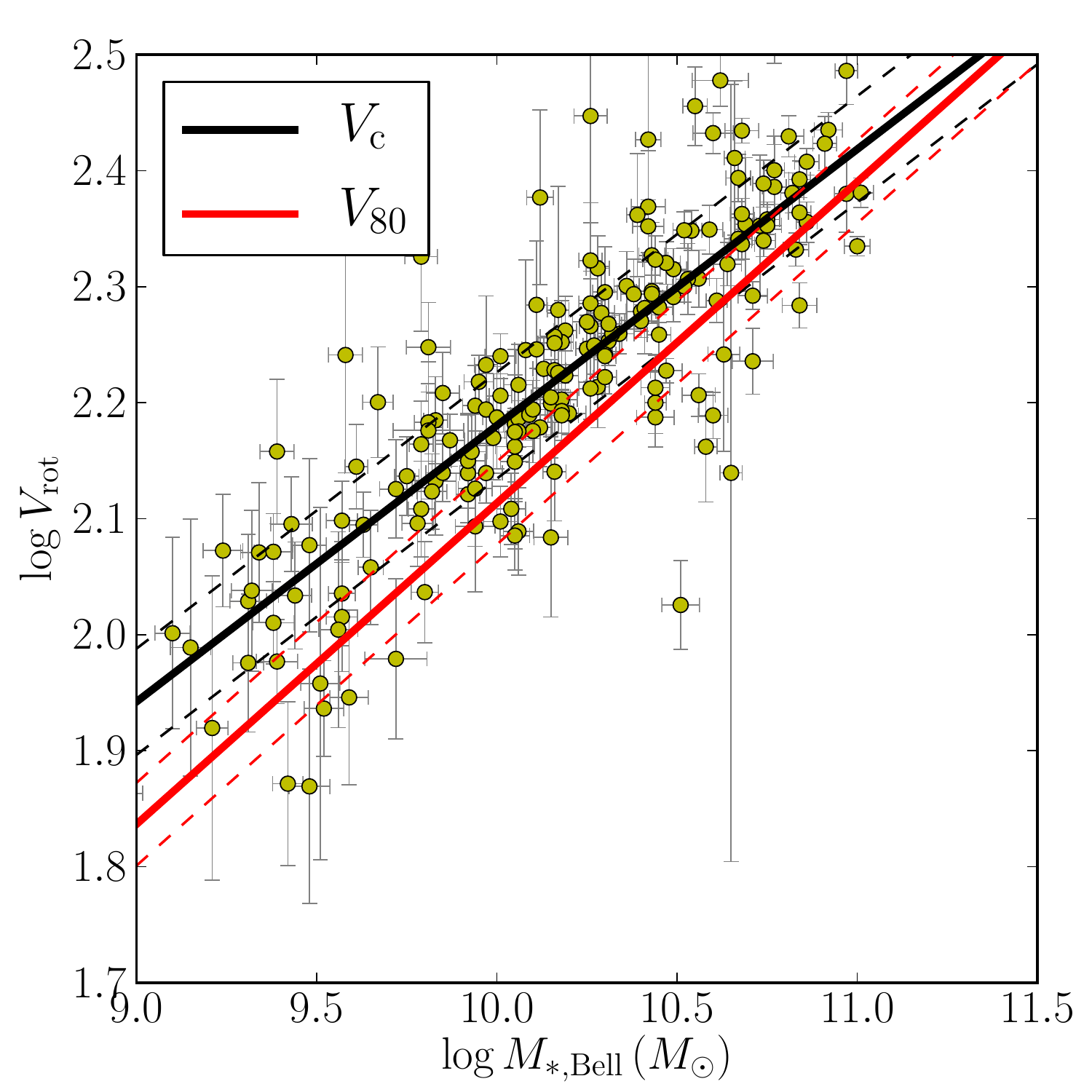}
\end{array}
$\eec
\caption{Calibrated $\mbell$ ITFRs with rotation velocities $V_{2.2}$ and $\vc$ (left and right panels, respectively) for the child disk sample (filled circles with 1-$\sigma$ error bars). In each panels, the best-fit ITFR is shown by the black solid line and the amount of intrinsic scatter is shown by the dashed lines (displaced by $\pm\tildesig$ from the mean relation). For comparison, the $\mbell$ ITFR with $\vrot=V_{80}$ is shown by red solid and dashed lines.}
\label{fig:tfcomp_vc}
\end{figure*}

The panels of Figure~\ref{fig:tfcomp_vc} compare the $\mbell$ ITFR with $\vrot=V_{80}$ (red solid and dashed lines, on both left and right panels), with the observed ITFRs with $\vrot=V_{2.2}$ and $\vc$ (black solid and dashed lines in the left and right panels, respectively). As expected from the systematic difference between the $V_{2.2}$ and $V_{80}$, the zero-point of the $V_{2.2}$ ITFR is 0.02 dex (5 per cent) lower than that of the $V_{80}$ ITFR. The slope of the ITFR is slightly steeper for the former, but only by $\sim 1\sigma$. The amounts of intrinsic and measured scatter are similar for the two relations. These results confirm our earlier assertion (in \S\ref{subsec:vrot_definition}) that the ITFR fits are relatively insensitive to the choice of rotation velocity amplitude definition, for reasonable choices for the optical (i.e., evaluation) radius. 

In contrast to $V_{80}$ and $V_{2.2}$, using $\vc$ as the rotation velocity involves extrapolation well beyond the observed rotation curve. This can lead to systematic overestimation of the disk rotation velocity, especially in galaxies with large turn-over radii (i.e., slowly-rising rotation curves). We find that the best-fit $\vc$ ITFR has a significantly higher zero-point, shallower slope, and larger intrinsic scatter than the $V_{80}$ ITFR. The right panel of Fig.~\ref{fig:tfcomp_vc} shows that the two relations coincide at the high-mass end. This is consistent with our observation that the highest $M_*$ galaxies have small turn-over radii (i.e., steeply-rising rotation curves), and consequently $\vc/V_{80}$ ratios close to unity (see Fig.~\ref{fig:mstr_fto}). 

\begin{table*} 
\begin{center}
\caption{Calibrated $M_i$, $\msyn(i,g-r)$, and $\mbell$ ITFR fits for the child disk sample (189 galaxies) with $\vrot=V_{80}$, $V_{2.2}$, and $\vc$.}
\begin{tabular}{llrrrrr}
\hline 
 \multicolumn{1}{c}{$y$} &
 \multicolumn{1}{c}{$x$} &
 \multicolumn{1}{c}{$x_{\rm p}$} &
 \multicolumn{1}{c}{$a$} &
 \multicolumn{1}{c}{$b$} &
 \multicolumn{1}{c}{$\tilde{\sigma}$} &
 \multicolumn{1}{c}{$\sigma$} \\
 \multicolumn{1}{c}{(1)} & 
 \multicolumn{1}{c}{(2)} & 
\multicolumn{1}{c}{(3)} & 
\multicolumn{1}{c}{(4)} & 
\multicolumn{1}{c}{(5)} &
\multicolumn{1}{c}{(6)} &
\multicolumn{1}{c}{(7)} \\
\hline
    $\log V_{80}$ &                  $M_i$ & $-20.558$ & $  2.142$\,($  0.005$) & $ -0.128$\,($  0.006$) & $  0.049$\,($  0.007$) &  $  0.066$ \\
              --- &   $M_{\rm syn}(i,g-r)$ & $-21.965$ & $  2.160$\,($  0.004$) & $ -0.107$\,($  0.004$) & $  0.035$\,($  0.006$) &  $  0.056$ \\
              --- &       $M_{\rm *,Bell}$ & $ 10.102$ & $  2.142$\,($  0.004$) & $ -0.111$\,($  0.004$) & $  0.036$\,($  0.005$) &  $  0.056$ \\
\hline
   $\log V_{2.2}$ &                  $M_i$ & $-20.558$ & $  2.120$\,($  0.004$) & $ -0.133$\,($  0.006$) & $  0.052$\,($  0.007$) &  $  0.065$ \\
              --- &   $M_{\rm syn}(i,g-r)$ & $-21.799$ & $  2.142$\,($  0.004$) & $ -0.107$\,($  0.004$) & $  0.035$\,($  0.005$) &  $  0.056$ \\
              --- &       $M_{\rm *,Bell}$ & $ 10.102$ & $  2.120$\,($  0.004$) & $ -0.116$\,($  0.004$) & $  0.036$\,($  0.005$) &  $  0.055$ \\
\hline
 $\log V_{\rm c}$ &                  $M_i$ & $-20.558$ & $  2.205$\,($  0.005$) & $ -0.112$\,($  0.006$) & $  0.053$\,($  0.007$) &  $  0.093$ \\
              --- &   $M_{\rm syn}(i,g-r)$ & $-21.799$ & $  2.142$\,($  0.004$) & $ -0.107$\,($  0.004$) & $  0.035$\,($  0.005$) &  $  0.056$ \\
              --- &       $M_{\rm *,Bell}$ & $ 10.102$ & $  2.205$\,($  0.004$) & $ -0.095$\,($  0.004$) & $  0.046$\,($  0.006$) &  $  0.087$ \\
\hline 
\end{tabular}
\label{tab:linfits_alt_vrot}
\end{center}
\begin{flushleft}
Notes. --- {Columns are the same as in Table~\ref{tab:itfr_absmag}. Results for the $y=\log V_{80}$ are the same as in Table~\ref{tab:itfr_obs3}.}
\end{flushleft}
\end{table*}%

\subsection{Alternative internal extinction corrections}
\label{subsec:alt_extcorr}

We also present fits to the $M_\lambda$ ITFRs for alternative internal extinction corrections. Our default internal extinction corrections are applied to the mean orientation of the child disk sample, and chosen so as to minimize the size, and therefore uncertainty, of the corrections (see \S\ref{subsec:int_ext_corr}). Table~\ref{tab:alt_extcorr} lists ITFR fit parameters for $M_\lambda^{i-0}$, with extinction corrections applied to face-on orientation (usually adopted in the literature, in particular by P07), as well as $M_\lambda^{\rm NC}$, with no extinction corrections at all.

For easy comparisons, we have chosen the pivot values $M_{\lambda,\rm p}$ to be that used for the default fit. As expected from the differences in the amount of extinction corrections, the normalization of the ITFR is lower by $\sim$0.04 dex for $M_\lambda^{i-0}$, and higher by $\sim$ 0.003 for $M_\lambda^{\rm NC}$. The amount of measured scatter is similar in all three cases; the amount of intrinsic scatter is slightly smaller for the extinction-corrected magnitudes than for the uncorrected ones (because of the additional uncertainty from the extinction corrections). 

As a test of our internal extinction corrections, we will show in \S\ref{subsec:correl_axisratio} that the velocity residuals in the $M_i^{\rm NC}$ ITFR are correlated with disk axis ratios, while those for the $M_i$ ITFR are not. 


\begin{table*}
\begin{center}
\caption{Calibrated  $M_\lambda$ ITFR fits for alternative internal extinction corrections for the child disk sample (189 galaxies).}
\begin{tabular}{llrrrrr}
\hline 
 \multicolumn{1}{c}{$y$} &
 \multicolumn{1}{c}{$x$} &
 \multicolumn{1}{c}{$x_{\rm p}$} &
 \multicolumn{1}{c}{$a$} &
 \multicolumn{1}{c}{$b$} &
 \multicolumn{1}{c}{$\tilde{\sigma}$} &
 \multicolumn{1}{c}{$\sigmeas$} \\
 \multicolumn{1}{c}{(1)} & 
 \multicolumn{1}{c}{(2)} & 
\multicolumn{1}{c}{(3)} & 
\multicolumn{1}{c}{(4)} & 
\multicolumn{1}{c}{(5)} &
\multicolumn{1}{c}{(6)} &
\multicolumn{1}{c}{(7)} \\
\hline
\multicolumn{7}{c}{With default internal extinction corrections (same as Table~\ref{tab:itfr_absmag})} \\ 
\hline
    $\log V_{80}$ &        $M_u$ & $-18.731$ & $  2.188$\,($  0.026$) & $ -0.080$\,($  0.009$) & $  0.114$\,($  0.011$) &  $  0.114$ \\
              --- &        $M_g$ & $-19.903$ & $  2.142$\,($  0.006$) & $ -0.129$\,($  0.008$) & $  0.068$\,($  0.008$) &  $  0.082$ \\
              --- &        $M_r$ & $-20.375$ & $  2.142$\,($  0.005$) & $ -0.130$\,($  0.007$) & $  0.056$\,($  0.007$) &  $  0.071$ \\
              --- &        $M_i$ & $-20.558$ & $  2.142$\,($  0.005$) & $ -0.128$\,($  0.006$) & $  0.049$\,($  0.007$) &  $  0.066$ \\
              --- &        $M_z$ & $-20.649$ & $  2.142$\,($  0.005$) & $ -0.119$\,($  0.005$) & $  0.049$\,($  0.007$) &  $  0.065$ \\
\hline
\multicolumn{7}{c}{With internal extinction corrections to face-on orientation} \\
\hline
    $\log V_{80}$ &       $M_u^{i-0}$ & $-18.731$ & $  2.142$\,($  0.034$) & $ -0.072$\,($  0.015$) & $  0.112$\,($  0.016$) &  $  0.112$ \\
              --- &       $M_g^{i-0}$ & $-19.903$ & $  2.079$\,($  0.007$) & $ -0.117$\,($  0.007$) & $  0.064$\,($  0.009$) &  $  0.082$ \\
              --- &       $M_r^{i-0}$ & $-20.375$ & $  2.087$\,($  0.006$) & $ -0.120$\,($  0.006$) & $  0.052$\,($  0.008$) &  $  0.071$ \\
              --- &       $M_i^{i-0}$ & $-20.558$ & $  2.098$\,($  0.005$) & $ -0.119$\,($  0.005$) & $  0.046$\,($  0.008$) &  $  0.066$ \\
              --- &       $M_z^{i-0}$ & $-20.649$ & $  2.111$\,($  0.005$) & $ -0.112$\,($  0.004$) & $  0.047$\,($  0.007$) &  $  0.065$ \\
\hline
\multicolumn{7}{c}{Without internal extinction corrections} \\
\hline
    $\log V_{80}$ &    $M_u^{\rm NC}$ & $-18.731$ & $  2.190$\,($  0.022$) & $ -0.081$\,($  0.011$) & $  0.114$\,($  0.017$) &  $  0.114$ \\
              --- &    $M_g^{\rm NC}$ & $-19.903$ & $  2.147$\,($  0.006$) & $ -0.123$\,($  0.008$) & $  0.072$\,($  0.009$) &  $  0.083$ \\
              --- &    $M_r^{\rm NC}$ & $-20.375$ & $  2.147$\,($  0.005$) & $ -0.126$\,($  0.006$) & $  0.060$\,($  0.009$) &  $  0.073$ \\
              --- &    $M_i^{\rm NC}$ & $-20.558$ & $  2.145$\,($  0.005$) & $ -0.127$\,($  0.005$) & $  0.052$\,($  0.008$) &  $  0.066$ \\
              --- &    $M_z^{\rm NC}$ & $-20.649$ & $  2.145$\,($  0.005$) & $ -0.119$\,($  0.005$) & $  0.049$\,($  0.008$) &  $  0.065$ \\
\hline
\end{tabular}
\label{tab:alt_extcorr}
\end{center}
\begin{flushleft}
Notes. --- {Columns are the same as in Table~\ref{tab:itfr_absmag}.}
\end{flushleft}
\end{table*}%

\subsection{Alternative fit directions}
\label{subsec:alt_directions}

We present calibrated forward fits to the $M_i$, $\msyn(i,g-r)$, and $\mbell$ TFRs in Table~\ref{tab:alt_directions_fwd}. In these fits, the photometric quantity ${\cal O}$ serves as the dependent variable, and $\vrot=V_{80}$ serves as the independent variable (see Eqs.~\ref{eq:tf_absmag_fwd}--\ref{eq:tf_bellmass_fwd}). 
We list the best-fit forward TFR parameters $a_{\rm fwd}$, $b_{\rm fwd}$, and ${\tildesig}_{\rm fwd}$, as well as the converted fit parameters $a_{\rm conv}$, $b_{\rm conv}$ and ${\tildesig}_{\rm conv}$ (defined after Eq.~\ref{eq:tf_bellmass_fwd} in \S\ref{subsec:tfr_model}). 

\begin{figure}
\bec
\includegraphics[width=3in]{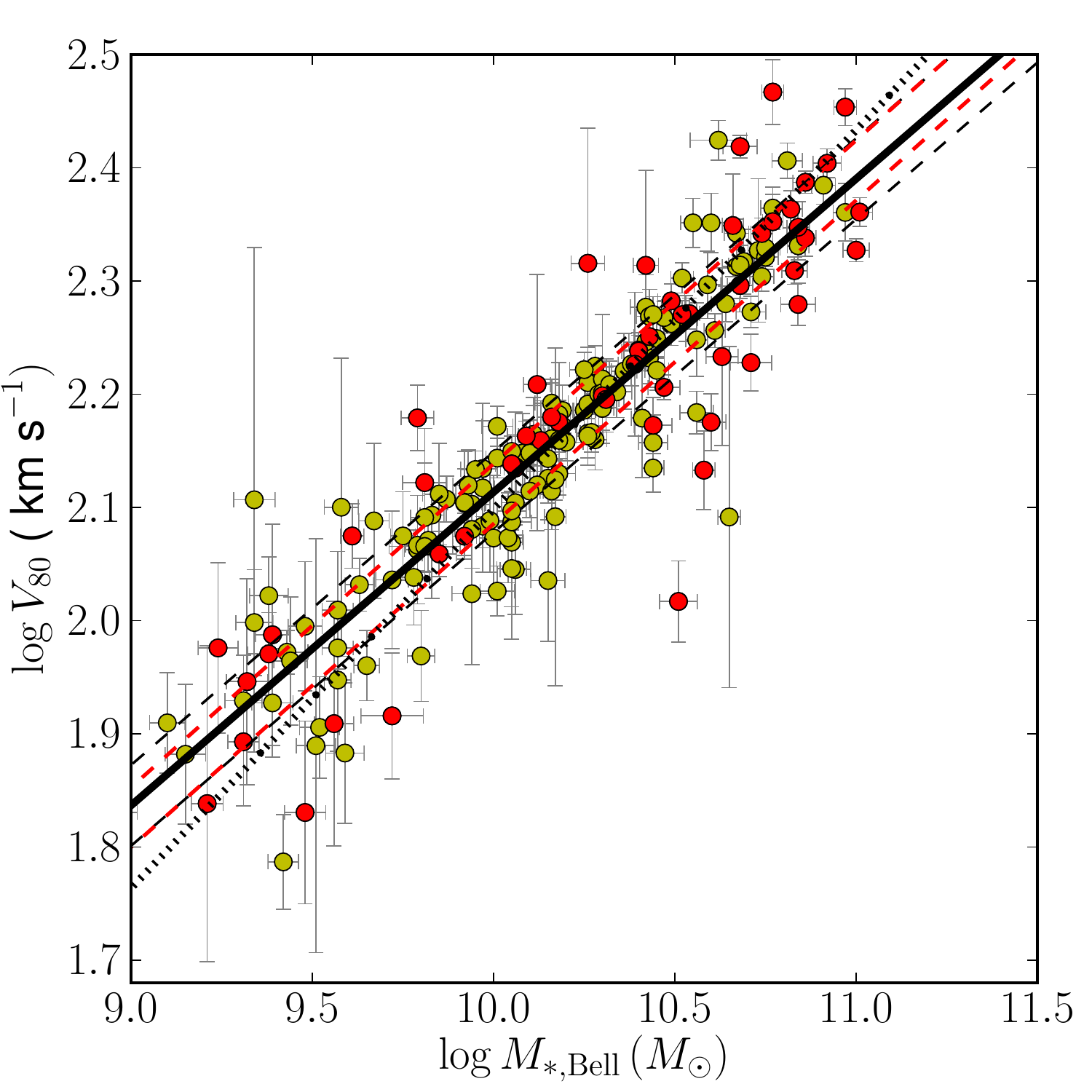}
\eec
\caption{Comparison of alternative fits to the $\mbell$ TFR for the child disk sample (189 galaxies). Galaxies with $D/T>0.9$ (137 galaxies) and $D/T\le 0.9$ are shown by yellow and red circles, respectively (shown with 1-$\sigma$ error bars). Solid and dashed black lines show the best-fit ITFR and $\pm 1\tildesig$ bounds; the dotted black line shows the best-fit forward TFR. Red dashed lines show $\pm 1\tildesig$ bounds for the best-fit ITFR for galaxies with $D/T>0.9$.}
\label{fig:linfits_comp_fwd_dtr}
\end{figure}

Compared with the inverse fits (shown in the lower half of Table~\ref{tab:alt_directions_fwd}, for easy comparisons), the forward fits have slightly larger intrinsic scatter (by $\sim$1$\sigma$), and significantly steeper slopes ($b_{\rm conv}<b$). Figure~\ref{fig:linfits_comp_fwd_dtr} shows the inverse and forward fits to the $\mbell$ TFR (black solid and dotted lines, respectively).

The steeper slopes from the forward fits can be (at least, partly) attributed to a Malmquist-type bias due to our applied absolute magnitude cuts. Based on a Monte-Carlo experiment performed by P07 (see their Appendix B), the effect of the cuts is to steepen the slope and reduce the intrinsic scatter. We do not attempt to correct for this bias and present the forward fit results at face value. Throughout the rest of the paper, we focus on inverse fits, which are not affected by this bias.

We have also performed orthogonal fits to the TFRs, in which offsets are measured in the direction orthogonal to the best-fit relation itself (rather than in the horizontal or vertical directions). We find that the best-fit parameters from the orthogonal fits are almost identical to those for the inverse fits. This is not surprising because the total measurement uncertainty is dominated by the rotation velocity uncertainty, i.e., $\sigma_i(\log\vrot)$ is much larger than $b\times \sigma_i({\cal O})$ (c.f. Eq.~\ref{eq:sigma_tot} and discussion afterwards). 

\begin{table*}
\begin{center}
\caption{Calibrated forward fits to the TFRs for the child disk sample (189 galaxies) with independent variable $x=\log V_{80}$ and dependent variable $y={\cal O}=\{M_i,\msyn(i,g-r),\log\mbell\}$.} 
\label{tab:alt_directions_fwd}
\begin{tabular}{lrrrrrrr}
\hline
 \multicolumn{1}{c}{${\cal O}$} &
 \multicolumn{1}{c}{$x_{\rm p}$} &
 \multicolumn{1}{c}{$a_{\rm fwd}$} &
 \multicolumn{1}{c}{$a_{\rm conv}$} & 
 \multicolumn{1}{c}{$b_{\rm fwd}$} &
  \multicolumn{1}{c}{$b_{\rm conv}$} &
 \multicolumn{1}{c}{${\tilde{\sigma}}_{\rm fwd}$} &
 \multicolumn{1}{c}{${\tilde{\sigma}}_{\rm conv}$} \\
 \multicolumn{1}{c}{(1)} & 
 \multicolumn{1}{c}{(2)} & 
\multicolumn{1}{c}{(3)} & 
\multicolumn{1}{c}{(4)} & 
\multicolumn{1}{c}{(5)} &
\multicolumn{1}{c}{(6)} &
\multicolumn{1}{c}{(7)} &
\multicolumn{1}{c}{(8)} \\
\hline 
\multicolumn{8}{c}{Forward fits to the TFR}\\
\hline
                 $M_i$ & $  2.142$ & $-20.602$\,($  0.033$) & $  2.131$\,($  0.006$) & $ -5.807$\,($  0.307$) & $ -0.172$\,($  0.009$) & $  0.344$\,($  0.041$) & $  0.059$\,($  0.007$) \\
         $M_{\rm syn}$ & $  2.160$ & $-21.983$\,($  0.032$) & $  2.115$\,($  0.004$) & $ -7.716$\,($  0.310$) & $ -0.130$\,($  0.005$) & $  0.312$\,($  0.046$) & $  0.041$\,($  0.006$) \\
      $M_{\rm *,Bell}$ & $  2.142$ & $ 10.054$\,($  0.033$) & $  2.133$\,($  0.004$) & $ -7.465$\,($  0.281$) & $ -0.134$\,($  0.005$) & $  0.307$\,($  0.043$) & $  0.041$\,($  0.006$) \\
\hline
\multicolumn{8}{c}{Inverse fits to the TFR with $\vrot=V_{80}$ (same as in Table~\ref{tab:itfr_obs3})} \\
\hline
                 $M_i$ & $-20.558$ &               $\ldots$ & $  2.142$\,($  0.005$) &               $\ldots$ & $ -0.128$\,($  0.006$) &               $\ldots$ & $  0.049$\,($  0.007$) \\
         $M_{\rm syn}$ & $-21.965$ &               $\ldots$ & $  2.160$\,($  0.004$) &               $\ldots$ & $ -0.107$\,($  0.004$) &               $\ldots$ & $  0.035$\,($  0.006$) \\
      $M_{\rm *,Bell}$ & $ 10.102$ &               $\ldots$ & $  2.142$\,($  0.004$) &               $\ldots$ & $ -0.111$\,($  0.004$) &               $\ldots$ & $  0.036$\,($  0.005$) \\
\hline
\end{tabular}
\end{center}
\begin{flushleft}
Notes. --- {Col. (1): ${\cal O}$ is the photometric quantity--- the dependent (independent) variable in the forward (inverse) fits. Col. (2): $x_{\rm p}$ is the pivot value for the independent variable--- $(\log V_{80})_{\rm p}$ for the forward fits and ${\cal O}_{\rm p}$ for the inverse fits. Cols. (3),(5), and (7): Best-fit forward TFR parameters--- $a_{\rm fwd}$, $b_{\rm fwd}$, and ${\tildesig}_{\rm fwd}$--- and their 1-$\sigma$ uncertainties, as defined in Eqs.~\ref{eq:tf_absmag_fwd}--\ref{eq:tf_bellmass_fwd}. Cols. (4),(6), and (8): Converted forward TFR parameters to be compared with the inverse fit parameters.}
\end{flushleft}
\end{table*}%

\subsection{Alternative sample: $D/T>0.9$} 
\label{subsec:alt_dtr}

For various reasons, disk-dominated galaxies are expected to form a tighter TFR than the general population. The lack of a significant bulge component makes the measurement of various photometric quantities cleaner (i.e., less noisy). In addition, late-type spirals have been found to have smaller intrinsic disk ellipticities than earlier types \citep{2006ApJ...641..773R}, leading to a smaller contribution to the scatter due to random viewing angles. Disk-dominated galaxies tend to be kinematically homogeneous systems as well. Figure~\ref{fig:linfits_comp_fwd_dtr} shows that galaxies in the child disk sample with $D/T>0.9$ (135 galaxies; yellow circles) indeed form a tighter TFR than the full sample. 

We have performed weighted ITFR fits to the disk-dominated subsample of galaxies with $D/T>0.9$. The red dashed lines in Fig.~\ref{fig:linfits_comp_fwd_dtr} mark $\pm 1\tildesig$ from the best-fit relation. We find $\tildesig=0.027\pm 0.009$ and $\sigmeas=0.049$, compared to $\tildesig=0.036\pm 0.005$ and $\sigmeas=0.056$ for the full child disk sample (Table~\ref{tab:itfr_obs3}). We conclude that inclusion of only the disk-dominated galaxies yield a tighter ITFR, at the $\sim$1$\sigma$ level.  

\subsection{Comparison with P07 fits}
\label{subsec:alt_P07}

To make a fair comparison with the results presented in P07, we have performed ITFR fits to the P07 galaxy sample using the same photometric quantity and rotation velocity amplitude definition used by P07. The photometric quantity used is absolute magnitude corrected for internal extinction to face-on orientation, $M_i^{i-0}$, and the rotation velocity amplitude definition used is $V_{80}$.  

First, we perform unweighted fits to P07 measurements (both for the $x$ and $y$ variables in the ITFR) and reproduce the results published in Table 4 of P07. Then, we perform weighted fits to our own measurements. Table~\ref{tab:alt_P07} lists the three different sets of best-fit ITFR parameters. The relations derived using P07 measurements and ours are consistent with one another within their reported intrinsic scatter, but the two deviate slightly at the bright end (our measurements yield a best-fit ITFR with a shallower slope and lower normalization). The slight differences can be traced to differences in the determination of $V_{80}$ (c.f. \S\ref{subsubsec:syst_fits}) and internal extinction corrections to $M_i$ in the two methods. 

\begin{table*}
\begin{center}
\caption{Comparison of ITFR fits for the P07 and child disk samples (with 157 and 189 galaxies, respectively).}
\begin{tabular}{llrrrrr}
\hline 
 \multicolumn{1}{c}{$y$} &
 \multicolumn{1}{c}{$x$} &
 \multicolumn{1}{c}{$x_{\rm p}$} &
 \multicolumn{1}{c}{$a$} &
 \multicolumn{1}{c}{$b$} &
 \multicolumn{1}{c}{$\tilde{\sigma}$} &
 \multicolumn{1}{c}{$\sigmeas$} \\
 \multicolumn{1}{c}{(1)} & 
 \multicolumn{1}{c}{(2)} & 
\multicolumn{1}{c}{(3)} & 
\multicolumn{1}{c}{(4)} & 
\multicolumn{1}{c}{(5)} &
\multicolumn{1}{c}{(6)} &
\multicolumn{1}{c}{(7)} \\
\hline
\multicolumn{7}{c}{ITFR parameters from Table~4 of P07} \\ 
\hline
$\log V_{80}$ &    $M_i^{i-0}$ & $-21.327$ &    $2.212(0.005)$ &   $-0.130(0.005)$ &    $0.061(0.005)$ &  $\ldots$ \\
\hline
\multicolumn{7}{c}{Unweighted ITFR fits for the P07 galaxy sample, using P07 measurements} \\
\hline
    $\log V_{80}$ &            $M_i^{i-0}$ & $-21.327$ & $  2.207$\,($  0.005$) & $ -0.127$\,($  0.005$) & $  0.058$\,($  0.005$) &  $  0.067$ \\
\hline
\multicolumn{7}{c}{Calibrated ITFR fits for the P07 galaxy sample, using our measurements} \\
\hline
    $\log V_{80}$ &            $M_i^{i-0}$ & $-21.327$ & $  2.195$\,($  0.007$) & $ -0.122$\,($  0.006$) & $  0.055$\,($  0.009$) &  $  0.102$ \\
\hline
\end{tabular}
\label{tab:alt_P07}
\end{center}
\begin{flushleft}
Notes. --- {Columns are the same as in Table~\ref{tab:itfr_absmag}.}
\end{flushleft}
\end{table*}%


\section{TFR residual correlations}
\label{sec:tfr_correl}

In this section, we study correlations between velocity residuals from the calibrated ITFRs (derived in \S\ref{sec:itfr_calib} and summarized in Table~\ref{tab:itfr_obs3}) and various galaxy properties, namely: disk axis ratios in \S\ref{subsec:correl_axisratio}, galaxy colours in \S\ref{subsec:correl_colour}, and disk sizes in \S\ref{subsec:correl_size}.

We calculate two kinds of correlation coefficients: the Pearson linear correlation coefficient $r$ and the Spearman rank correlation coefficient $\rho$. The former assesses how well the relationship between two variables can be described as linear ($r=\pm 1$ means there is a perfect positive/negative linear correlation), while the latter assesses how well the relationship can be described as monotonic, regardless of the form of the relationship. We also quote the two-tailed significance of the deviation of $\rho$ from zero, ${\rm Sig}(\rho)$. A small value (i.e., smaller than $0.01$) indicates a significant correlation. We also perform unweighted linear fits using a similar procedure as for the TFR fits (described in \S\ref{subsec:tfr_fit}) and quote best-fit slopes $b$. 

\subsection{Correlations with disk axis ratio}
\label{subsec:correl_axisratio}
To test our internal extinction corrections (defined in \S\ref{subsec:int_ext_corr}), we study the correlation between disk axis ratios $\qd$ and velocity residuals from the $\minc$ and $M_i$ ITFRs, i.e., before and after internal extinction corrections (c.f. Table~\ref{tab:alt_extcorr}). We expect to see a positive correlation for the $\minc$ ITFR residuals, because more highly-inclined galaxies tend to be more heavily affected by dust and artificially displaced from the TFR toward fainter luminosities. Hence, they tend to have larger (i.e., more positive) velocity residuals. If our internal extinction corrections are valid on average, we expect the correlation to decrease or disappear altogether in the $M_i$ ITFR residuals. 

\begin{figure*}
\bec
\begin{tabular}{cc}
\includegraphics[width=3in]{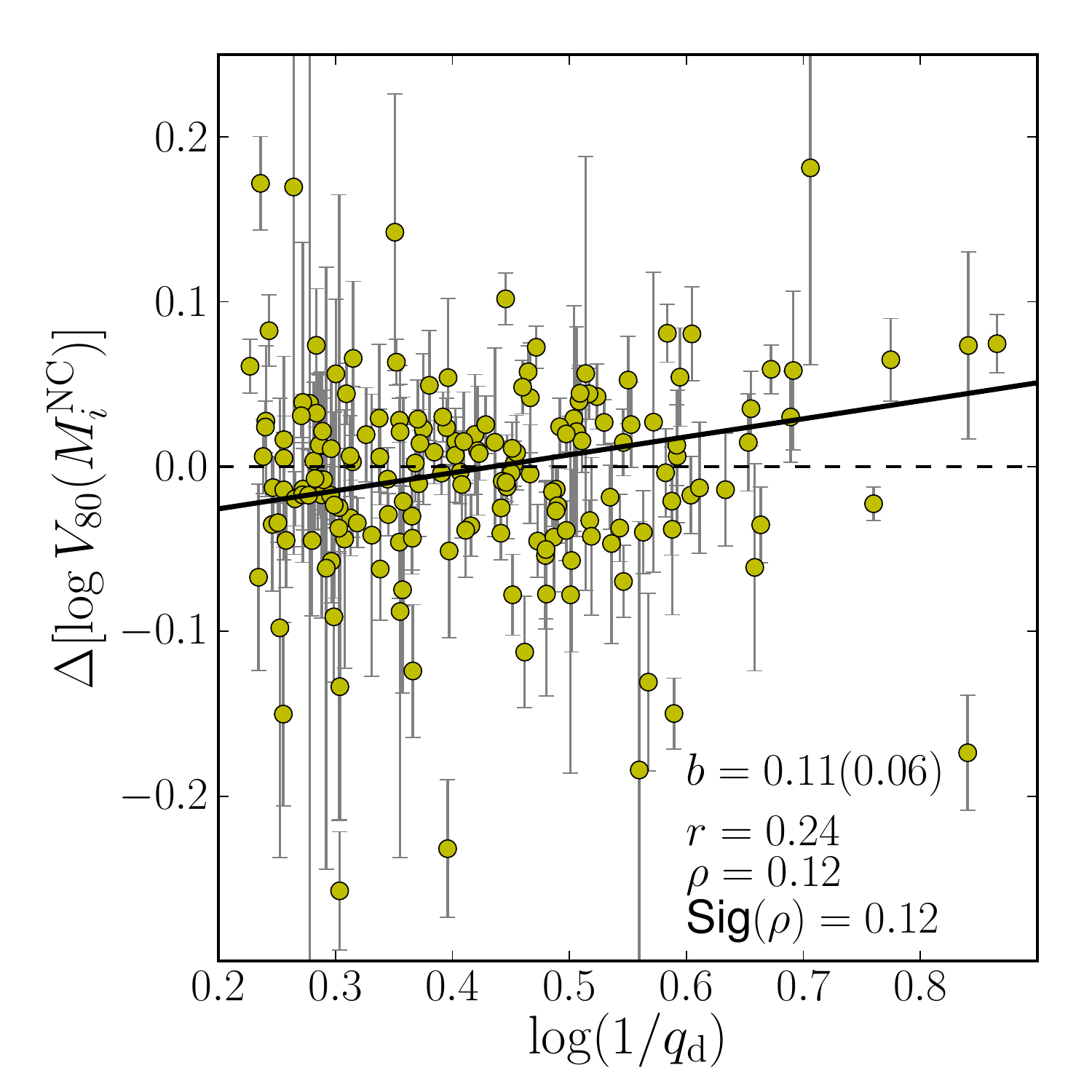} &
\includegraphics[width=3in]{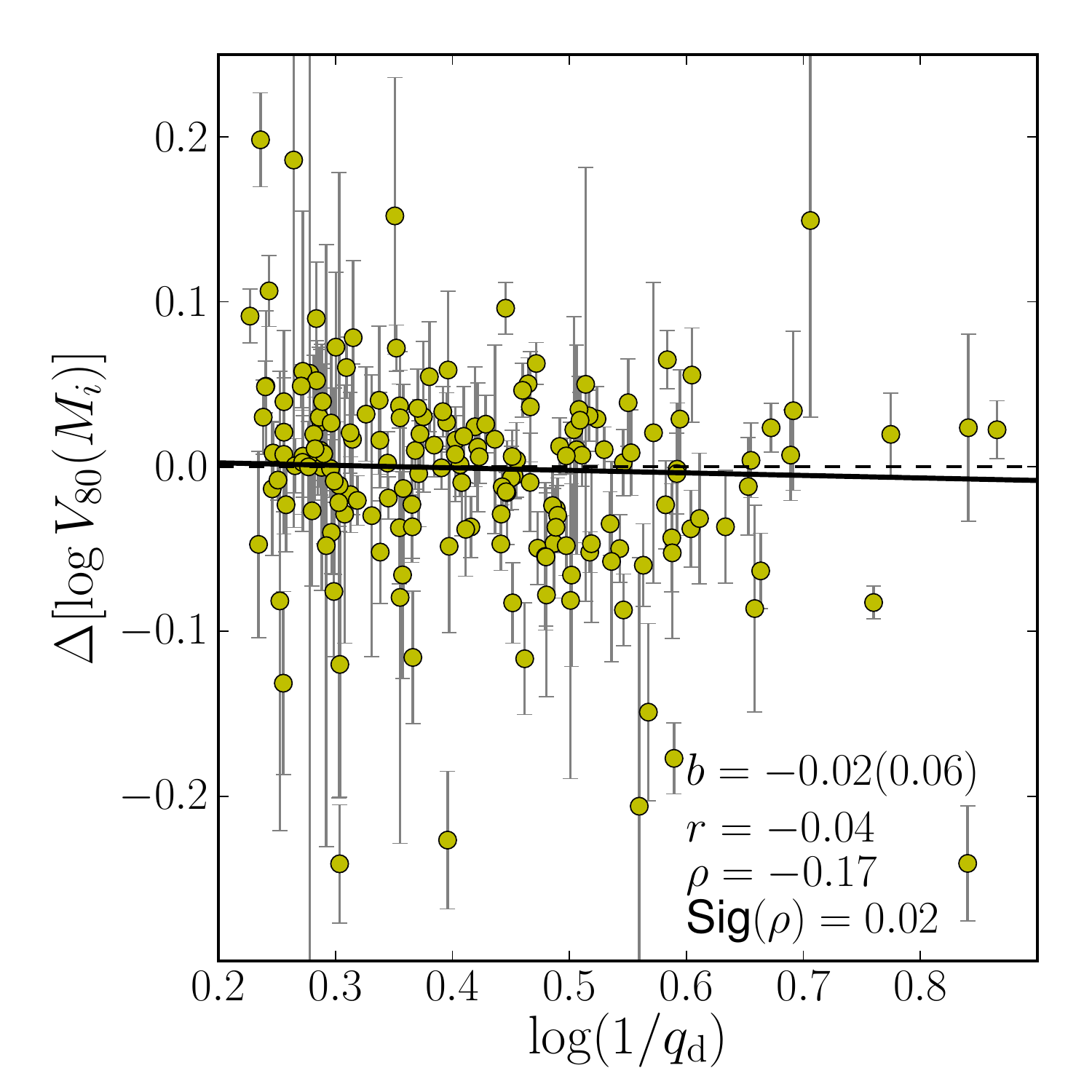}
\end{tabular}
\eec
\caption{Correlation between disk axis ratios $\qd$ and $\log V_{80}$ residuals from the $\minc$ and $M_i$ ITFRs (i.e., absolute magnitudes before and after internal extinction correction) (left and right panels, respectively). Filled circles are shown with 1-$\sigma$ error bars, for 179 (out of 189) galaxies in the child disk sample with $\qd<0.6$. Solid lines show the best-fit linear relations; horizontal dashed lines show the zero level. Labels show the best-fit slope $b$ and its 1-$\sigma$ uncertainty, the Pearson linear correlation coefficient $r$, the Spearman rank correlation $\rho$ and the two-tailed significance of its deviation from zero.}
\label{fig:resfits_axisratio}
\end{figure*}

Figure~\ref{fig:resfits_axisratio} shows $\log(1/\qd)$ versus velocity residuals $\Delta[\log V_{80}(M_i^{\rm NC})]$ and $\Delta[\log V_{80}(M_i)]$ (left and right panels, respectively). The best-fit linear relations for 179 (out of 189) galaxies in the child disk sample with $\qd<0.6$ are shown by solid lines on both panels. (Recall that our target selection cut is based on $\qiso$, not $\qd$; a number of child disk sample galaxies have $\qd$ slightly above 0.6.) For $\minc$, the best-fit slope is non-zero and positive, $b=0.11 \pm 0.06$, while for $M_i$, the best-fit slope is consistent with zero, $b=0.02 \pm 0.06$. The correlation coefficients indicate a weak positive correlation in the former case, and a weak negative correlation in the latter, albeit neither correlation is very significant. We conclude that the applied internal extinction corrections are adequate on average.
 
\subsection{Correlations with galaxy colour}
\label{subsec:correl_colour}

The observed colours of a galaxy encapsulate properties of its stellar population, in particular, its average stellar mass-to-light ratio $M_*/L$. In \S\ref{sec:itfr_calib}, we showed that by adding color information to the luminosity (i.e., using $M_*$ or $\msyn$, instead of $M_i$), we can reduce the intrinsic scatter in the ITFR by $\sim 30$ per cent (or 2$\sigma$). Now, we explicitly show that in doing so, we have successfully extracted most of the available information from galaxy colour to yield the tightest possible TFR. In other words, most of the correlation between galaxy colours and velocity residuals from the $M_i$ ITFR is removed when $\msyn$ is used instead. 

\begin{figure*}
\bec
\begin{tabular}{cc}
\includegraphics[width=3in]{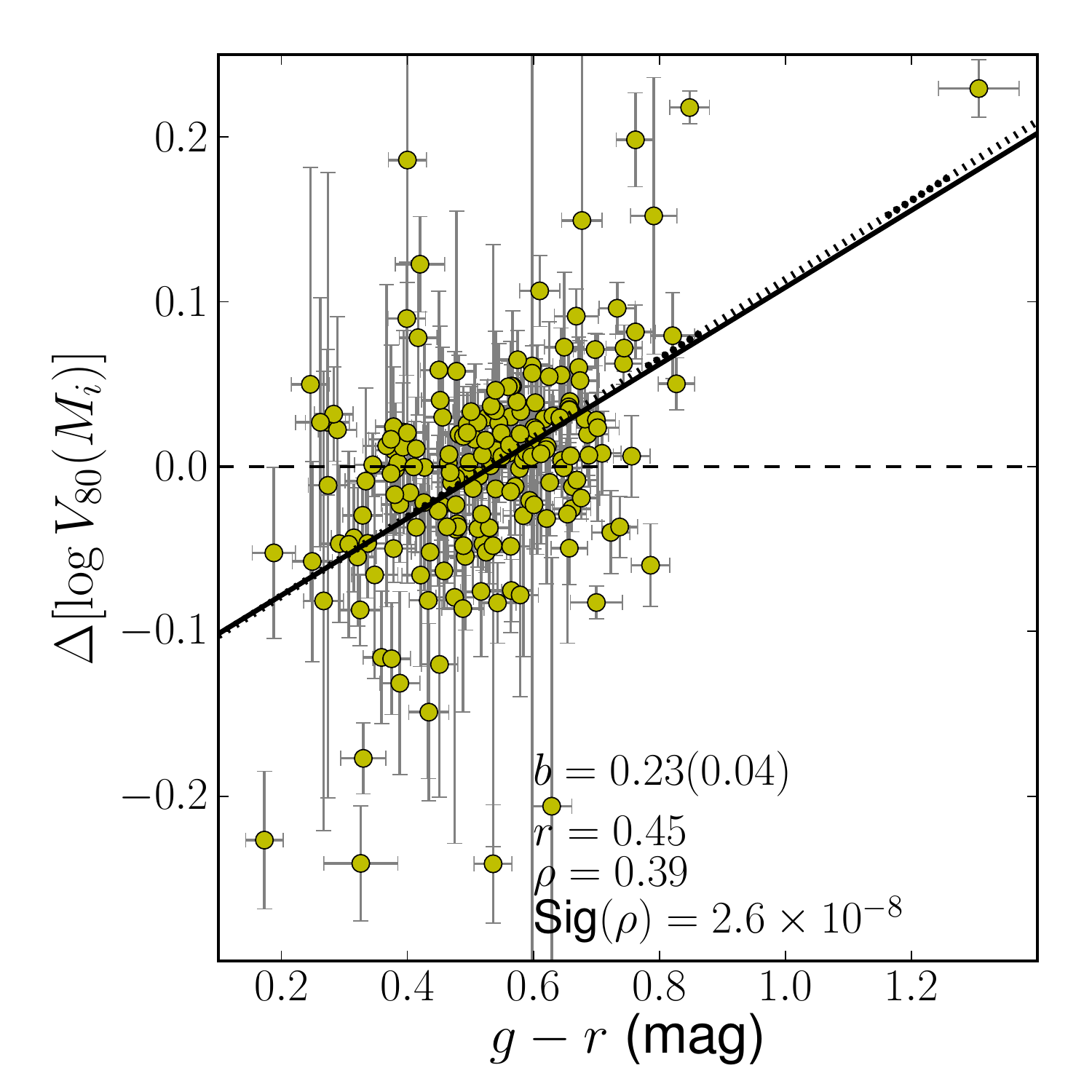} &
\includegraphics[width=3in]{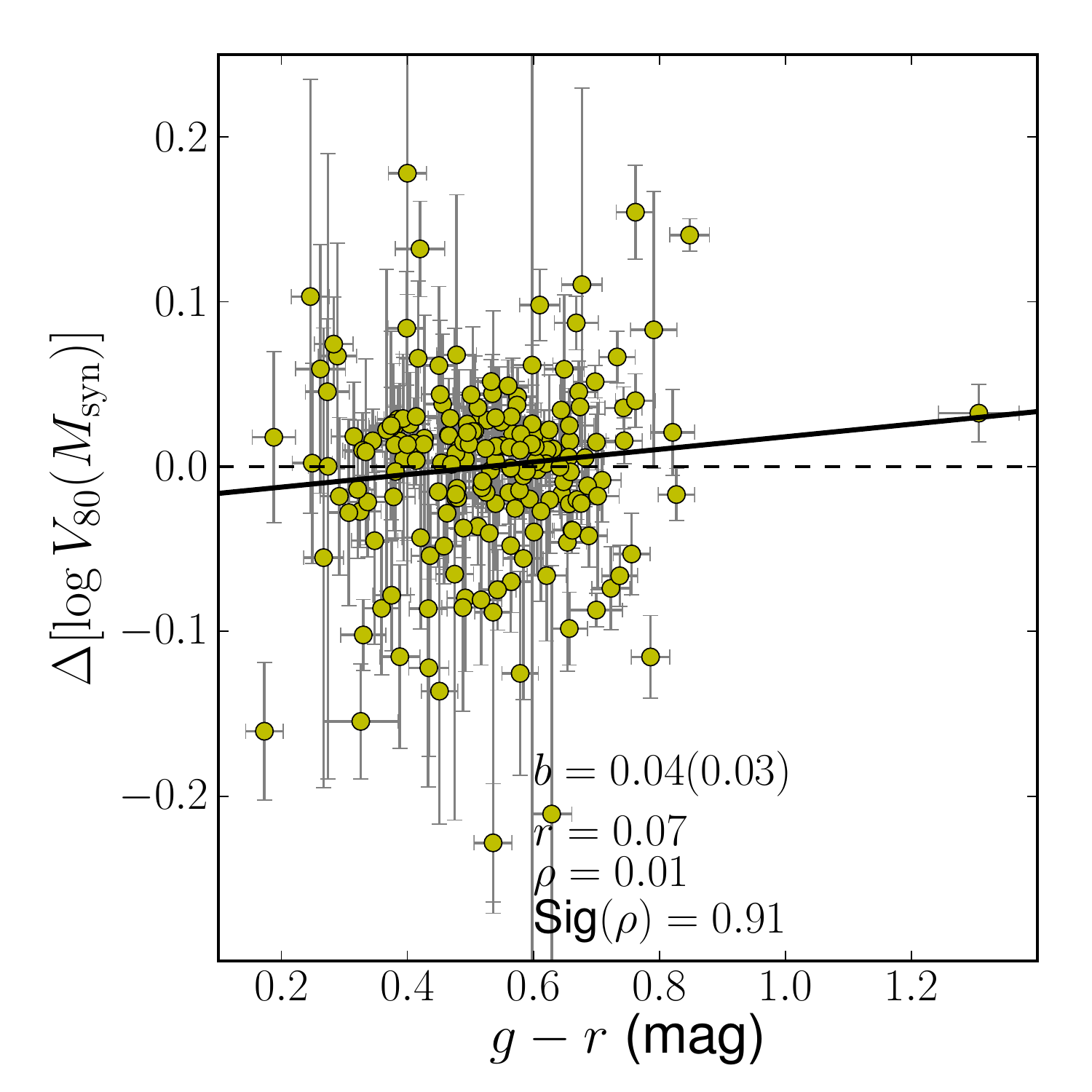}
\end{tabular}
\eec
\caption{Correlation between $g-r$ colours and $\log V_{80}$ residuals from the $M_i$ and $\msyn$ ITFRs (left and right panels, respectively). Filled circles are shown with 1-$\sigma$ error bars for the 189 galaxies in the child disk sample. Solid lines show the best-fit linear relations; horizontal dashed lines show the zero level. The dotted line in the left panel has a slope of 0.24, the predicted slope from the Bell et~al. (2003) scaling relation $M_*/L \propto (g-r)^{0.68}$. Labels show the best-fit slope $b$ and its 1-$\sigma$ uncertainty, the Pearson linear correlation coefficient $r$, the Spearman rank correlation $\rho$ and the two-tailed significance of its deviation from zero.}
\label{fig:resfits_color}
\end{figure*}

Figure~\ref{fig:resfits_color} shows the correlation between $g-r$ colour and $\Delta(\log V_{80})$ for 189 galaxies in the child disk sample. Left and right panels show residuals from the best-fit $M_i$ and $\msyn$ ITFRs, respectively. Best-fit linear relations are shown as solid lines in both panels. Labels show the best-fit slope $b$ and its 1-$\sigma$ uncertainty, the Pearson linear correlation coefficient $r$, the Spearman rank correlation $\rho$ and the two-tailed significance of its deviation from zero.

Clearly, we find a significant positive correlation between galaxy colours and $M_i$ ITFR residuals. The best-fit slope is $0.23\pm 0.04$, and the Spearman rank correlation coefficient $\rho=0.39$, with very high significance. On the other hand, we find no significant correlation for the $\msyn$ ITFR residuals. The best-fit slope $0.04 \pm 0.03$ is consistent with zero, and $\rho=0.01$. Indeed, the addition of colour information in $\msyn$ removes essentially all of the residual correlation. 

Notably, the reddest galaxy in the sample SDSSJ203523$-$06 is an extreme outlier from the $M_i$ TFR (with $g-r=1.31 \pm 0.42$ and $\Delta[\log V_{80}(M_i)]=0.22$), but lies within $2\tildesig$ of the $M_*$ and $\msyn$ ITFRs (with $\log V_{80}$ residuals $0.03$ and $0.05$, respectively). This edge-on galaxy is an extreme case, with a prominent dust lane, disk axis ratio $\qd=0.064$ (smallest in the sample) and $M_*/L_i = 9.7$ (largest in the sample; for comparison, the second largest value is 3.66, and the mean is 1.86), so it is gratifying to find that the color corrections seem to have worked in this case.

Now, we show that the observed correlation between $g-r$ colours and $M_i$ ITFR velocity residuals can be attributed to the variation in $M_*/L$ with colour. Redder galaxies have larger $M_*/L$, so they tend to be fainter than average for a given rotation velocity. Assuming the Bell et.~al. (2003) relation, $M_*/L_i \propto (g-r)^{0.86}$, we can predict the slope of the residual correlation to be
\beqa \nonumber
\frac{\Delta[\log V_{80}(M_i)]}{\Delta (g-r)} &\simeq& \frac{\Delta[\log V_{80}(M_*)]}{\Delta(\log M_*)} \times \frac{\Delta[\log (M_*/L_i)]}{\Delta (g-r)} \\ \nonumber
 &=& (-2.5)(-0.111 \pm 0.004)(0.86) \\ \nonumber
 &=& 0.24 \pm 0.09,
\eeqa
where we have used the best-fit slope of the $M_*$ ITFR from Table~\ref{tab:itfr_masses}. This predicted slope is very close to the best-fit value (dotted vs. solid lines in the left panel of Fig.~\ref{fig:resfits_color}).\footnote{If one instead assumes a pure gravitational disk model, in which $\vrot^2 \propto M_*$ at fixed scale length (as was done in P07), one would predict a different slope, equal to $0.86/2=0.43$, which is inconsistent with the observations.} 

Putting these results together, we form a fully consistent picture. The addition of colour information to $M_i$ via $M_*$ and $\msyn$ corrects for the variation in $M_*/L_i$ with colour. Doing so successfully removes most of the residual correlation with galaxy colour, and yields to significantly tighter ITFRs.

\subsection{Correlations with disk size}
\label{subsec:correl_size}

Recent observational studies of the TFR have consistently found a lack of correlation between TFR residuals and disk galaxy scale lengths \citep[P07;][]{2007ApJ...671..203C}. \citet{1999ApJ...513..561C} noted that this lack of correlation argues that the disk mass does not contribute most of the disk rotation velocity at the optical radius. However, Dutton et~al. (2007) noted that there can be other interpretations as well. Meanwhile, \citet{2007ApJ...671.1115G} emphasized that the observed lack of residual correlations plays a key role in their conclusions about the baryon fractions of disk galaxies. For a pure self-gravitating disk model, $\vrot^2 \propto \rd^{-1}$ at fixed $M_*$, so a strong negative correlation is expected with $\partial\log\vrot/\partial\log\rd=-0.5$. On the other hand, a strong positive correlation is expected for a pure NFW dark matter halo model, with an expected slope of $+0.5$ in the inner regions.

\begin{figure}
\bec
\includegraphics[width=3in]{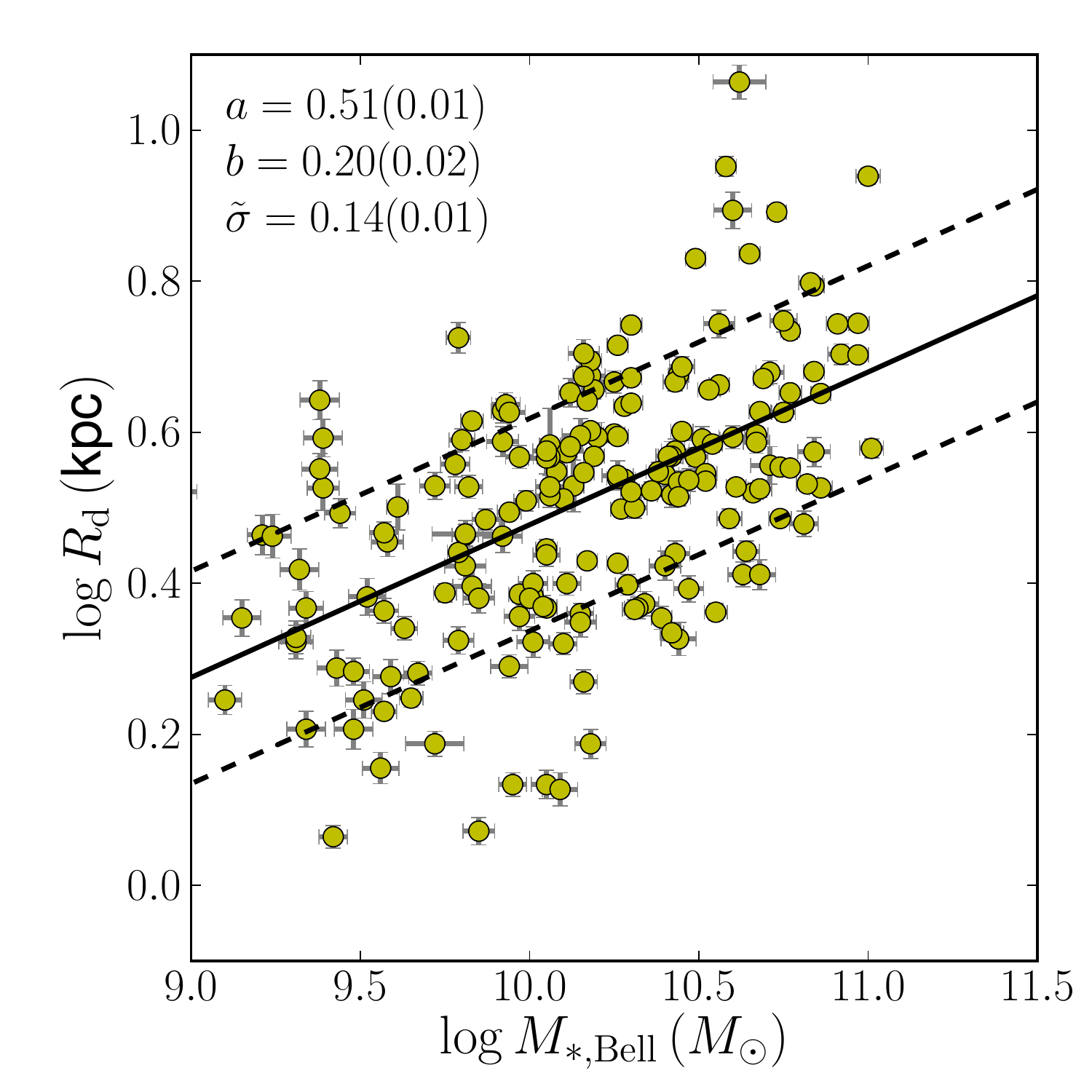}
\eec
\caption{Relation between stellar masses $\mbell$ and disk scale lengths $\rd$ for 189 galaxies in the child disk sample (filled circles with 1-$\sigma$ error bars). The best-fit relation $\log \rd = a + b (\log M_* - 10.17)$ is shown by the solid line, and the dashed lines are displaced from the mean relation by $\pm 1\tildesig$, the best-fit Gaussian intrinsic scatter. Best-fit parameters are listed in the upper left corner, together with their 1-$\sigma$ uncertainties.}
\label{fig:meanfits_rdphys}
\end{figure}

We study correlations between velocity residuals from the $M_*$ ITFR and disk size offsets from the mean size-mass relation, $\Delta (\log \rd) = \log (\rd/\bar{R}_{\rm d}(M_*))$. Figure~\ref{fig:meanfits_rdphys} shows the $\log M_*$--$\log \rd$ relation for 189 galaxies in the child disk sample (filled circles), and the best-fit mean relation $\bar{R}_{\rm d}(M_*)$ (solid line). Using a procedure similar to that used to fit the TFRs (c.f. \S\ref{subsec:tfr_fit}), we find  
\beqa \label{eq:meanfit_rdphys} \nonumber
\log \left(\frac{\bar{R}_{\rm d}}{\rm kpc}\right) &=& (0.51 \pm 0.010) 
	+ (0.20 \pm 0.02) \times \\ 
	&& \mbox{          } \left[\log \left(\frac{M_*}{M_\odot}\right) - 10.17\right],
\eeqa
with a best-fit Gaussian intrinsic scatter in $\log\rd$ of $\tildesig=0.14 \pm 0.01$ dex. 

\begin{figure}
\bec
\includegraphics[width=3in]{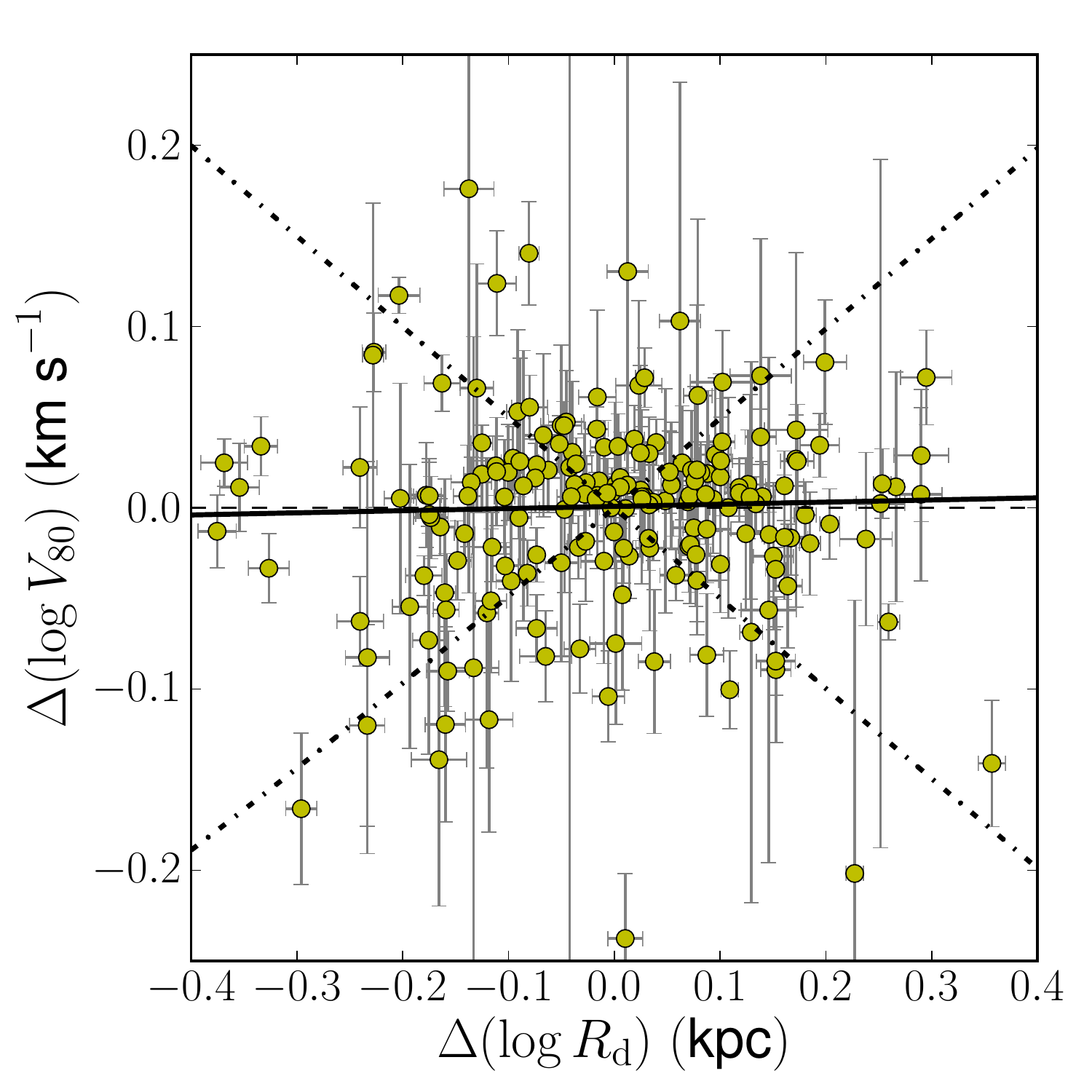} 
\caption{Correlation between velocity residuals from the $M_*$ ITFR, $\Delta(\log V_{80})$, and disk size offsets $\Delta(\log \rd)$, defined relative to the mean relation $\log \bar{R}_{\rm d}(M_*)$ (given by Eq.~\ref{eq:meanfit_rdphys}). The best-fit linear relation has a slope consistent with zero (solid line). Predicted trends for a pure self-gravitating disk model ($\mbox{slope}=-0.5$) and a pure NFW DM halo model ($\mbox{slope}=+0.5$) are also shown (dot-dashed lines).}
\label{fig:resfits_rdphys_all}
\eec
\end{figure}

\begin{figure}
\bec
\includegraphics[width=3in]{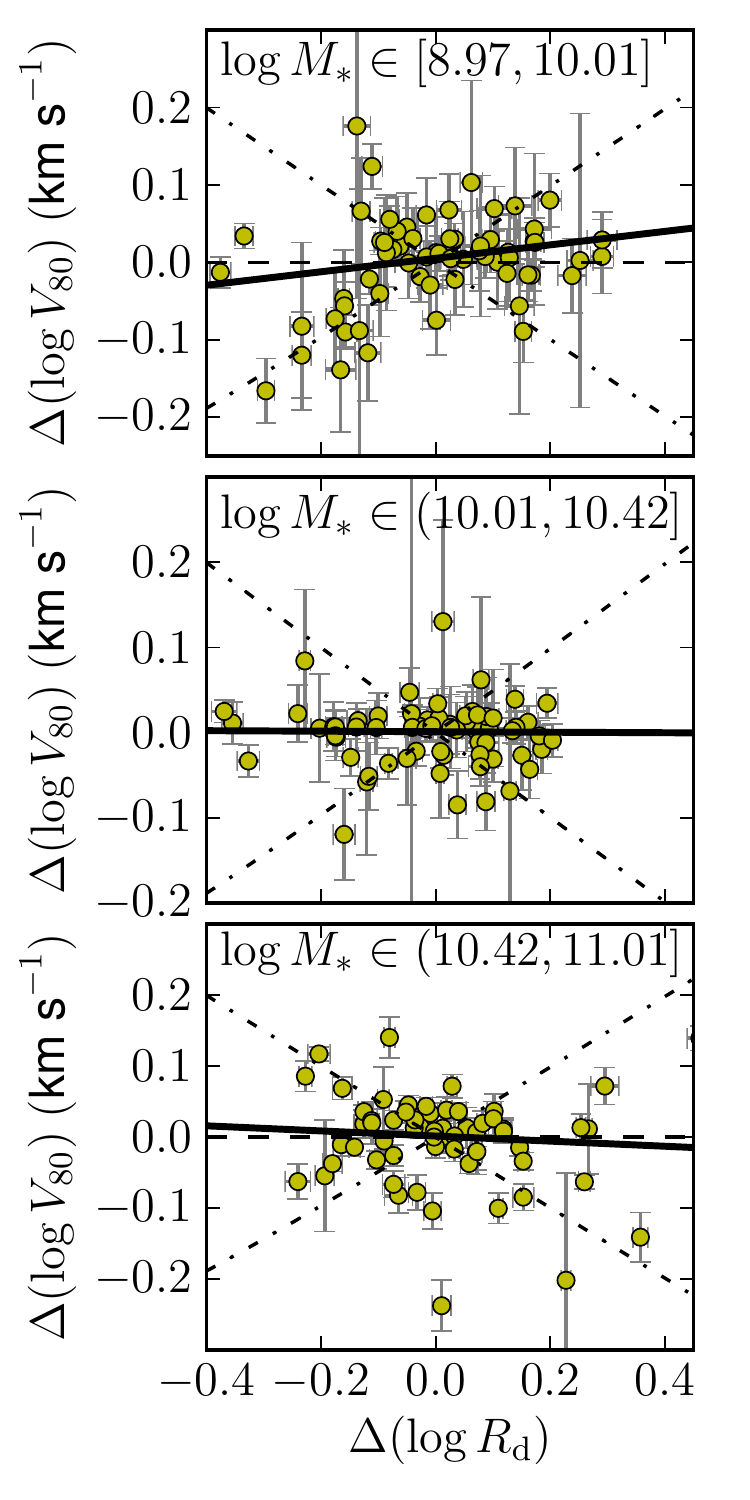}
\eec
\caption{Correlations between velocity residuals from the $M_*$ ITFR, $\Delta(\log V_{80})$, and disk size offsets $\Delta(\log \rd)$ for the child disk sample divided into three stellar mass bins, as labelled (with increasing $M_*$ from top to bottom). Solid lines show the best-fit linear relations, dashed horizontal lines show the zero level, and dot-dashed lines show the predicted trends for a pure self-gravitating disk model ($\mbox{slope}=-0.5$) and a pure NFW DM halo model ($\mbox{slope}=+0.5$).} 
\label{fig:resfits_rdphys_bins}
\end{figure}

Figure~\ref{fig:resfits_rdphys_all} shows the correlation between velocity residuals $\Delta[\log V_{80}(M_*)]$ and disk size offsets $\Delta (\log \rd)$ for the 189 galaxies in the child disk sample. Confirming the results of prior studies, we find no evidence for a correlation: $r=0.05$ and $\rho=0.04$. The best-fit linear relation has a slope consistent with zero, $b=0.012\pm 0.034$ (solid line).

Now, we go one step further and repeat the analysis for three bins in stellar mass (with 64, 64, and 61 galaxies in the low, intermediate, and high stellar mass bins, respectively). Figure~\ref{fig:resfits_rdphys_bins} shows the results for each stellar mass bin (as labeled; the legend here is similar to that in Fig.~\ref{fig:resfits_rdphys_all}). Although the best-fit slopes are not close to either of the predictions from the pure disk and pure DM models ($-0.5$ and $+0.5$, respectively; dot-dashed lines), we find a decreasing trend in the best-fit slope (changing sign from slightly positive to slightly negative) with increasing stellar mass: $b=0.087\pm 0.055$, $0.00\pm 0.03$, $-0.04\pm 0.08$, for the low, intermediate, and high stellar mass bins, respectively. The correlations are weak, but reflect the same trend with increasing stellar mass: $\rho=0.25$, $-0.08$, and $-0.15$, with corresponding ${\rm Sig}(\rho)=0.05$, $0.5$, and $0.2$, for the three bins, respectively. 

The observed trend indicates that the stellar mass (or baryon) fraction within the optical region of disk galaxies increases systematically with stellar mass, over the range of stellar masses we consider. In \S\ref{sec:mass_ratios}, we explicitly calculate stellar mass fractions and confirm the trend with stellar mass suggested by these residual correlations. In \S\ref{subsec:disc_relative}, we discuss the interpretation of these results.

\section{Dynamical-to-stellar mass ratios}
\label{sec:mass_ratios}

We calculate dynamical-to-stellar mass ratios within the optical radius $R_{80}$ for the 189 galaxies in the child disk sample, denoted by $\mdynratioopt \equiv (\mdynratio)(R_{80}) = \mdyn(R_{80})/M_*(R_{80}) \equiv \mdynopt/\mstropt$. We adopt an empirical definition that depends straightforwardly on directly-observed quantities 
\beq  \label{eq:mdynratio} 
\left(\frac{\mdyn}{M_*}\right)_{\rm opt} 
= \frac{V_{80}^2 R_{80}/G}{0.8 \mbell}-\left[{\cal K}(D/T)_{80}-1\right],
\eeq 
where  $G=4.3012 \times 10^{-6}\,\mbox{kpc} {(\kms)}^{2} M_\odot^{-1}$ is the gravitational constant, ${\cal K}=1.34$ is a geometrical factor that corrects for the flattened potential of the disk (assuming an exponential disk+NFW DM halo model; see Appendix for the derivation), and $(D/T)_{80}$ is the disk-to-total mass ratio within $R_{80}$, 
\beq \label{eq:dtr80}
(D/T)_{80} = \frac{1 - e^{-R_{80}/\rd}(1+R_{80}/\rd)}{0.8} \times {(D/T)}.
\eeq
For simplicity, we have assumed that the average stellar mass-to-light ratio within $R_{80}$ is equal to its global value, so that $\mstropt/M_* \approx 0.8$ by definition of $R_{80}$ (actually, the value of $M_*/L$ varies with radius, as indicated by color gradients along the disk). Note that $V_{80}^2 R_{80}/G$ is the dynamical mass within $R_{80}$ for a spherical mass distribution with a circular velocity equal to $V_{80}$, and that we have assumed a fixed Kroupa IMF in the definition of $\mbell$.

We also calculate dynamical-to-baryonic mass ratios within $R_{80}$ 
\beq \label{eq:mdynratiobar}
\left(\frac{\mdyn}{M_{\rm bar}}\right)_{\rm opt} = \left(\frac{\mdyn}{M_*}\right)_{\rm opt} \left(\frac{M_*}{M_{\rm bar}}\right),
\eeq
where $\mbar$ are baryonic mass estimates based on $u-r$ color-based gas-to-stellar mass ratios from \citet{2004ApJ...611L..89K} (c.f.~\S\ref{subsec:phot_mbar}). For simplicity, we have assumed that the average gas-to-stellar mass ratio within $R_{80}$ is equal to its global value. Actually, the radial extent of HI is usually larger than the optical radius \citep{1997A&A...324..877B}.

\begin{figure}
\bec
\includegraphics[width=3in]{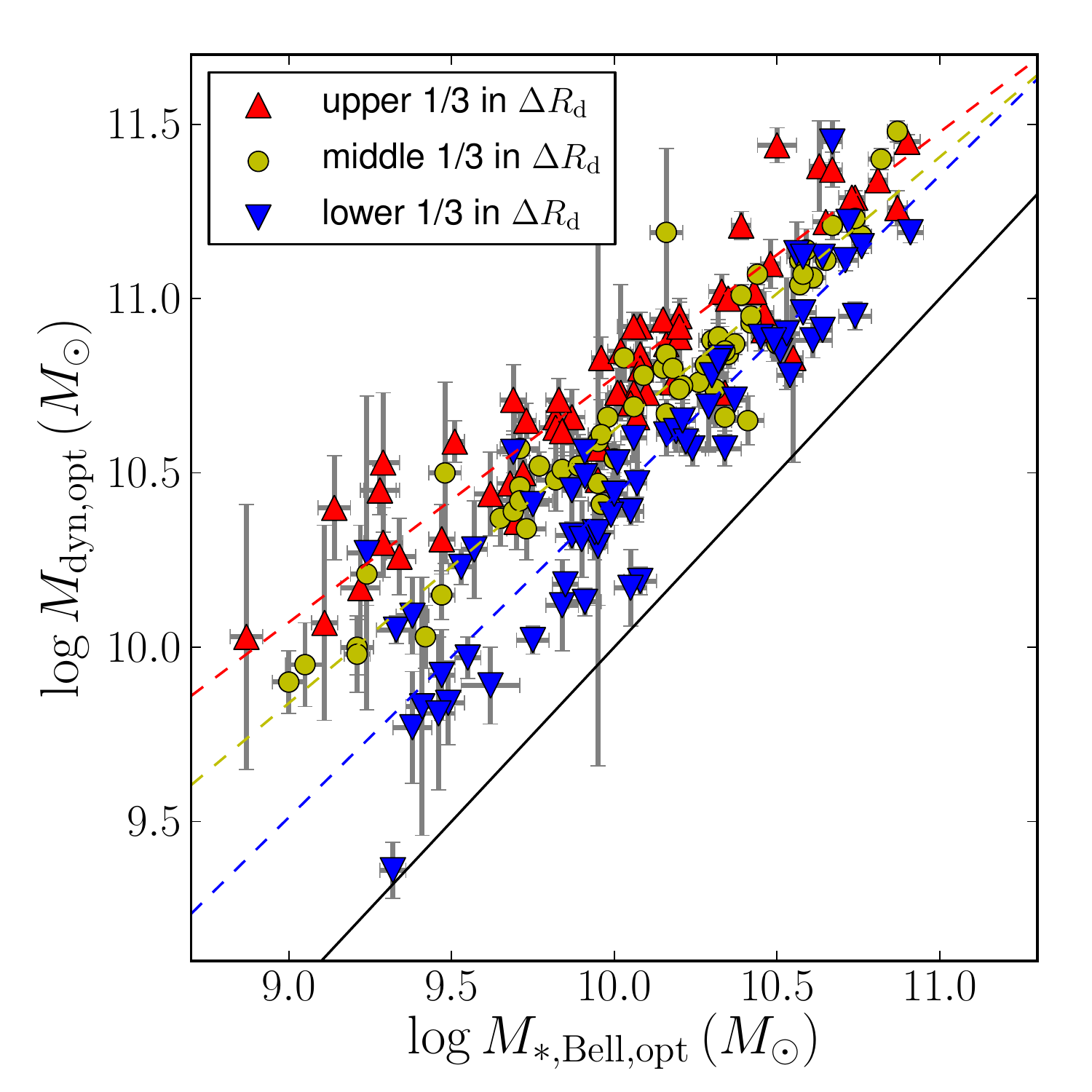}
\includegraphics[width=3in]{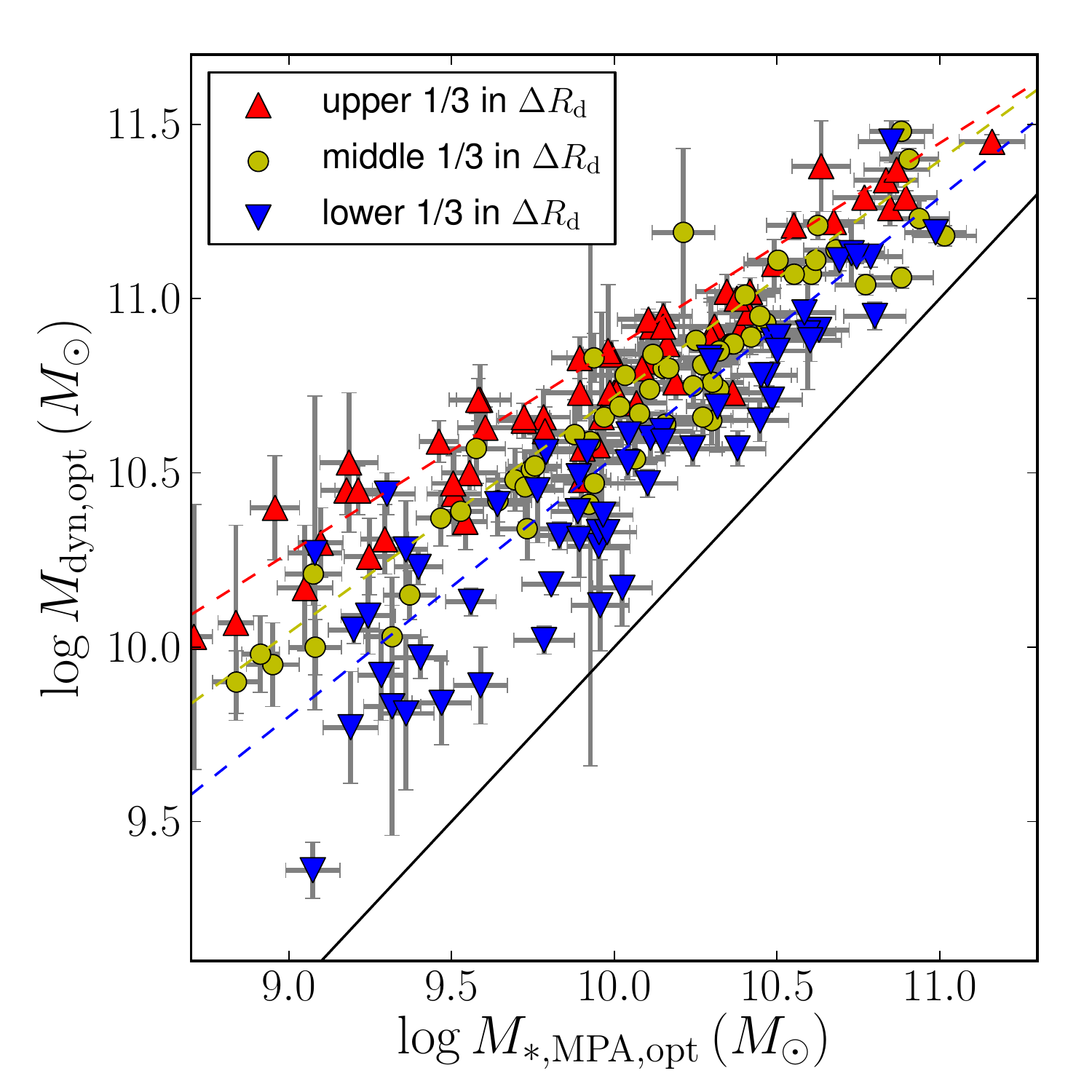}
\eec
\caption{Relation between dynamical mass $\mdynopt$ and stellar mass $\mstropt$ within the optical radius $R_{80}$ for 189 galaxies in the child disk sample. Top and bottom panels show different stellar mass estimates $\mbell$ and $\mmpa$, respectively. Different symbols show galaxies in three bins in disk size offsets $\Delta\rd$ (red triangles, yellow circles, blue inverted triangles, for the upper, middle, and lower bins, respectively), with corresponding best-fit relations shown by colored dashed lines. Solid lines shows $\mdynopt=\mstropt$ .} 
\label{fig:mdyn_mstr}
\end{figure}

Figure~\ref{fig:mdyn_mstr} shows the relation between $\mstropt$ and $\mdynopt$ for the 189 galaxies in the child disk sample, using stellar mass estimates $\mbell$ and $\mmpa$ (top and bottom panels, respectively). The scatter in this relation is clearly being driven by disk size (red triangles, yellow circles, blue inverted triangles show the upper, middle, and lower bins in disk size offset $\Delta\rd$, respectively). Best-fit linear relations for each disk size bin are shown by colored dashed lines and best-fit parameters are listed in Table~\ref{tab:mdyn}. The slopes of the relations are less than unity for both stellar mass estimates: $\mdynratioopt$ decreases with stellar mass. In subsequent figures, we will show only results for $\mbell$. 

\begin{figure}
\bec
\includegraphics[width=3in]{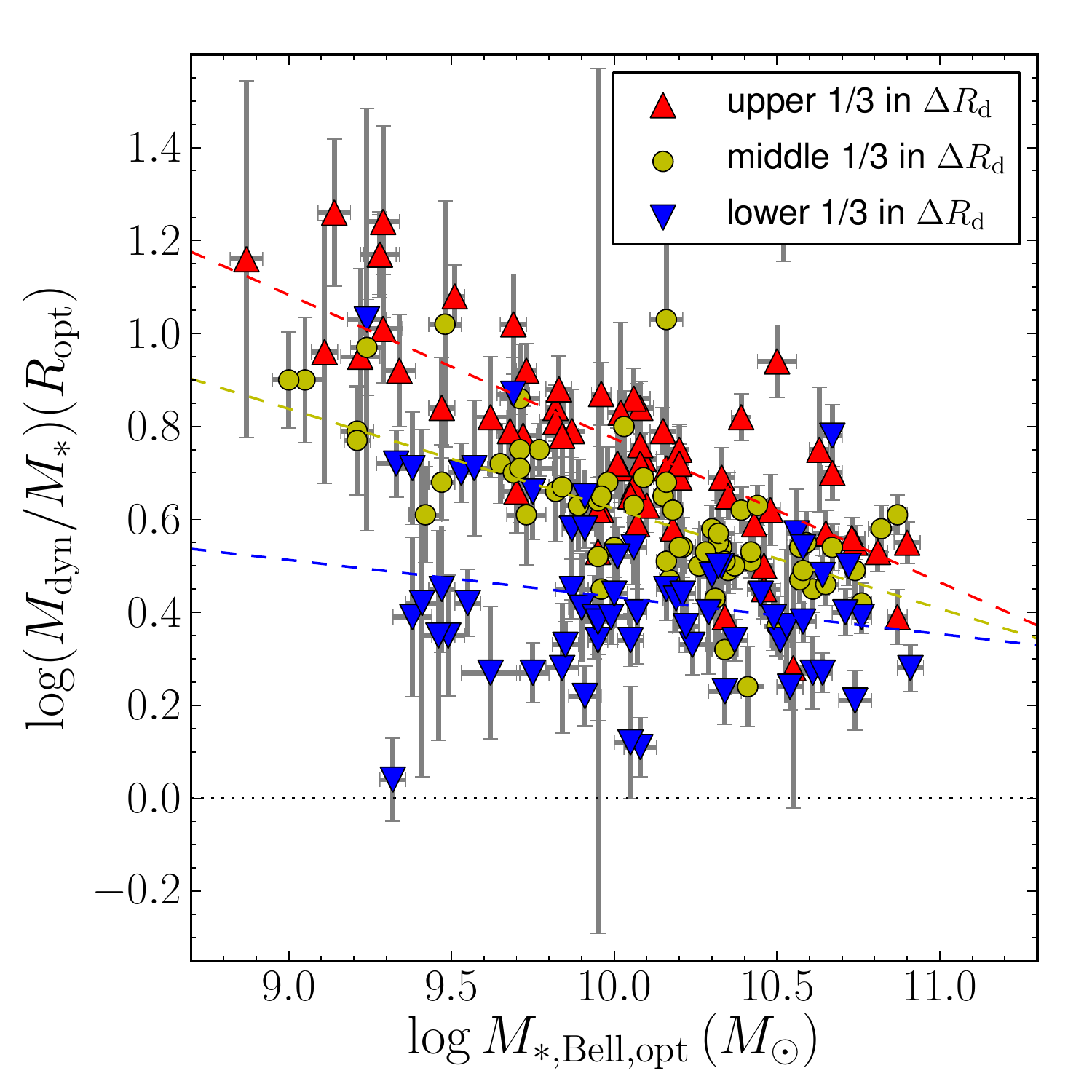}
\includegraphics[width=3in]{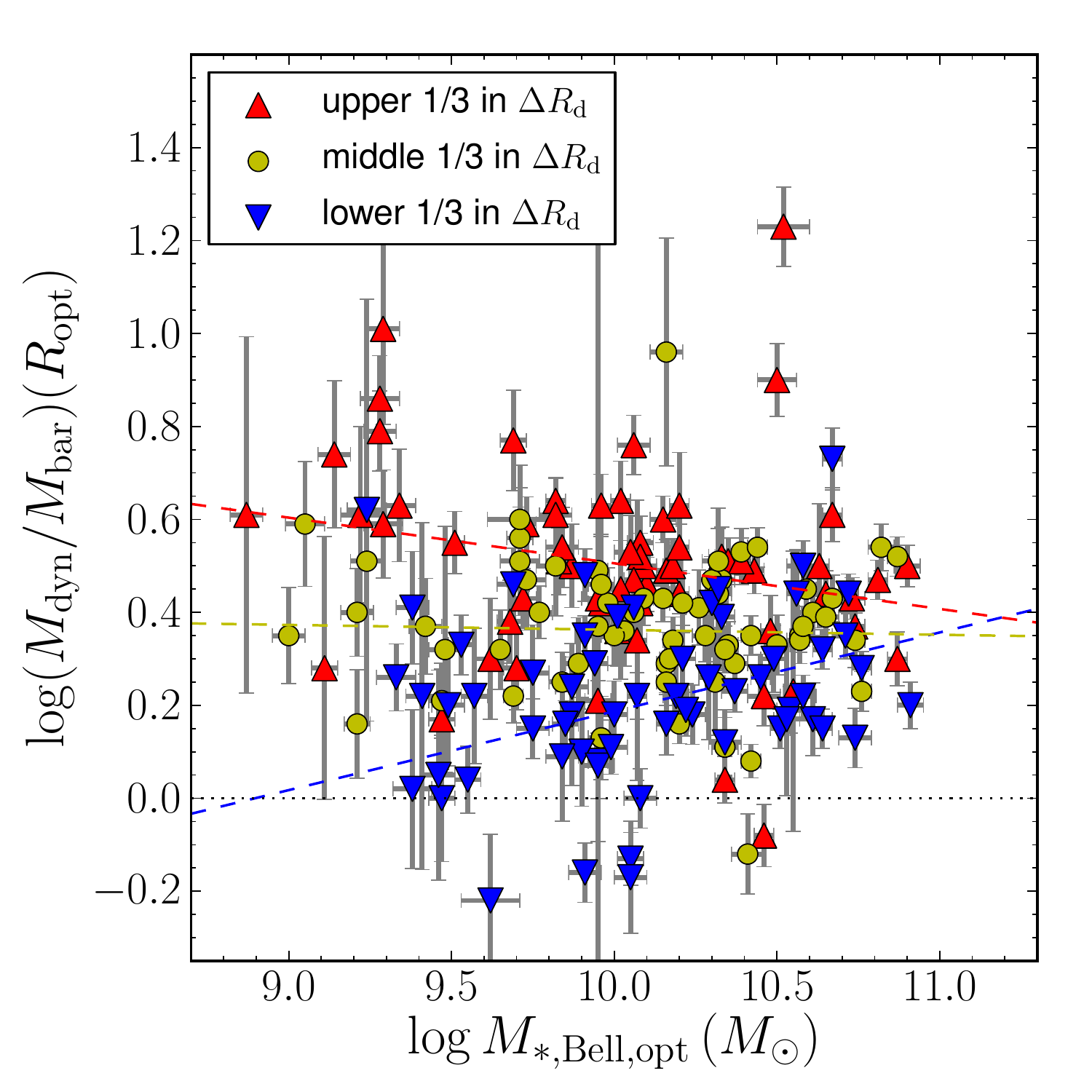}
\eec
\caption{Relation between stellar masses within $R_{80}$, $\mstropt$, and dynamical-to-stellar and dynamical-to-baryonic mass ratios within $R_{80}$ (top and bottom panels, respectively). Different symbols correspond to bins in disk size offset $\Delta\rd$ (red triangles, yellow circles, blue inverted triangles, for the upper, middle, and lower bins, respectively).}
\label{fig:mdynratio_mstr}
\end{figure}

\begin{figure}
\bec
\includegraphics[width=3in]{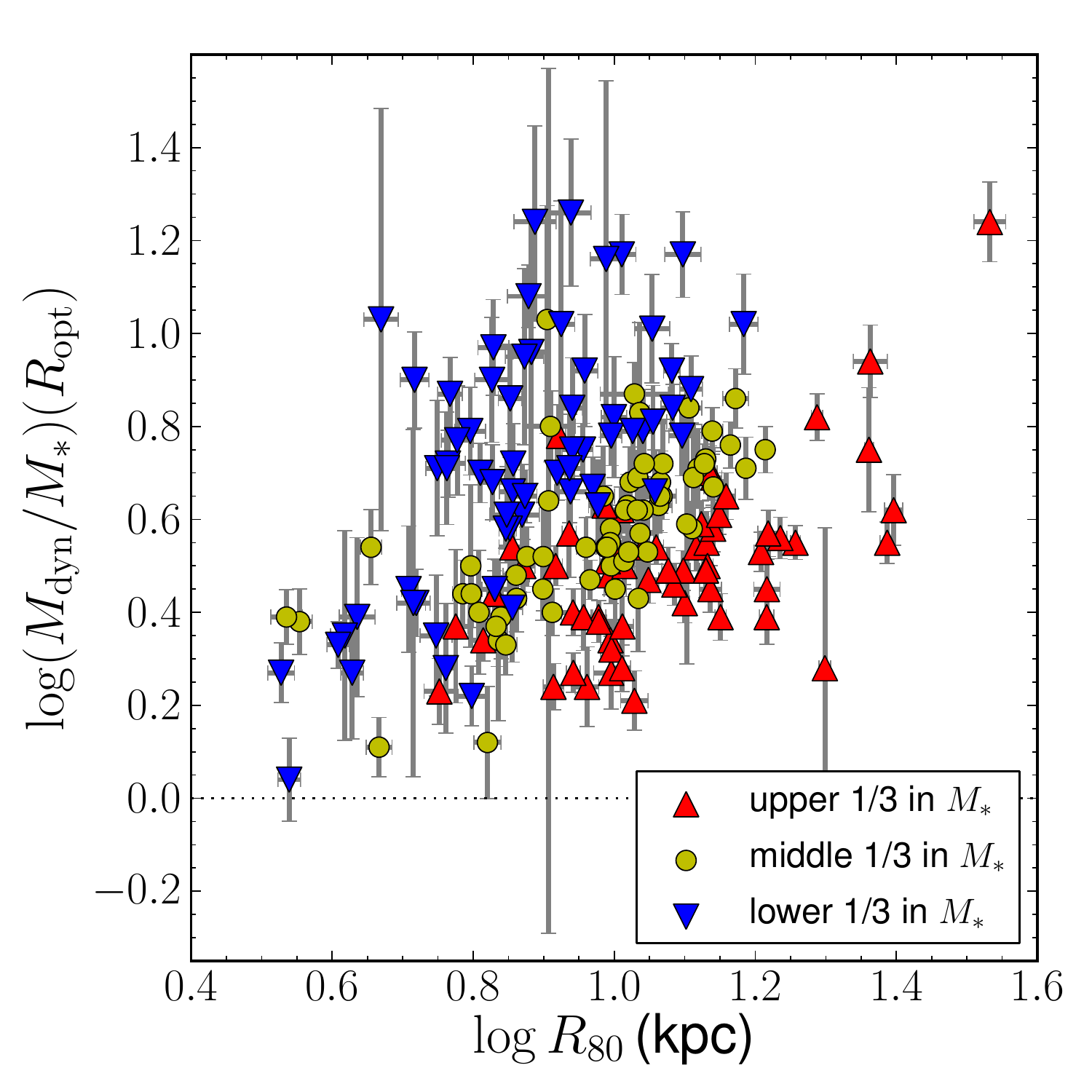}
\includegraphics[width=3in]{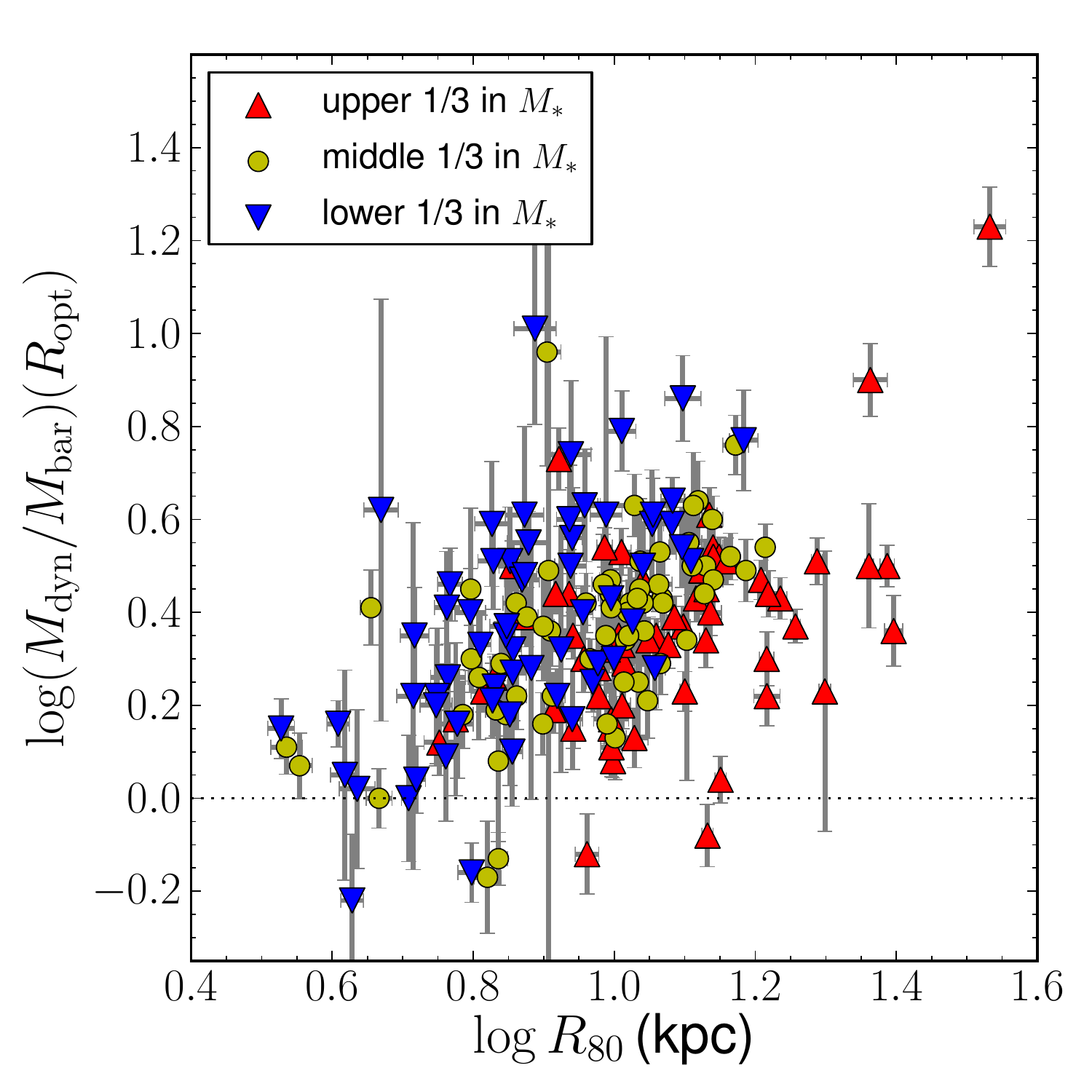}
\eec
\caption{Relation between $R_{80}$ and dynamical-to-stellar and dynamical-to-baryonic mass ratios within $R_{80}$ (top and bottom panels, respectively). Different symbols correspond to bins in stellar mass $\mbell$ (red triangles, yellow circles, blue inverted triangles, for the upper, middle, and lower bins, respectively).} 
\label{fig:mdynratio_ropt}
\end{figure}

Figure~\ref{fig:mdynratio_mstr} shows the relation between $\mdynratioopt$ (top panel) and $\mdynratiobaropt$ (bottom panel) and stellar masses $\mstropt$. Here, different symbols show bins in disk size offsets $\Delta\rd$. Figure~\ref{fig:mdynratio_ropt} shows the corresponding relations with $R_{80}$. Here, different symbols show bins in stellar mass $\mbell$. In both figures, red triangles, yellow circles, blue inverted triangles show the upper, middle, and lower bins, respectively. Best-fit relations are shown by dashed lines of corresponding colors in Fig.~\ref{fig:mdynratio_mstr} and the best-fit parameters are listed in Table~\ref{tab:mdyn}. 

We begin by discussing the trends in $\mdynratioopt$ (top panels of Figs.~\ref{fig:mdynratio_mstr} and \ref{fig:mdynratio_ropt}). Then, we discuss the trends in $\mdynratiobaropt$ (bottom panels of the same Figures). Our interpretation of these trends will be discussed in \S\ref{subsec:disc_relative}. 

We confirm that at a fixed stellar mass, larger (smaller) disks tend to have higher (lower) $\mdynratioopt$ \citep{2003A&A...412..633Z,2005ApJ...633..844P}. In addition, we find that while most disks show a clear trend in $\mdynratioopt$ with stellar mass, the smallest disks fall off that main relation and tend to have low $\mdynratioopt$ regardless of their stellar mass (blue inverted triangles and dotted line in the top panel of Fig.~\ref{fig:mdynratio_mstr} and leftmost points in top panel of Fig.~\ref{fig:mdynratio_ropt}).\footnote{This result is more apparent here than in  Fig.~6 of \citet{2005ApJ...633..844P} because of the addition of newly-observed galaxies with low stellar masses and small disks in our sample.} 

For stellar mass estimates $\mbell$, with a fixed Kroupa IMF, intermediate and large disks have $\mdynratioopt$ decreasing from $\approx 10$ to $3$ as $\mbell$ increases from $10^9$ to $10^{11} M_\odot$; while the smallest disks have $\mdynratioopt \approx 2.7$, regardless of stellar mass.\footnote{Since $\mmpa$ is systematically lower than $\mbell$, except at the highest mass end (c.f. Fig.~\ref{fig:mstr_comp} in \S\ref{subsubsec:comp_mstr}), the inferred $\mdynratioopt$ would be correspondingly higher.} 
At the high $M_*$ end, the ratios converge to the same low value, regardless of disk size. In other words, if we ignore the contribution of the gas mass, maximal disks (i.e., with $\mdynopt=\mstropt$) require a non-uniform increase in the IMF normalization varying from $\sim 0.4$ dex (for the highest $M_*$ and lowest $\rd$ galaxies) to $\sim 1$ dex (for the lowest $M_*$ galaxies with intermediate-to-high $\rd$). 

Now, looking at $\mdynratiobaropt$, we find the same dependence on disk size, at a fixed $\mstropt$, seen in the corresponding relation for $\mdynratioopt$ (bottom and top panels of Fig.~\ref{fig:mdynratio_mstr}). The relation between $\mdynratiobaropt$ and $R_{80}$ is actually tighter than the corresponding relation with $\mdynratioopt$ (bottom and top panel of Fig.~\ref{fig:mdynratio_ropt}), because the trend with stellar mass (at fixed $R_{80}$) seen in the latter is not found in the former.

We find that $\mdynratiobaropt$ is roughly constant with stellar mass, at a fixed $\Delta\rd$, unlike $\mdynratioopt$. This is a direct consequence of the sign and size of the trend of increasing gas mass fraction as stellar mass decreases (see Fig.~\ref{fig:mgas_mstr_ratio}). For the full child disk sample, the mean value of $\mdynratiobaropt \approx 2.6$, which corresponds to a 44 per cent baryon contribution to the mass inside $R_{80}$. Taking these baryonic masses at face value (i.e., assuming that the gas and stars have the same spatial extent), maximal disks (i.e., with $\mdynopt=M_{\rm bar,opt}$) require a non-uniform increase in the IMF normalization varying from $\sim 0.4$ dex (for the highest $M_*$ galaxies) to $\sim 0.9$ dex (for the lowest $M_*$ galaxies); the amounts are larger for the low-mass galaxies because they have higher gas mass fractions and the photometrically-derived gas masses are unaffected by changes in the IMF.

The scatter in $\mdynratiobaropt$ at fixed $\mstropt$ is smallest at the highest $M_*$ end, as is the case for $\mdynratioopt$. 
The measured scatter in $\mdynratiobaropt$ is 0.23 dex for the full sample, and 0.22, 0.15, and 0.21 dex for the bottom, middle, and top bins in disk size offsets, respectively (Table~\ref{tab:mdyn}). 

\begin{figure}
\bec
\includegraphics[width=3in]{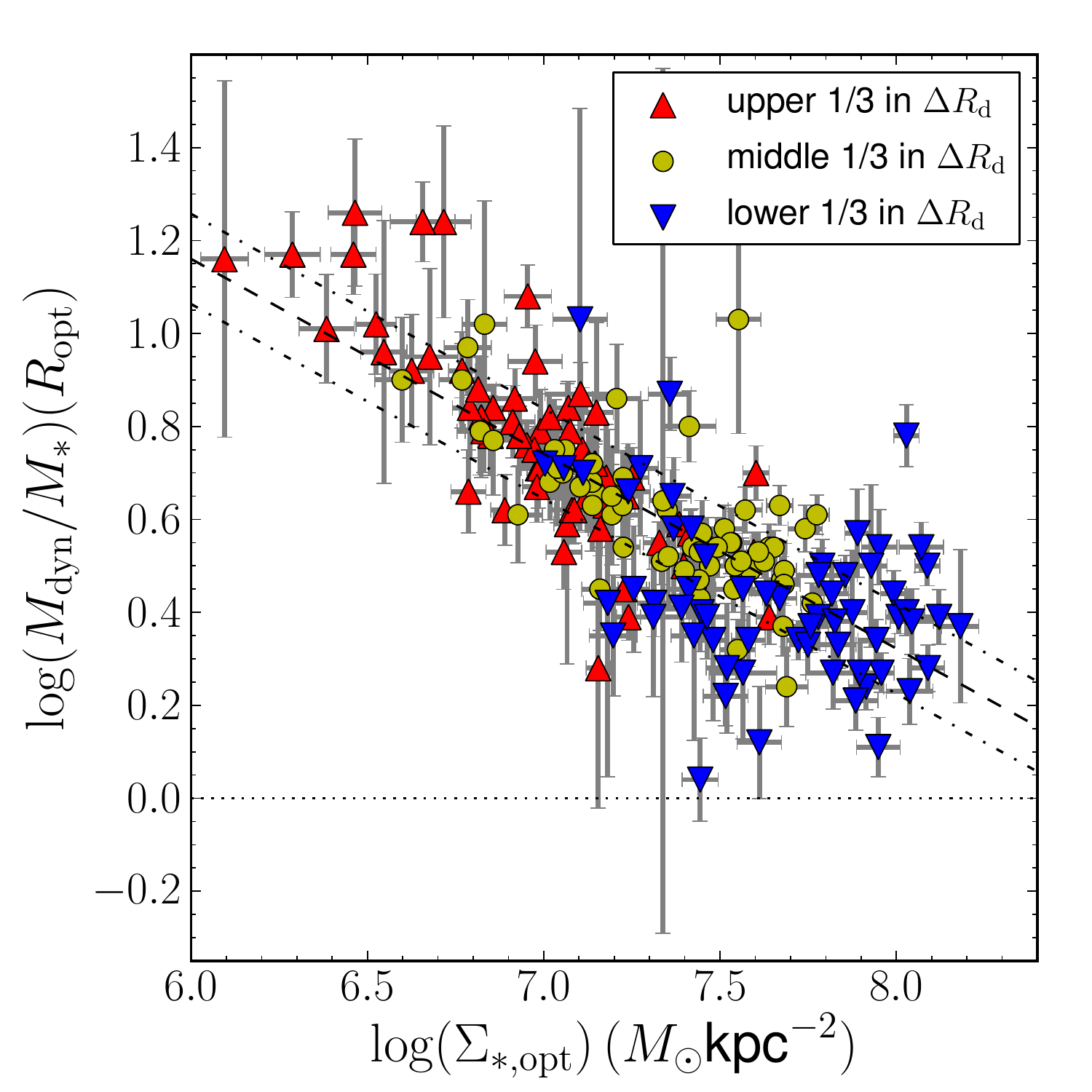}
\includegraphics[width=3in]{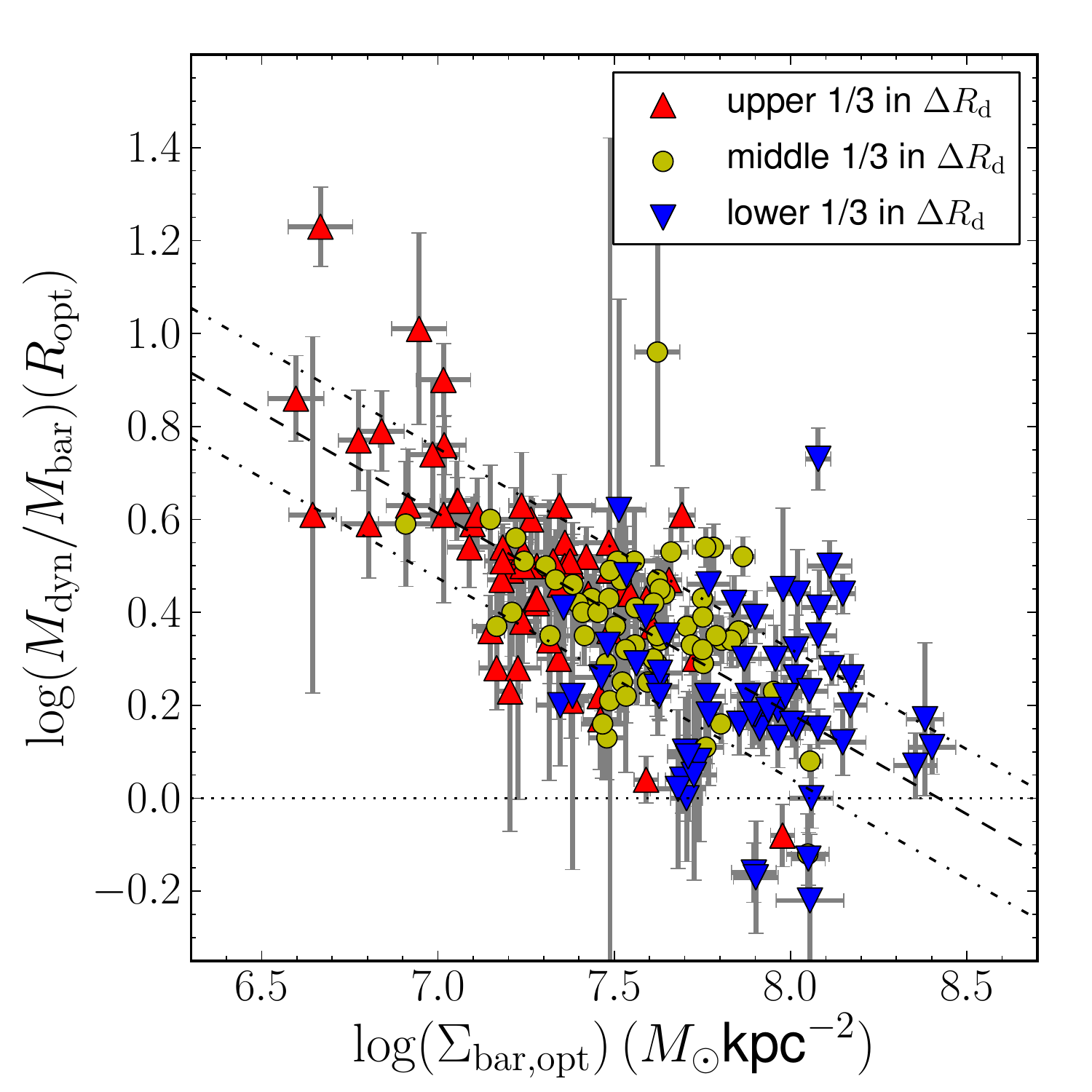}
\eec
\caption{Top panel: Relation between dynamical-to-stellar mass ratio $\mdynratioopt$ and stellar surface density within the optical radius $R_{80}$, $\sigmaopt$. Bottom panel: Relation between dynamical-to-baryonic mass ratio $\mdynratiobaropt$ and baryonic surface density within the optical radius $R_{80}$, $\sigmabar$. In both panels, red triangles, yellow circles, blue inverted triangles show upper, middle, and lower bins in disk size offset $\Delta \rd$, respectively. The dashed line shows the best-fit linear relation for the full sample and dot-dashed lines are displaced from this relation by $\pm\tildesig$ to show the amount of intrinsic scatter.} 
\label{fig:mdynratio_sigma}
\end{figure}

The correlation of both $\mdynratioopt$ and $\mdynratiobaropt$ with disk size at fixed stellar mass suggests that a combination of these two properties can lead to a tighter relation with $\mdynratio$. We confirm that the stellar surface density within $\ropt$, $\sigmaopt = \mstropt/(2\pi\ropt^2)$ forms a tighter relation with $\mdynratio$ than either $M_*$ or $\ropt$ alone \citep{2003A&A...412..633Z,2005ApJ...633..844P}. 


The top panel of Figure~\ref{fig:mdynratio_sigma} shows the relation between $\sigmaopt$ and $\mdynratioopt$ for the child disk sample. The best-fit relation is shown by the dashed line and best-fit parameters are listed in Table~\ref{tab:mdyn}. The bottom panel of Figure~\ref{fig:mdynratio_sigma} shows the relation between the baryonic surface density within $R_{80}$, $\sigmabar=\sigmaopt \times (M_{\rm bar}/M_*)$ and $\mdynratiobaropt$. As in Fig.~\ref{fig:mdynratio_mstr}, different symbols correspond to bins in disk size offsets $\Delta\rd$ (red triangles, yellow circles, blue inverted triangles show upper, middle, and lower bins, respectively). 

Comparing the upper panels of Figs.~\ref{fig:mdynratio_mstr} and \ref{fig:mdynratio_sigma}, we find that the smallest disks, which fall below the relation involving $M_*$ move to the right (toward higher $\sigmaopt$) by an amount that places it roughly on the $\sigmaopt$--$\mdynratioopt$ relation defined by the larger disks. 

The scatter in the  $\sigmabar$--$\mdynratiobaropt$ is comparable to the scatter in the $\sigmaopt$--$\mdynratioopt$ relation--- equal to 0.18 and 0.15 dex, respectively. The trends in $\mdynratiobaropt$ with $\mstropt$ and $R_{80}$ (lower panels of Figs.~\ref{fig:mdynratio_mstr} and \ref{fig:mdynratio_ropt}) indicate that the tightness of its relation with $\sigmabar$ is almost entirely driven by its dependence on disk size, rather than stellar mass. This observation has important consequences for our interpretation of these trends in \S\ref{subsec:disc_relative}.

Finally, to look for the tightest possible relation for $\mdynratioopt$, we explore other combinations of the form $\mstropt/\ropt^{\kappa}$, with the exponent $\kappa$ treated as an additional free parameter. Maximum likelihood fits yield $\kappa=1.7 \pm 0.3$, consistent with the definition of the stellar surface density (with $\kappa=2$). Therefore, we conclude that the stellar surface density is, indeed, a near-optimal photometric tracer of $\mdynratioopt$. 

\begin{table*}
\begin{center}
\caption{Fits to the $\mdyn-\mstropt$ relation and $\mdynratio$ relations with various physical parameters.}
\begin{tabular}{lllrrrrr}
\hline 
 \multicolumn{1}{c}{$y$} &
 \multicolumn{1}{c}{$x$} &
 \multicolumn{1}{c}{Bin} &
 \multicolumn{1}{c}{$x_{\rm p}$} &
 \multicolumn{1}{c}{$a$} &
 \multicolumn{1}{c}{$b$} &
 \multicolumn{1}{c}{$\tilde{\sigma}$} &
 \multicolumn{1}{c}{$\sigmeas$} \\
 \multicolumn{1}{c}{(1)} & 
 \multicolumn{1}{c}{(2)} & 
\multicolumn{1}{c}{(3)} & 
\multicolumn{1}{c}{(4)} & 
\multicolumn{1}{c}{(5)} &
\multicolumn{1}{c}{(6)} &
\multicolumn{1}{c}{(7)} &
\multicolumn{1}{c}{(8)} \\
\hline
      $\log \mdyn$ &   $\log M_{\rm *,Bell,opt}$ &                     all galaxies & $  10.07$ & $ 10.667$\,($  0.014$) & $  0.779$\,($  0.038$) & $  0.172$\,($  0.015$) &  $  0.202$ \\
                --- &                         --- &         lower 1/3 in $\Delta\rd$ & $  10.07$ & $ 10.495$\,($  0.023$) & $  0.921$\,($  0.061$) & $  0.137$\,($  0.018$) &  $  0.174$ \\
                --- &                         --- &        middle 1/3 in $\Delta\rd$ & $  10.07$ & $ 10.676$\,($  0.010$) & $  0.783$\,($  0.030$) & $  0.062$\,($  0.020$) &  $  0.114$ \\
                --- &                         --- &         upper 1/3 in $\Delta\rd$ & $  10.07$ & $ 10.822$\,($  0.020$) & $  0.703$\,($  0.044$) & $  0.125$\,($  0.028$) &  $  0.151$ \\
\hline
      $\log \mdyn$ &    $\log M_{\rm *,MPA,opt}$ &                     all galaxies & $   9.93$ & $ 10.661$\,($  0.015$) & $  0.650$\,($  0.029$) & $  0.169$\,($  0.016$) &  $  0.202$ \\
                --- &                         --- &         lower 1/3 in $\Delta\rd$ & $   9.93$ & $ 10.491$\,($  0.023$) & $  0.745$\,($  0.054$) & $  0.129$\,($  0.028$) &  $  0.180$ \\
                --- &                         --- &        middle 1/3 in $\Delta\rd$ & $   9.93$ & $ 10.669$\,($  0.013$) & $  0.677$\,($  0.030$) & $  0.000$\,($  0.029$) &  $  0.116$ \\
                --- &                         --- &         upper 1/3 in $\Delta\rd$ & $   9.93$ & $ 10.815$\,($  0.019$) & $  0.588$\,($  0.035$) & $  0.123$\,($  0.050$) &  $  0.141$ \\
\hline
\hline
  $\log \mdynratio$ &   $\log M_{\rm *,Bell,opt}$ &                     all galaxies & $  10.07$ & $  0.600$\,($  0.014$) & $ -0.225$\,($  0.037$) & $  0.168$\,($  0.015$) &  $  0.202$ \\
                --- &                         --- &         lower 1/3 in $\Delta\rd$ & $  10.07$ & $  0.428$\,($  0.023$) & $ -0.080$\,($  0.061$) & $  0.136$\,($  0.019$) &  $  0.174$ \\
                --- &                         --- &        middle 1/3 in $\Delta\rd$ & $  10.07$ & $  0.608$\,($  0.011$) & $ -0.215$\,($  0.030$) & $  0.057$\,($  0.028$) &  $  0.114$ \\
                --- &                         --- &         upper 1/3 in $\Delta\rd$ & $  10.07$ & $  0.753$\,($  0.018$) & $ -0.309$\,($  0.040$) & $  0.113$\,($  0.028$) &  $  0.150$ \\
\hline
  $\log \mdynratio$ &                 $\sigmaopt$ &                     all galaxies & $   7.33$ & $  0.603$\,($  0.009$) & $ -0.419$\,($  0.027$) & $  0.097$\,($  0.012$) &  $  0.145$ \\
\hline
\hline
$\log \mdynratiobar$ &  $\log M_{\rm *,Bell,opt}$ &                     all galaxies & $  10.07$ & $  0.358$\,($  0.017$) & $ -0.004$\,($  0.045$) & $  0.221$\,($  0.023$) &  $  0.232$ \\
                --- &                         --- &         lower 1/3 in $\Delta\rd$ & $  10.07$ & $  0.199$\,($  0.026$) & $  0.170$\,($  0.084$) & $  0.203$\,($  0.035$) &  $  0.215$ \\
                --- &                         --- &        middle 1/3 in $\Delta\rd$ & $  10.07$ & $  0.362$\,($  0.017$) & $ -0.011$\,($  0.041$) & $  0.127$\,($  0.018$) &  $  0.154$ \\
                --- &                         --- &         upper 1/3 in $\Delta\rd$ & $  10.07$ & $  0.499$\,($  0.026$) & $ -0.098$\,($  0.060$) & $  0.190$\,($  0.029$) &  $  0.206$ \\
\hline
$\log \mdynratiobar$ &           $\log \sigmabar$ &                     all galaxies & $   7.33$ & $  0.470$\,($  0.014$) & $ -0.432$\,($  0.045$) & $  0.140$\,($  0.013$) &  $  0.177$ \\
\hline
\end{tabular}
\label{tab:mdyn}
\end{center}
\begin{flushleft}
Notes. --- { 
Col. (1): $y$ is the dependent variable in the fit.
Col. (2): $x$ is the independent variable in the fit.
Col. (3): Subsample or bin used for the fit.
Col. (4): $x_{\rm p}$ is the pivot value for $x$.
Cols. (5-7): Best-fit parameters and their 1-$\sigma$ uncertainties: $a$ is the zero-point, $b$ is the slope, and $\tilde{\sigma}$ is the intrinsic Gaussian scatter.
Col. (8): $\sigma_{\rm meas}$ is the measured scatter in the mean relation, defined to be the RMS of the residuals $(\Delta y)_i = y_i - [a + b \times (x-x_{\rm p})]$.
Masses are in units of $M_\odot$, and $\sigmaopt$ and $\sigmabar$ are in units of $M_\odot\, {\rm kpc}^{-2}$.}
\end{flushleft}
\end{table*}%

\section{Summary and Discussion}
\label{sec:summ}

\subsection{Methodology}
\label{subsec:disc_meth}

Our sample selection and methodology are dictated by our primary goal of obtaining photometric estimates of disk rotation velocity for a large sample of galaxies with imaging data. To achieve this goal, we construct a TFR sample of 189 galaxies that is a fair subsample of a well-defined parent disk sample of $\sim$170~000 galaxies from the SDSS (\S\ref{sec:samp_sele}). 

To derive robust TFRs, we use rotation velocity amplitudes $V_{80}$ defined at $R_{80}$, the radius containing 80 per cent of the $i$-band galaxy light and approximately at the peak of the total (disk+DM halo) rotation curve for typical disk galaxies and their haloes (c.f. \S\ref{subsubsec:syst_fits} and \S\ref{subsec:alt_vrot}). We calculate and account for systematic errors in $V_{80}$, in addition to statistical errors, to downweight the contribution of those few galaxies with peculiar rotation curves to the TFR fits. We also perform weighted fits to the TFR to account for the small difference between the stellar mass functions of the TFR and parent disk samples; however, in practice, the two are similar enough that the weights do not significantly affect the fits.

\subsection{Tests of systematics}
\label{subsec:disc_syst}

We find that the TFR is robust to many systematic effects, including slit misalignments (\S\ref{subsubsec:syst_pos}), differences in observing instruments and conditions (\S\ref{subsubsec:syst_piz}), differences in analysis pipelines (i.e., the rotation curve extraction and fitting methods, and bulge-disk decomposition fits, which differ between this work and P07) (\S\ref{subsubsec:syst_fits} and \S\ref{subsec:alt_P07}), and differences in sample selection (i.e., between this work and P07) (\S\ref{subsec:alt_extcorr} and \S\ref{subsec:alt_P07}). We also verify that our internal extinction corrections successfully remove the weak correlation between disk axis ratios $\qd$ and velocity residuals from the (uncorrected) $\minc$ ITFR (\S\ref{subsec:correl_axisratio}). This indicates that even if they are uncertain on a galaxy-per-galaxy basis, they are adequate for the TFR sample on average. 

While there are many other potential sources of systematics that we have not explicitly tested for, we note that the observed small scatter in the TFR--- 0.056 dex of total scatter in $\log V_{80}$, corresponding to $0.20$ dex in $\log \mbell$ or 0.5 mag in $M_i$--- itself serves as an upper bound for their contribution. We find that the expected additional sources of scatter (e.g., intrinsic disk ellipticities) can explain essentially all of the observed scatter in the TFR, regardless of whether our measurement errors were overestimated (\S\ref{subsec:tfr_interpret}). The disk-dominated sample yields an even smaller total TFR scatter of $0.05$ dex in $\log V_{80}$, corresponding to 0.18 dex in $\log\mbell$ (or 0.44 in magnitude ``units''). For comparison, previous TFR studies based on pruned samples of spirals (for distance indicator work) found a range of observed total scatter $\sim 0.3$--0.4 mag \citep[][and references therein]{1995PhR...261..271S}.

\subsection{Optimal estimator of disk rotation velocity}
\label{subsec:disc_optimal}

One of the primary aims of this work is to identify and calibrate an optimal estimator of disk rotation velocity, i.e., one with minimal scatter in its ITFR. We confirm that the use of redder bands  \citep{1995PhR...261..271S}, or the addition of colour information \citep{2002AJ....123.2358K}, yield tighter ITFRs. It is interesting to note that $\msyn(\lambda,g-r)$ for different bands yielded not only ITFRs with identical scatter, but almost identical slopes and zero-points as well. We note that $\msyn$ can be interpreted as an extrapolation to some redder (i.e., infrared) band, which traces stellar mass better than any single optical band. We also find that the best-fit coefficients $\alpha$ for the colour terms in $\msyn(\lambda,g-r)$ are consistent with the corresponding coefficients $\alpha^{\rm Bell}$ from the $M_*/L$ fitting formulae of Bell et~al. (2003) (see Table~\ref{tab:itfr_synmag}). Then, in \S\ref{subsec:correl_colour}, we explicitly show that almost all of the correlation between $g-r$ colour and velocity residuals from the $M_i$ ITFR can be attributed to, and hence removed by, the colour dependence of the stellar mass-to-light ratio, $M_*/L_i \propto (g-r)^{0.86}$ (also from Bell et~al. 2003). 
 
Ultimately, we choose $\mbell$ to be our optimal estimator since it has a natural physical interpretation (unlike $\msyn$) and is straightforward to calculate (unlike $\mmpa$). We find (c.f. Table~\ref{tab:itfr_masses} and Fig.~\ref{fig:linfits_obs3}):
\beqa \nonumber
\log V_{80}(\mbell) &=& (2.142 \pm 0.004) + (0.278 \pm 0.010) \\ &\times& (\log \mbell-10.102),
\eeqa
with an intrinsic Gaussian scatter $\tildesig=0.036\pm 0.005$ in $\log V_{80}$ (using a Kroupa IMF to define $\mbell$). The associated uncertainty in this estimate is given by the measured scatter in the ITFR, $\sigmeas=0.056$ dex. 

Pruning the galaxy sample to include only disk-dominated galaxies is another way to obtain a tighter ITFR. We show in \S\ref{subsec:alt_dtr} that removing galaxies with $D/T\le 0.9$ yields a tighter $\mbell$ ITFR. We note that pruning a large imaging dataset with many fainter and smaller galaxies than those in our sample is nontrivial, but worth investigating. 

\subsection{Relative contributions of stars and DM}
\label{subsec:disc_relative}

The question of the relative contribution of stars and DM in the optical regions of disk galaxies remains unresolved. Observed rotation curves only weakly constrain the relative contributions of the different components: stellar and gas disks, bulge, and DM halo. To disentangle these components, we assume a fixed stellar IMF, then infer stellar and gas masses based on the galaxy luminosities and colours. We use our observations to confirm several known trends and point out interesting new ones that will help constrain the possible scenarios. In future work, we will aim to construct disk galaxy models that reproduce these observations.

In \S\ref{subsec:correl_size}, we study correlations between velocity residuals from the $M_*$ ITFR and disk size offsets $\Delta\rd$ for the child disk sample divided into three bins in $M_*$ (Figure~\ref{fig:resfits_rdphys_bins}). We find that the lowest stellar mass galaxies show a positive correlation between TFR residuals and disk size offsets, suggesting that they have a dominant NFW DM halo. Moreover, the sign of the correlation changes from positive to negative toward increasing stellar mass, suggesting a progression toward an increasing contribution of the stellar disk. (Recall that a pure NFW DM halo yields a slope of $+0.5$, while a pure exponential disk yields a slope of $-0.5$.) 

In \S\ref{sec:mass_ratios}, we use stellar mass estimates $\mbell$ and baryonic mass estimates $\mbar$ (defined in \S\ref{subsec:phot_mstr} and \S\ref{subsec:phot_mbar}, respectively) with a fixed Kroupa IMF, to calculate dynamical-to-stellar and dynamical-to-baryonic mass ratios within $R_{80}$, $\mdynratioopt$ and $\mdynratiobaropt$, respectively. We find that $\mdynratioopt$ decreases from $\approx$ 10 to 3 as $\mbell$ increases from $10^9$ to $10^{11} M_\odot$ (top panel of Fig.~\ref{fig:mdynratio_mstr}). This trend of an increasing (decreasing) contribution of stars (DM) to the potential on optical scales as stellar mass increases is consistent with that indicated by the size residual correlations. 

On the other hand, because low $M_*$ galaxies have higher gas mass fractions than high $M_*$ galaxies, $\mdynratiobaropt$ is roughly constant with stellar mass, with a mean of $2.6$ and standard deviation of 0.23 dex (bottom panel of Fig.~\ref{fig:mdynratio_mstr}). \citet{2008ApJ...682..861B} found a similar result from a sample of isolated dwarf galaxies with luminosities $M_r-5\log h=-14$ to $-21.5$ mag. 
At face value, it seems that this finding is inconsistent with the observed dependence of the sign of the size residual correlations on stellar mass (\S\ref{subsec:correl_size}). However, recall that we have not taken into account the difference in the spatial distributions of the gas and the stars. In fact, $\mdynratiobaropt$ and $\mdynratioopt$ bracket the more realistic scenario, in which only some fraction of the gas mass is within the optical radius. We defer detailed analysis and discussion to a future modelling paper.

Panels of Figure~\ref{fig:mdynratio_mstr} clearly show that the scatter in both $\mdynratioopt$ and $\mdynratiobaropt$ at fixed stellar mass is driven by differences in disk sizes: larger disks tend to have higher values in both ratios. As a consequence, both ratios form tighter relations with stellar (or baryonic) surface density than with stellar mass or disk size alone \citep{2003A&A...412..633Z,2005ApJ...633..844P}. In the disk galaxy models of \citet{2000MNRAS.315..457F}, the observed tight trend of $\mdynratioopt$ with $\sigmaopt$ arises from a correlation of star formation efficiency with surface density.\footnote{This has, in fact, motivated disk galaxy models in which the disk mass fraction (the ratio of the stellar mass to the total halo mass) is dependent on $\sigmaopt$ \citep[e.g.,][]{2007ApJ...671.1115G}.} 
Here, we propose an alternative explanation: the observed variation in $\mdynratioopt$ with disk size is actually tracing the slope of the (average) total mass profile of disk galaxies at these radii. In our own disk galaxy models, we will show that the observed trends are reproduced even if the disk mass fraction is independent of disk size or stellar surface density (R. Reyes, in prep.).


The reason this happens, qualitatively, is that if we assume a fixed DM halo for a given disk (ignoring realistic variation due to, e.g., scatter in halo concentration at fixed mass), then the larger disks will necessarily have more DM within $R_{80}$. This assumption is not necessarily true-- it is possible, for example, that a disk that is larger or smaller than the typical size for its stellar mass would have a different effect on the DM halo density profile.  Our results suggest that such a correlation between DM halo profile modifications and disk size at fixed $M_*$ are not of the right sign, or sufficient magnitude, to cause a deviation from our naive expectation that the disk size simply modulates how much of the dark matter the gas in the disk ``sees.''


\subsection{Applications to future work}
\label{subsec:disc_future}

The calibrated TFRs presented here can be used for a variety of purposes: for estimating disk rotation velocities for large photometric samples, connecting galaxies with their DM haloes, and studying models of galaxy formation and evolution. As mentioned above, we will use these observations to construct and constrain models of disk galaxy formation. In another paper, we will constrain $\vrot/\vvir$ as a function of stellar mass, using our calibrated ITFR to determine $\vrot$, and weak lensing measurements to determine $\vvir$. As discussed in \S\ref{sec:intr}, this measurement constrains the shape of the total mass profile, and can potentially indicate whether baryons have modified the gravitational potential well, in which sense, and by how much. 

\section*{Acknowledgments}
We thank M.A. Strauss for helpful comments on the paper. J.E.G. acknowledges the support of his NSF grant, AST0908368.

\bibliography{tullyfisher,sdss,rr,tf_intro}

\section*{Appendix}
In this Appendix, we derive the relation between dynamical-to-stellar mass ratio within a radius $R$ $(\mdynratio)(R)$ and the observed rotation velocity at that radius $\vrot$ (Eq.~\ref{eq:mdynratio}). First, we consider an idealized galaxy model with two components: an exponential stellar disk and a NFW DM halo (we neglect the bulge for now, for simplicity). The disk has an infinitesimal thickness, scale length $\rd$, central surface mass density $\Sigma_0$, and a surface mass density distribution, $\Sigma_{\rm d}(R)=\Sigma_0 \exp(-R/\rd)$. The mass interior to $R$ is given by
\beqa \nonumber 
M_{\rm d}(R) &=& 2\pi \Sigma_0 \rd^2 \left[1-\exp(-R/\rd)\left(1+R/\rd\right)\right] \\ \label{eq:def_F}
&\equiv& 2\pi \Sigma_0 \rd^2 {\cal F}(R/\rd).
\eeqa
and the rotation curve $V_{\rm c,d}(R)$ is given by \citep{1970ApJ...160..811F} 
\beqa \nonumber
V_{\rm c,d}^2(R) &=& 4\pi G \Sigma_0 \rd y^2 [I_0(y) K_0(y) - I_1(y) K_1(y)] \\ \label{eq:def_B}
&\equiv& 4\pi G \Sigma_0 \rd {\cal B}(y), 
\eeqa
where $y\equiv R/(2\rd)$, $G=4.3012 \times 10^{-6}$ is the gravitational constant in units of kpc (km/s)$^2 M_\odot^{-1}$, and $I_i$ and $K_i$ are the modified Bessel functions of the first and second kind. The disk rotation curve will peak at $R_{\rm disk, max} = 2.15 \rd$ and the peak velocity is about 15\% higher than that of a spherical mass distribution with the same interior mass. 

The NFW halo has an interior mass $M_{\rm NFW}(R)$ and a rotation curve $V_{\rm c,NFW}(R)$. If one (incorrectly) assumed that the observed (i.e., total) rotation velocity $\vrot$ arises from a spherical mass distribution, one would infer a dynamical mass $M_{\rm sph}$ interior to $R$ given by
\beqa
M_{\rm sph}(R) &\equiv& \frac{R\vrot^2(R)}{G} = \frac{R(V_{\rm c,NFW}^2 + V_{\rm c,d}^2)}{G} \\ \label{eq:def_K}
	&=& M_{\rm NFW}(R) + {\cal K}(R/\rd) M_{\rm d}(R),
\eeqa
where we have introduced a geometric correction factor 
\beqa
{\cal K}(R/\rd) &\equiv& \frac{V_{\rm c,d}^2(R)}{V_{\rm c,d,sph}^2(R)} 
	= \frac{4\pi G \Sigma_0 \rd {\cal B}(y)}{G M_{\rm d}(R)/R} \\
	&=& \frac{4\pi G \Sigma_0 \rd {\cal B}(y)}{2\pi G \Sigma_0 \rd^2 {\cal F}(R/\rd)/R} \\
	&=& \frac{2 {\cal B}(y)}{{\cal F}(R/\rd)} \frac{R}{\rd}.
\eeqa
Here, $V_{\rm c,d,sph}$ is the rotation curve for a spherical mass distribution with the same interior mass as the exponential disk, and ${\cal F}$ and ${\cal B}$ are functions defined in Eqs.~\ref{eq:def_F} and \ref{eq:def_B}, respectively. We find that ${\cal K}$ peaks at around $\sim 3\rd$, with a value of 1.3, and then falls off slowly and asymptotes to 1. For $R/\rd = 2.2$, $({\cal F},{\cal B},{\cal K})=(0.645,0.1935,1.319)$; for $R/\rd=3$, the values are (0.8009,0.1795,1.345).

The dynamical-to-stellar mass ratio is therefore related to its ``spherical'' counterpart by
\beqa
\frac{M_{\rm dyn}(R)}{M_{\rm d}(R)} &=& 1+ \frac{M_{\rm NFW}(R)}{M_{\rm d}(R)} \\
	&=& \frac{M_{\rm sph}(R)}{M_{\rm d}(R)} + \left[ 1-{\cal K}\left(\frac{R}{\rd}\right) \right].
\eeqa
In other words, if one assumes a spherical mass distribution, one would falsely attribute a larger dynamical mass to the stellar disk, and therefore overestimate the total-to-stellar mass ratio by a constant additive term ${\cal K}(R/\rd)-1 \approx 0.3$ for $R/\rd=2-3$. 

We now generalize the above results for a model galaxy with an additional bulge component. Since the bulge mass distribution is approximately spherical, it can be lumped together with the halo mass in Eq.~\ref{eq:def_K}. Defining $M_{\rm star}\equiv M_{\rm d}+M_{\rm bulge}$, we have
\beqa \nonumber
\frac{M_{\rm dyn}(R)}{M_{\rm star}(R)} &=&  \frac{[M_{\rm NFW}(R)+M_{\rm bulge}(R)]+M_{\rm d}(R)}{M_{\rm star}(R)} \\ \nonumber
 &=& \frac{M_{\rm sph}(R)}{M_{\rm star}(R)} \\ 
 &+& \left(\frac{M_{\rm d}(R)}{M_{\rm star}(R)}\right)\times  \left[1- {\cal K}\left(\frac{R}{\rd}\right) \right].
\eeqa
In practice, we adopt the radius $R=R_{80}$ and $M_{\rm d}(R_{80})/M_{\rm star}(R_{80}) = (D/T)_{80}$, the disk-to-total light ratio interior to $R_{80}$ (Eq.~\ref{eq:dtr80}). The correction is suppressed in galaxies with significant bulges, because the contribution of the disk is smaller than what was assumed when all the light was attributed to the disk.

\end{document}